\documentclass[10pt]{iopart}

\usepackage{collapsecite}

\usepackage{iopams}
\usepackage{graphics}
\usepackage{epsfig}

\def\sech{{\rm sech}}
\def\csch{{\rm csch}}
\def\rhon{\rho}

\begin{document}

\title[Nonlinear Waves in Bose-Einstein Condensates]
{Nonlinear Waves in Bose-Einstein Condensates:\\
Physical Relevance and Mathematical Techniques}

\author{R.~Carretero-Gonz\'alez$^{1}$\footnote{URL: {\tt http://rohan.sdsu.edu/$\sim$rcarrete/}}, 
D.J.~Frantzeskakis$^2$, and P.G. Kevrekidis$^3$
}
\address{$^1$
Nonlinear Dynamical Systems Group{\footnote{URL: {\tt http://nlds.sdsu.edu/}}},
and Computational Science Research Center\footnote{URL: {\tt http://www.csrc.sdsu.edu/csrc/}},
Department of Mathematics and Statistics,
San Diego State University, San Diego CA, 92182-7720.
}

\address{$^2$
Department of Physics, University of Athens, Panepistimiopolis,
Zografos, Athens 15784, Greece.
}

\address{$^3$
Department of Mathematics and Statistics, University of
Massachusetts, Amherst MA 01003-4515.
}


\begin{abstract}
The aim of the present review is to introduce the reader to some of
the physical notions and of the mathematical methods that are relevant to
the study of nonlinear waves
in Bose-Einstein Condensates (BECs). 
Upon introducing the general framework, we discuss the prototypical
models that are relevant to this setting for different dimensions
and different potentials confining the atoms. We analyze some of
the model properties and explore their typical wave solutions
(plane wave solutions, bright, dark, gap solitons, as well as vortices).
We then offer a collection of mathematical methods that can be used
to understand the existence, stability and dynamics of nonlinear waves
in such BECs, either directly or 
starting from different types of limits (e.g., the linear
or the nonlinear limit, or the discrete limit of the corresponding equation).
Finally, we consider some special topics involving more recent developments,
and experimental setups in which there is still considerable 
need for developing mathematical as well as computational tools. 
\end{abstract}

\vspace{2pc}

\vspace{2pc}
\noindent{To appear in {\it Nonlinearity}, 2008}

\maketitle

{
\tableofcontents
}

\section*{Abbreviations}
\begin{itemize}
\item AC: Anti-Continuum (Limit)
\item BEC: Bose-Einstein condensate
\item BdG: Bogoliubov-de Gennes (Equations)
\item cqNLS: cubic-quintic NLS
\item DNLS: Discrete Nonlinear Schr{\"o}dinger (Equation)
\item EP: Ermakov-Pinney (Equation)
\item GP: Gross-Pitaevskii (Equation)
\item KdV: Korteweg-de Vries (Equation)
\item LS: Lyapunov-Schmidt (Technique)
\item MT: Magnetic Trap
\item NLS: Nonlinear Schr{\"o}dinger (Equation)
\item NPSE: Non-polynomial Schr{\"o}dinger Equation
\item ODE: Ordinary Differential Equation
\item OL: Optical Lattice
\item PDE: Partial Differential Equation
\item RPM: Reductive Perturbation Method
\item TF: Thomas-Fermi
\end{itemize}

\section{Introduction\label{Sec:intro}}

The phenomenon of Bose-Einstein condensation is
a quantum phase transition originally predicted by Bose 
\cite{chap01:bose} and Einstein \cite{chap01:einsteinA,chap01:einsteinB} in 1924. 
In particular, it was shown that below a critical transition temperature $T_c$, 
a finite fraction of particles of a boson gas (i.e., whose particles obey the Bose statistics) {\it condenses} 
into the same quantum state, known as the {\it Bose-Einstein condensate} (BEC).
Although Bose-Einstein condensation is known to be a fundamental phenomenon, connected, e.g., to 
superfluidity in liquid helium and superconductivity in metals (see, e.g., Ref.~\cite{chap01:becgss}), 
BECs were experimentally realized 70 years after their theoretical prediction: 
this major achievement took place in 1995, when different species of dilute alkali vapors confined in a 
magnetic trap (MT) were cooled down to extremely low temperatures  
\cite{chap01:anderson,chap01:davis,chap01:bradley}, and has already been recognized 
through the 2001 Nobel prize in Physics \cite{chap01:rmpnlA,chap01:rmpnlB}. 
This first unambiguous manifestation of a macroscopic quantum state in a many-body system 
sparked an explosion of activity, as reflected by the publication of several thousand papers 
related to BECs since then. Nowadays there exist more than fifty experimental BEC groups around 
the world, while an enormous amount of theoretical work has followed and driven 
the experimental efforts, with an impressive impact on many branches of Physics.

From a theoretical standpoint, and for experimentally relevant conditions, the static and dynamical properties 
of a BEC can be described by means of an effective mean-field equation known as the Gross-Pitaevskii
(GP) equation \cite{chap01:gross,chap01:pittaevskii}. This is a variant of the famous nonlinear
Schr{\"o}dinger (NLS) equation \cite{chap01:sulem} (incorporating an external potential used to confine the condensate), 
which is known to be a universal model describing the evolution of complex field envelopes in nonlinear dispersive media \cite{dodd}. 
As such, the NLS equation is a key model appearing in a variety of physical contexts, 
ranging from optics \cite{has1,ab1,has2,chap01:kiag}, to fluid dynamics and plasma physics \cite{infeld}, while it has 
also attracted much interest from a mathematical viewpoint \cite{chap01:sulem,chap01:ablowitz,chap01:bourgain}. 
The relevance and importance of the NLS model is not limited to the case of conservative systems and the 
theory of solitons \cite{dodd,infeld,segur,zakhbook,acnewell}; in fact, the NLS equation is directly connected to 
dissipative universal models, such as the complex Ginzburg-Landau equation \cite{hoh1,aranson}, 
which have been studied extensively in the context of pattern formation \cite{hoh2} (see also Ref.~\cite{nonlinsc} 
for further discussion and applications). 

In the case of BECs, the nonlinearity in the GP (NLS) model is introduced by 
the interatomic interactions, accounted for through an effective mean-field. Importantly, the mean-field approach, 
and the study of the GP equation, 
allows the prediction and description of important, and experimentally relevant, 
nonlinear effects and nonlinear waves, such as solitons and vortices. 
These, so-called, {\it matter-wave} solitons and vortices can be viewed as fundamental nonlinear excitations of BECs,
and as such have attracted considerable attention. Importantly, they have also been observed in many elegant experiments 
using various relevant techniques. These include, among others, phase engineering of the condensates in order to create vortices
\cite{chap01:vort1,chap01:williams} or dark matter-wave solitons in them \cite{chap01:denschl,chap01:dutton,chap01:dark1,chap01:dark,engels}, 
the stirring (or rotation) of the condensates providing angular momentum creating vortices
\cite{chap01:vort2,chap01:vort3} and vortex-lattices \cite{chap01:latt1,chap01:latt2,chap01:latt3},
the change of scattering length (from repulsive to attractive via Feshbach resonances) 
to produce bright matter-wave solitons and soliton trains \cite{chap01:bright1,chap01:bright2,chap01:bright3,chap01:njp2003b} 
in attractive condensates, or set into motion a repulsive BEC trapped in a 
periodic optical potential referred to as optical lattice 
to create gap matter-wave solitons \cite{chap01:gap}. As far as vortices and vortex lattices are concerned, it should be noted
that their description and connection to phenomena as rich and profound as superconductivity and
superfluidity, were one of the themes of the Nobel prize in Physics in 2003.

The aim of this paper is to give an overview of some physical and mathematical aspects of the theory of BECs. 
The fact that there exist already a relatively large number of reviews 
\cite{pw,chap01:reviewA,chap01:leg,fetterrev,And,bs,chap01:elieb} 
and textbooks \cite{chap01:becgss,book1,book2,asp,BECBOOK} devoted in the Physics of BECs, and given  
the space limitations of this article, will not allow us to be all-inclusive. Thus, this 
review naturally entails a personalized perspective on BECs, with a special emphasis on the nonlinear
waves that arise in them. In particular, our aim here is to present an overview of both the physical setting and, 
perhaps more importantly, of the mathematical techniques from dynamical systems and nonlinear dynamics 
that can be used to address the dynamics of nonlinear waves in such a setting. 
This manuscript is organized as follows. 

Section~\ref{Sec:GPE} is devoted to the mean-field description of BECs, 
the GP model and its properties. In particular, we present the GP equation and discuss its variants in 
the cases of repulsive and attractive interatomic interactions and how to
control them via Feshbach resonances. We also describe the ground state properties of BECs and their
small-amplitude excitations via the Bogoliubov-de Gennes equations. Additionally, we present 
the types of the external confining potential and how their form leads to 
specific types of simplified mean-field descriptions. 

Section~\ref{Sec:dim_reduc}
describes the reduction of the spatial dimensionality of the BEC
by means of effectively suppressing one or two transverse
directions. This can be achieved  by ``tightening'' 
the external confining potential (usually a harmonic magnetic trap)
along these directions. 
We introduce the basic nonlinear structures (dark and bright
solitons) that are ubiquitous to one-dimensional settings.
The different types of nonlinearities that arise
from different approximations due to the dimensionality reduction
are discussed.
We also present the dimensionality reduction in the
presence of external periodic potentials generated by
the optical lattices (which are created as interference patterns of
multiple laser beams) and the discrete limit, the
discrete nonlinear Schr\"odinger equation, that they
entail for strong potentials.

Section~\ref{Sec:math}
deals with the mathematical methods used to describe nonlinear waves in BECs.
The presentation concerns four categories of methods, depending on the particular features of the model at hand. 
The first one concern ``direct'' methods, which analyze the nonlinear mean-field models  
directly, without employing techniques based on some appropriate, physically relevant and mathematically tractable limit. 
Such approaches include, for example the method of moments, self-similarity and rescaling methods, 
or the variational techniques among others. 
The second one will concern methods that make detailed use of the
understanding of the {\it linear limit} of the problem (e.g., the linear Schr\"{o}dinger equation 
in the presence of a parabolic, periodic, or a double-well potential). The third category of the 
mathematical methods entails 
perturbation techniques from the {\it fully nonlinear limit} of the system (e.g., the integrable
NLS equation), while the fourth one concerns discrete systems (relevant to BECs trapped in strong optical lattices), 
where perturbation methods from the so-called anti-continuum limit are extremely helpful. 

Finally, in Sec.~\ref{Sec:special}
we present some special topics that have recently attracted much physical interest, both theoretical and 
experimental. These include multicomponent and spinor condensates described by systems of coupled GP equations, shock waves,
as well as nonlinear structures arising in higher-dimensions, such as vortices and vortex lattices in BECs, 
and multidimensional solitons (including dark and bright ones). We also 
briefly discuss the manipulation of matter-waves 
by means of various techniques based on the appropriate control of the external potentials. In that same context, the effect of disorder on the matter-waves is studied.
Finally, we touch upon 
the description of BECs beyond mean-field theory, presenting relevant theoretical models that have recently 
attracted attention.


\label{intro}

\section{The Gross-Pitaevskii (GP) mean-field model\label{Sec:GPE}}

\subsection{Origin and fundamental properties of the GP equation}
We consider a sufficiently dilute ultracold atomic gas, composed by $N$ interacting bosons of mass $m$ 
confined by an external potential $V_{\rm ext}({\bf r})$. Then, the many-body Hamiltonian of the system 
is expressed, in second quantization form, through the boson annihilation and creation field operators, 
${\hat \Psi}({\bf r},t)$ and ${\hat \Psi}^{\dagger}({\bf r},t)$, as \cite{chap01:reviewA,book2}, 
\begin{eqnarray}
\hskip-2.35cm
{\hat H} &=&  \int  d{\bf r} {\hat \Psi}^{\dagger}({\bf r}, t) {\hat H}_{0}
{\hat \Psi}({\bf r}, t)
+ \frac{1}{2} \int d{\bf r} d{\bf r'} {\hat \Psi}^{\dagger}({\bf r}, t)
{\hat \Psi}^{\dagger}({\bf r'}, t) V({\bf r}-{\bf r}'){\hat \Psi}({\bf r}', t) {\hat \Psi}({\bf r}, t),
\label{H}
\end{eqnarray}
where ${\hat H}_{0}= - (\hbar^{2}/2m) \nabla^2 + V_{\rm ext}({\bf r})$ is the ``single-particle'' operator 
and $V({\bf r}-{\bf r}')$ is the two-body interatomic potential. 
The mean-field approach is based on the so-called Bogoliubov approximation, first formulated by Bogoliubov in 1947 \cite{chap01:bog}, 
according to which the condensate contribution is separated from the boson field operator as 
$
{\hat \Psi}({\bf r}, t) = \Psi ({\bf r}, t) + {\hat \Psi}'({\bf r}, t).
$
In this expression, the complex function 
$\Psi ({\bf r}, t) \equiv \langle  {\hat \Psi}({\bf r}, t) \rangle $ (i.e., the expectation value of
the field operator), is commonly known as the {\it macroscopic wavefunction of the condensate}, 
while ${\hat \Psi}'({\bf r'}, t)$ describes the
non-condensate part, which, at temperatures well below $T_c$, is actually negligible
(for generalizations accounting for finite temperature effects 
see Sec.~\ref{Sec:beyond}). Then, the above prescription 
leads to a nontrivial zeroth-order theory for the BEC wavefunction 
as follows: First, from the Heisenberg evolution equation
$i \hbar (\partial {\hat \Psi} /\partial t) = [ {\hat \Psi}, {\hat H}]$
for the field operator ${\hat \Psi}({\bf r}, t)$,
the following equation is obtained:
\begin{eqnarray}
i \hbar \frac{\partial}{\partial t} {\hat \Psi}({\bf r}, t)
= \left[ {\hat H}_{0} 
+ \int d{\bf r'} {\hat \Psi}^{\dagger}({\bf r'}, t)
V({\bf r'}-{\bf r}){\hat \Psi}({\bf r'}, t)  \right] {\hat \Psi}({\bf r}, t).
\label{chap01:hem}
\end{eqnarray}
Next, considering the case of a dilute ultracold gas with binary collisions at low energy, characterized by the $s$-wave scattering length $a$, 
the interatomic potential can be replaced by an effective delta-function interaction potential,
$
V({\bf r'}-{\bf r}) = g \delta({\bf r'}-{\bf r})
$
\cite{chap01:reviewA,chap01:leg,book1,book2}, with the coupling constant (i.e., the nonlinear coefficient) $g$ given by
$g = {4 \pi \hbar^2 a}/{m}$.
%
%
Finally, employing this effective interaction potential, 
and replacing the field operator
${\hat \Psi}$ with the classical field $\Psi$, Eq.~(\ref{chap01:hem}) yields the GP equation,
\begin{equation}
i \hbar \frac{\partial}{\partial t} \Psi({\bf r}, t) = \left[ -\frac{{\hbar}^2}{2 m} \nabla^2
+ V_{{\rm ext}} ({\bf r}) + g |\Psi({\bf r}, t)|^2 \right] \Psi({\bf r}, t).
\label{chap01:peq1}
\end{equation}
The complex function $\Psi$ in the GP Eq.~(\ref{chap01:peq1}) can be expressed in terms of the density
$\rhon({\bf r}, t) \equiv |\Psi ({\bf r}, t)|^2$, and phase $S({\bf r}, t)$ of the condensate as
%
$\Psi({\bf r}, t) = \sqrt{\rhon({\bf r}, t)} \exp \left[iS({\bf r}, t)\right]$.
%
Note that the current density 
${\bf j} = \frac{\hbar}{2mi} (\Psi^{\ast}\nabla \Psi - \Psi \nabla \Psi^{\ast})$
(asterisk denotes complex conjugate), assumes a hydrodynamic form ${\bf j}=\rhon{\bf v}$,
with an atomic velocity 
%
${\bf v}({\bf r}, t) = \frac{\hbar}{m} \nabla S({\bf r}, t)$.  
%
The latter is irrotational (i.e., $\nabla\! \times\! {\bf v} = 0$), which is a typical feature 
of {\it superfluids}, and satisfies the famous Onsager-Feynman quantization condition $\oint_{C} d{\bf l} \cdot {\bf v} = (\hbar/m) \cal{N}$, 
where $\cal{N}$ is the number of {\it vortices} enclosed by the contour $C$ (the circulation is obviously zero for a simply connected geometry).

For time-independent external potentials, the GP model possesses two integrals of motion, namely, 
the total number of atoms, 
%
\begin{equation}
N = \int |\Psi({\bf r}, t)|^2 d{\bf r}, 
\label{chap01:N}
\end{equation}
%
%
%
%
and the energy of the system,
\begin{equation}
E = \int d{\bf r} \left[ \frac{\hbar^2}{2m} |\nabla \Psi|^2 + V_{\rm ext} |\Psi|^2 + \frac{1}{2}g |\Psi|^4 \right],
\label{chap01:E}
\end{equation}
with the three terms in the right-hand side representing, respectively, the kinetic energy, the
potential energy and the interaction energy. 

A time-independent version of the GP Eq.~(\ref{chap01:peq1}) can readily be obtained upon expressing the condensate wave function as
$\Psi({\bf r}, t)=\Psi_{0}({\bf r}) \exp(-i\mu  t/\hbar)$, where
$\Psi_0$ is a function normalized to the number of atoms
($N=\int d{\bf r}\, |\Psi_{0}|^{2}$) and $\mu = \partial E / \partial N$ is the 
chemical potential. Substitution of the above expression into the GP Eq.~(\ref{chap01:peq1}) yields the
following steady state equation for $\Psi_{0}$:
\begin{equation}
\left[-\frac{\hbar^2}{2m} \nabla^2 + V_{\rm ext}({\bf r})+g |\Psi_{0}|^{2}({\bf r})\right]\Psi_{0}({\bf r})
=\mu\Psi_{0}({\bf r}).
\label{chap01:statGPE}
\end{equation}
Equation~(\ref{chap01:statGPE}) is useful for the derivation of stationary solutions of the 
GP equation, including the {\it ground state} of the system (see Sec.~\ref{Sec:GPE}.5).


\subsection{The GP equation vs.~the full many-body quantum mechanical problem}

It is clear that the above mean-field approach and the analysis of the pertinent GP Eq.~(\ref{chap01:peq1})
is much simpler than a treatment of the full many-body Schr{\"o}dinger equation. However, a quite important 
question is if the GP equation can be derived rigorously from a self-consistent treatment of the respective 
many-body quantum mechanical problem. Although the GP equation is known from the early 1960s, this problem was 
successfully addressed only recently for the stationary GP Eq.~(\ref{chap01:statGPE}) in Ref.~\cite{chap12:Lieb1}. 
In particular, in that work it was proved that the GP energy functional describes correctly the energy and the 
particle density of a trapped Bose gas to the leading-order in the small parameter $\bar{\rhon}|a|^{3}$, where $\bar{\rhon}$ 
is the average density of the gas.%
\footnote{When ${N}|a|^{3} \ll 1$, the Bose-gas is called ``dilute'' or ``weakly-interacting''. In fact, 
the smallness of this dimensionless parameter is required for the derivation of the GP Eq.~(\ref{chap01:peq1});  
in typical BEC experiments this parameter takes values ${N}|a|^{3} < 10^{-3}$ \cite{chap01:reviewA}.}
The above results were proved in the limit where the number of particles ${N} \to \infty$ and the 
scattering length $a \to 0$, such that ${N} a$ is fixed. 
Importantly, although Ref.~\cite{chap12:Lieb1} referred to the full three-dimensional (3D) Bose gas, extensions 
of this work for lower-dimensional settings were also reported (see the review \cite{chap01:elieb} and references therein).

The starting point of the analysis of Ref.~\cite{chap12:Lieb1} is the effective Hamiltonian of ${N}$ identical bosons.
Choosing the units so that $\hbar=2m=1$, this Hamiltonian is expressed as (see also Ref.~\cite{book1}), 
\begin{equation}
\label{qmHam} 
H = \sum_{j=1}^{{N}} \left[ - \nabla^2_j+ V_{\rm ext}({\bf r}_j) \right] + 
\sum_{i < j} v(|{\bf r}_i - {\bf r}_j|),
\end{equation}
where $v(|{\bf r}|)$ is a general interaction potential assumed to be spherically symmetric and decaying faster than 
$|{\bf r}|^{-3}$ at infinity.
Then, denoting the quantum-mechanical ground-state energy of the 
Hamiltonian (\ref{qmHam}) (which depends on the number of 
particles ${N}$ and the dimensionless%
\footnote{The dimensionless two-body scattering length is 
obtained from the solution $u(r)$ of the 
zero-energy scattering equation $-u''(r) + \frac{1}{2} v(r) u(r) = 0$ with $u(0) = 0$ 
and is given, by definition, as $\tilde{a} = \lim\limits_{r \to \infty} \left( r - u(r)/u'(r) \right)$ 
(see also Refs.~\cite{book1,book2}).}
scattering length $\tilde{a}$) 
by $E_{\rm QM}({N},\tilde{a})$, the main theorem proved in Ref.~\cite{chap12:Lieb1} is as follows:
\begin{itemize}
\item The GP energy is the dilute limit of the quantum-mechanical energy: 
\begin{equation}
\label{convergence-1} \forall \tilde{a}_1 > 0 : \quad \lim_{n \to \infty}
\frac{1}{{N}} E_{\rm QM}\left({N},\frac{\tilde{a}_1}{n}\right) = E_{\rm
GP}(1,\tilde{a}_1),
\end{equation}
where $E_{\rm GP}({N},\tilde{a})$ is the energy of a solution of the dimensionless stationary 
GP Eq.~(\ref{chap01:statGPE}) (in units such that $\hbar=2m=1$), 
and the convergence is uniform on bounded intervals of $\tilde{a}_1$.
\end{itemize}

The above results (as well as the ones in Ref.~\cite{chap01:elieb}) were proved for stationary
solutions of the GP equation, and, in particular, for the ground state solution. More recently, 
the time-dependent GP Eq.~(\ref{chap01:peq1}) was also analyzed within a similar asymptotic 
limit (${N} \to \infty$) in Ref.~\cite{Erdos}. In this work, it was proved that the
limit points of the $k$-particle density matrices of $\Psi_{{N},t}$ 
(which is the solution of the ${N}$-particle 
Schr\"{o}dinger equation) satisfy asymptotically the GP equation (and the associated hierarchy of equations) 
with a coupling constant given by $\int v(x)dx$, where $v(x)$ describes the interaction potential. 

Thus, these recent rigorous results justify (under certain conditions) the use of the mean-field approach and 
the GP equation as a quite relevant model for the description of the static and dynamic properties of BECs.


\subsection{Repulsive and attractive interatomic interactions. Feshbach resonance}

Depending on the BEC species, the scattering length $a$ [and, thus, the nonlinearity coefficient $g$ 
in the GP Eq.~(\ref{chap01:peq1})] may take either positive or negative values, accounting for 
repulsive or attractive interactions between the atoms, respectively. Examples of repulsive (attractive) BECs are formed by atomic vapors of  
$^{87}$Rb and $^{23}$Na ($^{85}$Rb and $^{7}$Li) ones, which are therefore described by a GP model with 
a defocusing (focusing) nonlinearity in the language of nonlinear optics \cite{chap01:sulem,chap01:kiag}. 

On the other hand, it is important to note that during atomic collisions, the atoms can stick together 
and form bound states in the form of molecules. If the magnetic moment of the molecular state is different from 
the one of atoms, one may use an external (magnetic, optical or dc-electric) field to controllably 
vary the energy difference between the atomic and molecular states. Then, at a so-called {\it Feshbach resonance}
(see, e.g., Ref.~\cite{chap01:Weiner03} for a review), the energy of the molecular state becomes equal to the one of 
the colliding atoms and, as a result, long-lived molecular states are formed. This way, as the aforementioned 
external field is varied through the Feshbach resonance, the scattering length is significantly increased, 
changes sign, and finally is decreased. Thus, Feshbach resonance is a quite effective mechanism that can be used to manipulate 
the interatomic interaction (i.e., the magnitude and {\em sign} of the scattering length).  

Specifically, the behavior of the scattering length near a Feshbach resonant magnetic field $B_0$ is typically of the
form \cite{chap01:faA,chap01:faB},
\begin{equation}
a(B)=\hat{a}\left( 1-\frac{\Delta}{B-B_{0}} \right),
\label{chap01:aFR}
\end{equation}
where $\hat{a}$ is the value of the scattering length far from
resonance and $\Delta$ represents the width of the resonance. 
Feshbach resonances were studied in a series of elegant experiments 
performed with sodium 
\cite{chap01:feshbachNaA,chap01:feshbachNaB} and rubidium 
\cite{chap01:feshbachRbA,chap01:feshbachRbB} condensates. 
Additionally, they have been used in many important experimental investigations, including, among others, 
the formation of bright matter-wave solitons \cite{chap01:bright1,chap01:bright2,chap01:bright3,chap01:njp2003b}. 

\subsection{The external potential in the GP model}

The external potential $V_{\rm ext}({\bf r})$ in the GP Eq.~(\ref{chap01:peq1}) is used to 
trap and/or manipulate the condensate. In the first experiments, the BECs were confined 
by means of magnetic fields \cite{chap01:rmpnlA,chap01:rmpnlB}, 
while later experiments demonstrated that an optical confinement of BECs is also 
possible \cite{chap01:stamp,chap01:barrett}, utilizing the so-called optical dipole traps \cite{chap01:odt1,chap01:odt2}. 
While magnetic traps are typically harmonic (see below), the shape of 
optical dipole traps is extremely flexible and controllable, as the dipole potential 
is directly proportional to the light intensity field \cite{chap01:odt2}. 
An important example is the special case of periodic optical potentials called 
{\it optical lattices} (OLs), 
which have been used to reveal novel physical phenomena in BECs \cite{chap01:morsch,blochol,chap01:markus}.

In the case of the ``traditional'' magnetic trap, the external potential has the harmonic form:
\begin{equation}
V_{\rm MT}({\bf r}) = \frac{1}{2} m(\omega_x x^2 + \omega_y y^2 + \omega_z z^2),
\label{chap01:peq2}
\end{equation}
where, in general, the trap frequencies $\omega_x, \omega_y, \omega_z$ along the three directions are different.
On the other hand, the optical lattice is generated by a pair of laser beams forming 
a standing wave which induces a periodic potential. 
For example, a single periodic 1D standing wave 
of the form $E(z,t)=2E_0 \cos(kz) \exp(-i \omega t)$ can be created by the superposition of the two 
identical beams, $E_{\pm}(z,t) = E_{0} \exp[i(\pm kz - \omega t)]$, having the same polarization, 
amplitude $E_0$, wavelength $\lambda=2\pi / k$, and frequency $\omega$. Since the dipole potential 
$V_{\rm dip}$ is proportional to the intensity $I \sim |E(z,t)|^2$ of the light field \cite{chap01:odt2}, 
this leads to an optical lattice of the form $V_{\rm dip} \equiv V_{\rm OL} = V_0 \cos^2 (kz)$.
In such a case, the lattice periodicity is $\lambda/2$ and the lattice height is given by
$V_0 \sim I_{\rm max} / \Delta \omega$, where $I_{\rm max}$ is the maximum intensity of the light
field and $\Delta \omega \equiv \omega-\omega_{\rm o}$ is the detuning of the lasers from the
atomic transition frequency $\omega_{\rm o}$. Note that atoms are trapped at the nodes (anti-nodes)
of the optical lattice for blue- (red-) detuned laser beams, or $\Delta \omega >0$ ($\Delta \omega <0$).
In a more general 3D setting, the optical lattice potential can take the following form:
\begin{equation}
\hskip-1cm
V_{\rm OL}({\bf r}) = V_0 \left[\cos^2 \left({k_{x}} x + \phi_x \right)
+ \cos^2 \left( {k_{y}} y + \phi_y \right)
+ \cos^2 \left({k_{z}} z + \phi_z \right) \right],
\label{chap01:peq3}
\end{equation}
%
where $k_{i}=2\pi/\lambda_{i}$ ($i \in \{x, y, z\}$), $\lambda_{i}=\lambda / [2 \sin(\theta_{i}/2)]$,
$\theta_i$  are the (potentially variable) angles between the laser beams 
\cite{chap01:morsch,chap01:markus} and $\phi_{i}$ are arbitrary phases.

It is also possible to realize experimentally an ``optical superlattice'', characterized by two different periods.
In particular, as demonstrated in Ref.~\cite{chap01:peil}, such 
a superlattice can be formed by the sequential creation of two lattice structures using four laser beams.
A stationary 1D superlattice can be described as
%
$V(z)=V_1 \cos(k_1 z)+ V_2 \cos(k_2 z)$,
%
where $k_i$ and $V_i$ denote, respectively, the wavenumbers and amplitudes of the sublattices.
The experimental tunability of these parameters provides precise and flexible
control over the shape and time-variation of the external potential.

The magnetic or/and the optical dipole traps can be experimentally combined
either together, or with other potentials; an example concerns far off-resonant laser beams, that can
create effective repulsive or attractive localized potentials, for blue-detuned or red-detuned
lasers, respectively.
Such a combination of a harmonic trap with a repulsive localized 
potential located at the center of the harmonic trap is the {\it double-well} potential, as, e.g., the one 
used in the seminal interference experiment of Ref.~\cite{chap01:interf}.
Double-well potentials have also been created by a combination of a
harmonic and a periodic optical potential \cite{chap01:albiez}. Finally, other combinations, including, e.g.,
linear ramps of (gravitational) potential $V_{{\rm ext}} = m g z$ have also been experimentally applied
(see, e.g., Refs.~\cite{chap01:albiez,chap01:kasevich}).
Additional recent possibilities include
the design and implementation of external potentials,
offered, e.g., by the so-called {\it atom chips}
\cite{chap01:FolmanA,chap01:surfA,chap01:surfB} (see also the review \cite{chap01:FolmanB}). 
Importantly, the major flexibility for the creation of a wide variety 
of shapes and types  of external potentials
(e.g., stationary, time-dependent, etc), has inspired 
many interesting applications (see, for example, Sec.~\ref{Sec:special}.4).

\subsection{Ground state}

The ground state of the GP model of Eq.~(\ref{chap01:peq1}) can
readily be found upon expressing the condensate wave function as
$\Psi({\bf r}, t)=\Psi_{0}({\bf r}) \exp(-i\mu  t/\hbar)$.
%
%
If $g=0$,
Eq.~(\ref{chap01:statGPE}) reduces to the usual Schr\"{o}dinger equation
with potential $V_{\rm ext}$.
Then, for a harmonic external trapping potential
[see Eq.~(\ref{chap01:peq2})], the ground state of the system is obtained when letting all
non-interacting bosons
occupy the lowest single-particle state; there, $\Psi_0$ has the Gaussian profile
\begin{equation}
\Psi_{0}({\bf r})= \sqrt{N}\left( \frac{m \omega_{\rm ho}}{\pi \hbar}\right)^{3/4}
\exp\left[-\frac{m}{2\hbar} (\omega_x^2 x^2 + \omega_y^2 y^2 + \omega_z^2 z^2) \right],
\label{chap01:gaussian}
\end{equation}
where $\omega_{\rm ho} = (\omega_x \omega_y \omega_z)^{1/3}$ is the 
geometric mean of the confining frequencies.

For repulsive interatomic forces ($g>0$, or scattering length
$a>0$), if the number of atoms of the condensate is sufficiently
large so that $N a /a_{\rm ho} \gg 1$, the atoms are pushed
towards the rims of the condensate, resulting in slow spatial
variations of the density.
Then the kinetic energy (gradient) term
is small compared to the
interaction and potential energies and becomes significant only close to
the boundaries.
Thus,
the Laplacian kinetic energy term in  Eq.~(\ref{chap01:statGPE}) can safely
be neglected. This results in the, so-called, Thomas-Fermi (TF) approximation
\cite{chap01:reviewA,book2,book1}
for the system's ground state density profile:
\begin{equation}
\rhon({\bf r})=|\Psi_{0}({\bf r})|^2= g^{-1} \left[ \mu-V_{\rm ext}({\bf r}) \right],
\label{chap01:TF}
\end{equation}
in the region where $\mu>V_{\rm ext}({\bf r})$, and $\rhon=0$ outside.
%
For a spherically symmetric harmonic magnetic
trap ($V_{\rm ext}=V_{\rm MT}$ with $\omega_{\rm ho}=\omega_x=\omega_y=\omega_z$),
the radius $R_{\rm TF}=(2\mu/m)^{1/2}/\omega_{\rm ho}$ for which $\rhon(R_{\rm TF})=0$, is the so-called
Thomas-Fermi radius determining the size of the condensed cloud.
Furthermore, the normalization condition connects $\mu$ and $N$ through
the equation $\mu = (\hbar \omega_{\rm ho}/2) (15 N a/a_{\rm ho})^{2/5}$, where 
$a_{\rm ho}=(\hbar / m \omega_{\rm ho})^{1/2}$ is the harmonic oscillator length.

For attractive interatomic forces ($g<0$, or $a<0$), the density tends to increase at the trap center, while
the kinetic energy tends to balance this increase. However, if the number of atoms $N$ in the condensate 
exceeds a critical value, i.e., $N>N_{\rm cr}$, the system is subject to {\it collapse} in 2D or 
3D settings \cite{chap01:sulem,book1,book2}. 
Collapse was observed experimentally in both cases of the attractive $^{7}$Li \cite{chap01:colexp1}
and $^{85}$Rb condensate \cite{chap01:colexp2}. In these experiments, it was demonstrated that during collapse 
the density grows and, as a result, the rate of collisions (both elastic and inelastic) is increased; these collisions 
cause atoms to be ejected from the condensate in an energetic explosion, which leads to a loss of mass that results in 
a smaller condensate. It should be noted that the behavior of BECs close to collapse can be significantly affected 
by effects such as inelastic two- and three-body collisions that are not included in the original GP equation; 
such effects are briefly discussed below (see Sec.~3.4).

The critical number of atoms necessary for collapse in a spherical
BEC
is determined  by the equation
${N_{cr} |a|}/{a_{\rm ho}}=0.575$ \cite{chap01:rup},
%
%
where $|a|$ is the absolute value of the scattering length. 
Importantly, collapse may not occur in a quasi-1D setting (see Sec.~\ref{Sec:dim_reduc}.1), 
provided that the number of atoms does not exceed the critical value given by the equation
${N_{cr} |a|}/a_{r} 
=0.676$,
%
%
with 
$a_{r}$ being the transverse harmonic oscillator
length \cite{chap01:vpg,chap01:gamA,chap01:gamB}.

\subsection{Small-amplitude linear excitations}

We now consider small-amplitude excitations of the condensate, which can be found 
upon linearizing the time-dependent GP equation around the ground state.
Specifically, solutions of Eq.~(\ref{chap01:peq1}) can be sought in the form
\begin{equation}
\Psi({\bf r}, t)= e^{-i\mu t/\hbar} \left[ \Psi_{0}({\bf r})+ \sum_{j} \left( u_{j}({\bf r}) e^{-i\omega_{j} t}
+\upsilon_{j}^{\ast}({\bf r}) e^{i \omega_{j} t} \right) \right],
\label{chap01:ansatz_so}
\end{equation}
where $u_j$, $\upsilon_j$ are small (generally complex) perturbations, describing the components of the 
condensate's (linear) response to the external perturbations that oscillate at frequencies 
$\pm \omega_j$ [the latter are (generally complex) eigenfrequencies].
Substituting Eq.~(\ref{chap01:ansatz_so}) into Eq.~(\ref{chap01:peq1}), and keeping only the linear terms
in $u_j$ and $\upsilon_j$, the following set of equations is derived
\begin{eqnarray}
&&\left[\hat{H}_{0}-\mu+ 2g\, |\Psi_{0}|^{2}({\bf r})\right] u_{j}({\bf r})+g\, \Psi_{0}^{2}({\bf r}) \upsilon_{j}({\bf r})
= \hbar\, \omega_{j}\, u_{j}({\bf r}), 
\nonumber\\[1.0ex]
&&\left[\hat{H}_0-\mu+2g\, |\Psi_{0}|^{2}({\bf r})\right] \upsilon_{j}({\bf r}) + g\, \Psi_{0}^{\ast 2} ({\bf r}) u_{j}({\bf r})
= -\hbar\, \omega_{j}\, \upsilon_{j}({\bf r}),
\label{chap01:BdG}
\end{eqnarray}
where $\hat{H}_0 \equiv -(\hbar^2 / 2m) \nabla^2 + V_{\rm ext}({\bf r})$.
Equations~(\ref{chap01:BdG}) are
known as the {\it Bogoliubov-de Gennes} (BdG) equations. These equations can also be derived using a purely 
quantum-mechanical approach \cite{chap01:reviewA,chap01:leg,book1,book2} and can 
be used, apart from 
the ground state, for any
state (including solitons and vortices) with the function $\Psi_0$ being modified accordingly. 

The BdG equations are 
intimately connected to the stability of the state $\Psi_0$.
Specifically, suitable combinations of Eqs.~(\ref{chap01:BdG}) yield
\begin{equation}
(\omega_{j}-\omega_{j}^{\ast}) \int (|u_{j}|^2 - |\upsilon_j|^2)d{\bf r} =0.
\label{chap01:res}
\end{equation}
%
This equation can be satisfied in two different ways: First, if $\omega_{j}-\omega_{j}^{\ast}=0$, i.e.,
if the eigenfrequencies $\omega_{j}$ are real; 
if this is true for all $j$, the fact that Im$\{\omega_j\}=0$ shows that the state $\Psi_0$ is
{\it stable}. Note that, in this case, one can use the normalization
condition for the eigenmodes $u_j$, $\upsilon_j$ of the form
%
$
\int (|u_{j}|^2 - |\upsilon_j|^2)d{\bf r} =1.
$
%
On the other hand, occurrence of imaginary or complex eigenfrequencies $\omega_{j}$ (i.e., if $\omega_{j}-\omega_{j}^{\ast} \ne 0$ 
or Im$\{\omega_j \} \ne 0$),
indicates {\it dynamical instability} of the state $\Psi_0$; in such a case, 
Eq.~(\ref{chap01:res}) is satisfied only if
%
$\int |u_{j}|^2 d{\bf r} = \int |\upsilon_{j}|^2 d{\bf r}$.
%

%
In the case of a uniform gas (i.e, for $V_{\rm ext}({\bf r})=0$
and $\Psi_0^2=\rho=$const.), 
the amplitudes $u$ and $\upsilon$ are plane waves 
$\sim e^{i {\bf k} \cdot {\bf r}}$ 
(of wavevector ${\bf k}$) and the BdG 
Eqs.~(\ref{chap01:BdG}) lead to a dispersion relation, known as the {\it Bogoliubov spectrum}:
\begin{equation}
(\hbar \omega)^2=\left(\frac{\hbar^2{\bf k}^2}{2m}\right)\left(\frac{\hbar^2 {\bf k}^2}{2m}+2g \rho\right).
\label{chap01:Bogoliubov}
\end{equation}
For small momenta $\hbar {\bf k}$, Eq.~(\ref{chap01:Bogoliubov}) yields the phonon 
dispersion relation $\omega=c q$, where
\begin{equation}
c=\sqrt{{g\rhon}/{m}} 
\label{chap01:sound}
\end{equation}
is the {\it speed of sound}, while, for large momenta, the spectrum provides the free particle energy
$\hbar^2 {\bf k}^2 / (2m)$; 
the ``crossover'' between the two regimes occurs when the 
excitation wavelength is of the order of the 
healing length [see Eq.~(\ref{chap01:heal})].

In the case of 
attractive interatomic interactions ($g<0$), the speed of sound becomes imaginary, which indicates that long
wavelength perturbations grow or decay exponentially in time. This effect is directly connected to the
{\it modulational instability}, which leads to delocalization in momentum space and, in turn, to
localization in position space and the formation of solitary-wave structures. 
Modulational instability is responsible for the formation of bright matter-wave solitons 
\cite{chap01:bright1,chap01:bright2,chap01:bright3},  
as was analyzed by various theoretical works 
(see, e.g., Refs.~\cite{chap01:ourpraA,chap01:ourpraB,chap01:ourpraC} 
and the reviews \cite{chap01:pandim,chap01:rab}).


\section{Lower-dimensional BECs, solitons, and reduced mean-field models \label{Sec:dim_reduc}}

\subsection{The shape of the condensate and length scales}

In the case of the harmonic trapping potential (\ref{chap01:peq2}), the flexibility over the choice of the 
three confining frequencies $\omega_j$ ($j \in \{x, y, z \}$) may be used to control the shape of the condensate: 
if $\omega_x = \omega_y \equiv \omega_{r} \approx \omega_z$
the trap is isotropic and the BEC is almost spherical, while the cases $\omega_z < \omega_r$ or
$\omega_r < \omega_z$ describe anisotropic traps in which the BEC is, respectively, 
``cigar shaped'', or ``disk-shaped''. The strongly anisotropic cases 
with $\omega_z \ll \omega_r$ or $\omega_r \ll \omega_z$ are particularly interesting 
as they are related to effectively quasi one-dimensional (1D) and quasi two-dimensional (2D) BECs, respectively. 
Such lower dimensional BECs have 
been studied theoretically 
\cite{ldbec1a,ldbec1b,ldbec2,ldbec3,ldbec4,ldbec5,ldbec6} 
(see also Ref.~\cite{chap01:elieb} for a rigorous mathematical analysis) and have 
been realized experimentally in optical and magnetic traps \cite{chap01:q1db}, 
in optical lattice potentials \cite{chap01:olpA,chap01:olpB,chap01:ol1A,chap01:ol1C}
and surface microtraps \cite{chap01:surfA,chap01:surfB}. 

The confining frequencies of the harmonic trapping potential set characteristic length
scales for the spatial size of the condensate through the characteristic harmonic oscillator lengths
$a_{j} \equiv (\hbar/m\omega_j)^{1/2}$. Another important length scale, 
introduced by the effective mean-field nonlinearity, is the healing length, 
which is the distance over which the kinetic energy and the
interaction energy balance:  
if the BEC density grows from $0$ to $\rhon$ over the distance $\xi$, the kinetic energy, $\sim \hbar^2 / (2m\xi^2 )$, 
and interaction energy, $\sim 4\pi \hbar^2 a\rhon/m$, become equal at the value of $\xi$ given by 
\cite{chap01:reviewA,book1,book2}
\begin{equation}
\xi = (8 \pi \rhon a)^{-1/2}.
\label{chap01:heal}
\end{equation}
%
Note that the name of $\xi$ is coined by the fact that it is actually the distance over 
which the BEC wavefunction $\Psi$ ``heals'' over defects. 
Thus, the spatial widths of nonlinear excitations,
such as dark solitons and vortices in BECs, are of ${\cal O}(\xi)$.


\subsection{Lower-dimensional GP equations}

Let us assume that $\omega_z \ll \omega_x = \omega_y \equiv \omega_{r}$. Then,
if the transverse harmonic oscillator length 
$a_{r} \equiv \sqrt{\hbar/m\omega_{r}} < \xi$, the transverse confinement
of the condensate is so tight that the dynamics of such a
cigar-shaped BEC can be considered to be effectively 1D. This
allows for a reduction of the fully 3D GP equation to an
effectively 1D GP model, which can be done for 
sufficiently small trapping frequency ratios $\omega_z / \omega_r$.
It should be stressed, however, that such a reduction should be only considered
as the 1D limit of a 3D mean-field theory and {\it not} as a
genuine 1D theory (see, e.g., Ref.~\cite{chap01:elieb} for a rigorous mathematical 
discussion). Similarly, 
a disk-shaped BEC 
with $a_{z} < \xi$ 
and sufficiently small frequency ratios $\omega_r / \omega_z$, 
can be described by an effective 2D GP model. 
Below, we will focus on  
cigar-shaped BECs and briefly discuss the case of disk-shaped ones.

Following Refs.~\cite{chap01:vpg,chap01:gp1dA,chap01:gp1dB} 
(see also Ref.~\cite{chap01:rab}), we assume a quasi-1D setting with $\omega_{z} \ll \omega_{r}$ 
and decompose the wavefunction $\Psi$ in a longitudinal (along $z$) and a transverse
[on the $(x,y)$ plane] component; then, we seek for solutions of Eq.~(\ref{chap01:peq1}) in the form
\begin{equation}
\Psi({\bf r}, t)= \psi (z,t)\, \Phi(r;t), 
\label{chap01:psi}
\end{equation}
where $\Phi(r;t) = \tilde{\Phi}_{0}(r) \exp(-i \gamma t)$, $r^2 \equiv x^2+y^2$, while the chemical potential 
$\gamma$ and the transverse wavefunction $\tilde{\Phi}(r)$  
are involved in the auxiliary problem for the transverse quantum harmonic oscillator,
\begin{equation}
\frac{\hbar ^{2}}{2m} \bigskip \nabla_{r}^{2} \tilde{\Phi}_{0} - 
\frac{1}{2}m \omega_{r}^{2} r^2 \tilde{\Phi}_{0} + \gamma \tilde{\Phi}_{0} = 0,
\label{chap01:qho}
\end{equation}
where $\nabla_{r}^{2} \equiv \partial^2 /\partial x^2 +
\partial^2 /\partial y^2$. Since the considered system is 
effectively 1D, it is natural to assume that the transverse condensate wavefunction $\Phi(r)$ remains in the
ground state; in such a case $\tilde{\Phi}_{0}(r)$ takes the form
$\tilde{\Phi}_{0}(r)= \pi^{-1/2} a_r^{-1} \exp(-r^{2}/2 a_{r}^{2})$
[note that when considering the reduction from 3D to 2D the transverse wave
function 
takes the form $\tilde{\Phi}_{0}(r)= \pi^{-1/4} a_{r}^{-1/2} \exp(-r^{2}/2 a_{r}^{2})$]. 
Then, substituting Eq.~(\ref{chap01:psi}) into Eq.~(\ref{chap01:peq1}) and averaging
the resulting equation in the $r$-direction (i.e., multiplying by
$\Phi^{\ast}$ and integrating with respect to $r$), we
finally obtain the following 1D GP equation,
\begin{equation}
i \hbar \frac{\partial}{\partial t} \psi(z,t) = \left[ - \frac{\hbar^{2}}{2m} \frac{\partial^{2} }{\partial z^{2}}
+ V(z) + g_{1D} |\psi(z,t)|^{2} \right] \psi(z,t),
\label{chap01:dimgpe}
\end{equation}
where the effective 1D coupling constant is given by 
$g_{1D} = g/2\pi a_{r}^{2}=2 a \hbar \omega_{r}$ and $V(z)=(1/2)m \omega_{z}^{2} z^{2}$. 
On the other hand, in the 
2D case of disk-shaped BECs, the respective $(2+1)$-dimensional NLS equation has the form of Eq.~(\ref{chap01:dimgpe}), 
with $\partial_{z}^{2}$ being replaced by the Laplacian $\nabla_{r}^{2}$, 
the effectively 2D coupling constant is $g_{2D}=g/\sqrt{2\pi} a_{z} = 2\sqrt{2\pi} a a_{z} \hbar \omega_{z}$, while 
the potential is $V(x,y)=(1/2)m\omega_{r}^{2}(x^2 + y^2)$.
Note that such dimensionality reductions based on the averaging method are commonly used in other disciplines, 
as, e.g., in nonlinear fiber optics \cite{chap01:kiag}. 

A similar reduction can be performed if, additionally, an optical
lattice potential is present.
In this case, it is possible
(as, e.g., in the experiment of Ref.~\cite{chap01:gap}) to tune $\omega_z$ so that it
provides only a very weak trapping along the $z$-direction; 
this way, the shift in the potential trapping energies over the wells where the BEC 
is confined can be made practically 
negligible. In such a case, the potential in Eq.~(\ref{chap01:dimgpe})
is simply the 1D optical lattice $V(z) = V_0 \cos^2 (k z)$. Similarly, 
in the quasi-2D case, 
an ``egg-carton potential'' $V(x,y)=V_0 \left[\cos^2 (k_x x)+ \cos^2 (k_y y)\right]$
is 
relevant 
for disk-shaped condensates. 

We note in passing that the dimensionality reduction of the GP equation 
can also be done self-consistently, using multiscale expansion techniques 
\cite{chap01:VVK2A,chap01:VVK2B}. It is also worth mentioning that more 
recently a rigorous derivation of the 1D GP equation was presented in  
Ref.~\cite{abdallah}, using energy and Strichartz estimates, as well 
as two anisotropic Sobolev inequalities.

%
%


\subsection{Bright and dark matter-wave solitons}

The 1D GP Eq.~(\ref{chap01:dimgpe}) can be reduced to the following dimensionless form, 
\begin{equation}
i \frac{\partial}{\partial t} \psi(z,t) = \left[ -\frac{\partial^{2} }{\partial z^{2}}
+ V(z) + g |\psi(z,t)|^{2} \right] \psi(z,t), 
\label{chap01:adimgpe}
\end{equation}
where the density $|\psi|^2$, length, time and energy are respectively measured in units of 
$4\pi |a| a_{r}^{2}$, $a_{r}$, $\omega_{r}^{-1}$ and $\hbar\omega_{r}$, while 
the coupling constant $g$ is rescaled to unity (i.e., $g=\pm 1$ for repulsive and attractive interatomic interactions respectively). 
In the case of a homogeneous BEC ($V(z)=0$), Eq.~(\ref{chap01:adimgpe}) becomes the ``traditional'' 
completely integrable NLS equation. The latter, is well-known (see, e.g., Ref.~\cite{segur}) to 
possess an infinite number of conserved quantities (integrals of motion), with the lowest-order ones  
being the number of particles: $$N=\int_{-\infty}^{-\infty}|\psi|^{2}dz,$$ 
the momentum: $$P=(i/2)\int_{-\infty}^{-\infty}\left(\psi \psi^{\ast}_{z} -\psi^{\ast}\psi_{z} \right),dz$$ 
and the energy: $$E=(1/2)\int_{-\infty}^{-\infty} \left(|\psi_{z}|^{2}+g|\psi|^{4}\right)dz,$$ 
where the subscripts denote partial derivatives.
%
%

The type of soliton solutions of the NLS equation depends on the parameter $g$. In particular, 
for {\it attractive} BECs ($g=-1$), the NLS equation possesses a {\it bright} soliton solution of the following form 
\cite{chap01:zsb},
\begin{equation}
\psi_{\rm bs}(z,t)=\eta\, {\rm sech}[\eta (z-vt)]\exp[i(kz-\omega t)],
\label{chap01:bs}
\end{equation}
where $\eta$ is the amplitude and inverse spatial width of the soliton, while $k$, $\omega$ and  
$v\equiv \partial \omega/\partial k=k$ are the soliton wavenumber, 
frequency, and velocity, respectively. 
The frequency and wavenumber of the soliton are connected through the ``soliton dispersion relation'' 
$\omega=\frac{1}{2}(k^{2}-\eta^{2})$, which 
implies that 
the allowable region in the $(k,\omega)$ plane for bright solitons is located {\it below} the parabola 
$\omega=\frac{1}{2}k^{2}$, corresponding to the ``elementary excitations'' 
(i.e., the linear wave solutions) of the NLS equation.

Introducing the solution (\ref{chap01:bs}) into the integrals of motion $N$, $P$ and $E$
it is readily found that 
%
\begin{equation}
N=2\eta, \,\,\,\ P=2\eta k,  \,\,\,\ E=\eta k^{2}-\frac{1}{3}\eta^{3}.
\label{ib}
\end{equation}
These equations imply that the bright 
soliton behaves as a classical particle with effective mass $M_{\rm bs}$, 
momentum $P_{\rm bs}$ and energy $E_{\rm bs}$, respectively given by
$M_{\rm bs}=2\eta$, $P_{\rm bs}=M_{\rm bs}v$, and $E_{\rm bs}=\frac{1}{2}M_{\rm bs}v^{2}-\frac{1}{24}M_{\rm bs}^{3}$, 
where it is reminded that $v=k$. Notice that in the equation for the energy, the first 
and second terms in the right hand side are, respectively, the kinetic energy and the binding energy of the 
quasi-particles associated with the soliton \cite{djk}. Differentiating the soliton energy and 
momentum over the soliton velocity, the following relation is found,  
\begin{equation}
\frac{\partial E_{\rm bs}}{\partial P_{\rm bs}}=v, 
\label{pnbs}
\end{equation}
which underscores the particle-like nature of the bright soliton.

On the other hand, for {\it repulsive} BECs ($g=+1$), the NLS equation admits a dark soliton 
solution, 
which in this case lives on the nonzero background
$\psi=\psi_{0}\exp[i(k z-\omega t)]$. 
The dark soliton may be expressed 
as \cite{chap01:zsd}, 
\begin{equation}
\psi(z,t)=\psi_{0}\left( \cos \varphi \tanh \zeta +i\sin \varphi \right)\exp[i(kz-\omega t)],
\label{chap01:ds}
\end{equation}
where $\zeta \equiv \psi_{0} \cos \varphi \left( z-vt \right)$, 
$\omega=(1/2)k^{2}+\psi_{0}^{2}$,  
while the remaining 
parameters $v$, $\varphi$ and $k$, are connected through the relation $v=\psi_{0}\sin \varphi + k$.
Here, $\varphi$ is the so-called ``soliton phase angle'', or, simply, the phase shift of 
the dark soliton ($|\varphi |<\pi /2$), which describes the {\it darkness} of the soliton 
through the relation, $|\psi|^{2}=1-\cos^{2} \varphi {\rm sech}^{2}\zeta$; this way, the 
limiting cases $\varphi=0$ and $\cos\varphi \ll 1$ correspond to the so-called {\it black} 
and {\it gray} solitons, respectively. The amplitude and velocity of 
the dark soliton  are given by $\cos \varphi $ and $\sin \varphi $ respectively; thus, 
the black soliton, $\psi=\psi_{0}\tanh (\psi_{0}x)\exp (-i\mu t)$, is a stationary dark soliton ($v=0$), 
while the gray soliton moves with a velocity close to the speed of sound ($v \sim c \equiv \psi_0$ in our units). 
The dark soliton solution (\ref{chap01:ds}) has two independent parameters, 
for the background ($\psi_{0}$ and $k$) and one for the soliton  ($\varphi$). 
In fact, it should be mentioned that in both the bright and the dark soliton,
there is also a freedom 
in selecting 
the initial location of the solitary wave $z_0$ (in the above formulas, $z_0$
has been set equal to zero)
\footnote{
Recall that the underlying model, namely the completely integrable NLS equation, 
has infinitely many symmetries, including translational and Galilean invariances.
}.
Also, it should be noted that as in this case the dispersion relation implies that $\omega>k^{2}$, 
the allowable region in the $(k,\omega)$ plane for dark solitons is located {\it above} the 
parabola $\omega=\frac{1}{2}k^{2}$. 

As the integrals of motion of the NLS equation refer to {\it both} 
the background and the dark soliton, 
the integrals of motion for the dark soliton are {\it renormalized} 
so as to extract the 
contribution of the background \cite{uz,igor,KY}. In particular, the renormalized momentum and energy 
of the dark soliton (\ref{chap01:ds}) read (for $k=0$):  
\begin{eqnarray}
P_{\rm ds}&=&-2v(c^{2}-v^{2})^{1/2}+2c^{2} \tan^{-1}\left[ \frac{(c^{2}-v^{2})^{1/2}}{v} \right], 
\\
E_{\rm ds}&=&\frac{4}{3}(c^{2}-v^{2})^{3/2}.
\label{PDSs}
\end{eqnarray}
Upon differentiating the above expressions over the soliton velocity $v$,  it can readily be found that 
\begin{equation}
\frac{\partial E_{\rm ds}}{\partial P_{\rm ds}}=v, 
\label{pnds}
\end{equation}
which shows that, similarly to the bright soliton, the dark soliton effectively behaves like a classical particle. 
Note that, usually, dark matter-wave solitons are considered in the simpler case 
where the background is at rest, i.e., $k=0$; then, the frequency $\omega$ actually plays  
the role of a normalized one-dimensional chemical potential, namely $\mu \equiv \psi_{0}^{2}$, 
which is determined by the number of atoms of the condensate. 
Moreover, it should be mentioned 
that in the case of a harmonically confined condensate, i.e., for $V(z)=\frac{1}{2}\Omega^2 z^2$ (with $\Omega=\omega_{z}/\omega_{r}$ being the 
normalized trap strength) in Eq.~(\ref{chap01:adimgpe}), the background of the dark soliton is actually 
the ground state of the BEC which can be approximated by the 
Thomas-Fermi cloud
[see Eq.~(\ref{chap01:TF})]; 
thus, the ``composite'' wavefunction (containing both the background and the dark soliton) 
can be approximated e.g. by the form $\psi=\psi_{\mathrm{TF}}(z) \exp(-i \psi_{0}^{2}t) \psi_{\rm ds}(z,t)$, where $\psi_{\rm ds}(z,t)$ is 
the dark soliton of Eq.~(\ref{chap01:ds}). 

Both types of matter-wave solitons, namely the bright and the dark ones, have been observed in a series of experiments. 
In particular, the formation of quasi-1D bright solitons and bright soliton trains has been observed 
in $^{7}$Li \cite{chap01:bright1,chap01:bright2} and $^{85}$Rb \cite{chap01:bright3} atoms upon tuning 
the interatomic interaction within the stable BEC from repulsive to attractive via the Feshbach resonance mechanism 
(discussed in Sec.~3.3). On the other hand, quasi-1D dark solitons were 
observed in $^{23}$Na \cite{chap01:denschl,chap01:dutton} and $^{87}$Rb \cite{chap01:dark1,chap01:dark,engels} atoms
upon employing quantum-phase engineering techniques or by dragging a moving impurity (namely a laser beam) through the condensate. 
Note that multidimensional solitons (and vortices), as well as many interesting applications based on 
the particle-like nature of matter-wave solitons highlighted here will be discussed in 
Sec.~\ref{Sec:special}.


\subsection{Mean-field models with non-cubic nonlinearities} 

Mean-field models with non-cubic nonlinearities have also been derived and used in various 
studies, concerning either the effect of dimensionality on the dynamics of cigar-shaped BECs, 
or the effect of the three-body collisions irrespectively of the dimensionality of the system. 

Let us first discuss the former case, i.e., consider a condensate confined in a highly anisotropic 
trap with, e.g., $\omega_z \ll \omega_r$ and examine the effect of the deviation from one-dimensionality 
on the longitudinal condensate dynamics. Following Refs.~\cite{chap01:cqnlsC,sinha,chap01:cqnlsD,delgado1,delgado2}, 
one may factorize the wavefunction as per Eq.~(\ref{chap01:psi}), but with the transverse wavefunction $\Phi$ 
depending also on the longitudinal variable $z$. Then, one may employ an {\it adiabatic approximation} to 
separate the fast transverse and slow longitudinal dynamics (i.e., neglecting derivatives of $\Phi$ with respect to the 
slow variables $z$ and $t$). This way, assuming that the external potential is separable, 
$V_{\rm ext}({\bf r}) = U(r)+V(z)$, the following system of equations is obtained from the  
3D GP Eq.~(\ref{chap01:peq1}), 
\begin{equation}
i \hbar \frac{\partial \psi}{\partial t} = - \frac{\hbar^{2}}{2m} \frac{\partial^{2} \psi}{\partial z^{2}}
+ V(z)\psi + \mu_{r}(\rhon) \psi, 
\label{long}
\end{equation}
\begin{equation}
\mu_{r}(\rhon) \Phi = - \frac{\hbar^{2}}{2m} \nabla_{r}^{2}\Phi + U(r)\Phi + g \rhon |\Phi|^{2}\Phi, 
\label{transv}
\end{equation}
where the transverse local chemical potential $\mu_{r}(\rhon)$ (which depends on the longitudinal density 
$\rhon(z,t) = |\psi(z,t)|^2$) is determined by the ground state solution of Eq.~(\ref{transv}). 
An approximate solution of the above system of Eqs.~(\ref{long})-(\ref{transv}) was found in Ref.~\cite{chap01:cqnlsC} 
(see also Refs.~\cite{sinha,chap01:cqnlsD}) as follows. As the system is close to 1D, 
it is natural to assume that the transverse wave function is {\it close} to the ground state of the 
transverse harmonic oscillator, and can be expanded 
in terms of the radial eigenmodes $\tilde{\Phi}_{j}$, i.e., 
$\Phi(r;z)= \tilde{\Phi}_{0}(r)+\sum_{j} C_{j}(z) \tilde{\Phi}_{j}(r)$.
Accordingly, expanding the chemical potential $\mu_{r}(\rhon)$ in terms of the density as 
$\mu_{r}(\rhon) = \hbar \omega_{r} + g_{1}\rhon - g_{2} \rhon^2 + \cdots$, 
the following NLS equation 
is obtained:
\begin{equation}
i\hbar \frac{\partial \psi}{\partial t}=
\left[-\frac{\hbar^{2}}{2m}\frac{\partial^{2}}{\partial z^{2}} 
+ V(z) + f(\rhon) \right]\psi, 
\label{gengpe}
\end{equation}
with the nonlinearity function given by 
\begin{equation}
f(\rhon)= g_{1} \rhon - g_{2} \rhon^{2}. 
\label{cqn}
\end{equation}
It is clear that Eq.~(\ref{gengpe}) is a {\it cubic-quintic} NLS (cqNLS) equation, 
with the coefficient of the linear and quadratic term being given by Ref.~\cite{chap01:cqnlsC} 
$g_{1} = g_{1D}= 2 a \hbar \omega_{r}$ and $g_{2} = 24 \ln(4/3)a^2 \hbar \omega_r$, respectively. 
In the effectively 1D case discussed in Sec.~3.2, this cqNLS equation 
is reduced to the 1D GP Eq.~(\ref{chap01:dimgpe}). Note that the cqNLS model has been 
used in studies of the dynamics of dark \cite{chap01:cqnlsC} and bright \cite{sinha,chap01:cqnlsD} 
matter-wave solitons in elongated BECs.

The transverse chemical potential $\mu_r$ of an elongated condensate 
was also derived recently by Mu\~{n}oz Mateo and Delgado \cite{delgado1} as a function of the longitudinal density $\rhon$. 
This way, the same authors presented in the recent work of Ref.~\cite{delgado2} the effective 1D NLS Eq.~(\ref{gengpe}), 
but with the nonlinearity function given by 
\begin{equation}
f(\rhon)=\sqrt{1+4aN\rhon}, 
\label{mmd}
\end{equation}
where $a$ is the scattering length and $N$ is the number of atoms.\footnote{Note 
that the number of atoms $N$ appears in the nonlinearity function $f(\rhon)$ 
due to the fact that the wavefunction is now normalized to $1$ rather than to $N$, as in the GP Eq.~(\ref{chap01:dimgpe}).}
Note that in the weakly-interacting limit, $4aN\rhon \ll 1$, the resulting NLS equation 
has the form of Eq.~(\ref{chap01:dimgpe}), with the same coupling constant $g_{1D}$. This model, which was originally 
suggested in Ref.~\cite{gerbier}, predicts accurately ground state properties of the condensate, such as the chemical potential, 
the axial density profile and the speed of sound.

Other approaches to the derivation of effective lower-dimensional mean-field models have also been 
proposed in earlier works, leading (as in the case of Refs.~\cite{delgado1,delgado2,gerbier}) to NLS-type equations 
with {\it generalized nonlinearities} \cite{chap01:npse,chap01:btm,kam,mass,zhang}. Among these models, 
the so-called {\it non-polynomial Schr\"{o}dinger equation} (NPSE)
has attracted considerable attention. The latter was 
obtained 
by Salasnich {\it et al.} \cite{chap01:npse} by 
employing the following ansatz for the transverse wavefunction, 
$\Phi(r;t)=\left[\exp\left(-r^2/2\sigma^{2}(z,t)\right)\right]/[\pi^{1/2}\sigma(z,t)]$; 
then, the variational equations related to the minimization of the action functional (from which the 
3D GP equation can be derived as the associated Euler-Lagrange equation) led to Eq.~(\ref{gengpe}) 
with a nonlinearity function 
\begin{eqnarray}
f(\rhon)
=\frac{gN}{2\pi a_{r}^{2}} \frac{\rhon}{\sqrt{1+2aN\rhon}}+\frac{\hbar \omega_{r}}{2} 
\left(\frac{1}{\sqrt{1+2aN\rhon}} +\sqrt{1+2aN\rhon} \right), 
\label{salnl}
\end{eqnarray}
and to the following equation for the transverse width $\sigma$: $\sigma^{2}=a_{r}^{2}\sqrt{1+2aN\rhon}$. 
Note that in the weakly interacting limit of $aN\rhon \ll 1$, Eq.~(\ref{gengpe}) 
becomes again equivalent to the 1D GP Eq.~(\ref{chap01:dimgpe}), 
while the width $\sigma$ becomes equal to the transverse harmonic oscillator length $a_r$. 
The NPSE (\ref{gengpe}) has been found to predict accurately static and dynamic properties of cigar-shaped BECs 
(such as the density profiles, the speed of sound, and the collapse threshold of attractive BECs) \cite{sal2}, while its 
solitonic solutions have been derived in Ref.~\cite{chap01:npse}. 
Generalizations of the NPSE model in applications 
involving time-dependent potentials \cite{mass,ricardo_nic}, or the description of spin-1 atomic 
condensates \cite{zhang}, have also been presented. Moreover, it is worth mentioning that the 
NPSE has been found to predict accurately the BEC dynamics in recent experiments \cite{chap01:albiez}. 
%


On the other hand, as mentioned above, mean-field models with non-cubic 
nonlinearities, and particularly the 
cqNLS equation in a 1D, 2D or a 3D setting,  
may have a different physical interpretation, namely to take into account three-body interactions. 
In this context, and in the most general case, the coefficients $g_1$ and $g_2$ in Eq.~(\ref{cqn}) 
are complex, with the imaginary parts describing {\it inelastic} two- and three-body collisions, 
respectively \cite{mo}. As concerns the three-body collision process, 
it occurs at interparticle distances of order of the characteristic radius of 
interaction between atoms and, generally, results in the decrease of the density that can be achieved in traps. 
Particularly, the rate of this process is given by $(d\rhon/dt) = - L\rhon^3$ \cite{book1}, 
where $\rhon$ is the density and $L$ is the 
loss rate, which is of order of $10^{-27}$--$10^{-30}$ cm$^{6}$s$^{-1}$ for various species of alkali atoms \cite{ko}.
Accordingly, the decrease of the density is equivalent to the term $-(L/2)|\Psi|^4 \Psi$ in the time dependent GP equation, 
i.e., to the above mentioned quintic term. 

The cqNLS model has been studied in various works, mainly in the context of attractive BECs 
(scattering length $a<0$, or $g_{1}<0$). Particularly, in studies concerning collapse, both cases with real 
$g_2$ \cite{chap01:cqnlsB}, and with imaginary $g_2$ \cite{chap01:cqnlsA,saue} were considered. 
Additionally, relevant lower-dimensional (and in particular 1D) models were also analyzed in Refs.~\cite{lm1,lm2}. 
The latter works present also realistic values for these 1D cubic-quintic NLS models, including also estimations for 
the three-body collision  parameter $g_2$ (see also Refs.~\cite{abd2,xue,abd3} in which the coefficient $g_2$ was assumed to be real).
Additionally, {\it periodic potentials} were also considered in such cubic-quintic models and various properties 
and excitations of the BECs, such as ground state and localized excitations \cite{abd2}, or band-gap structure 
and stability \cite{xue}, were studied. Moreover, studies of modulational instability in the continuous model with 
the dissipative quintic term \cite{zoidiss}, or the respective 
discrete model with a conservative quintic
term \cite{abd3} were also reported.


\subsection[Weakly- and strongly-interacting 1D BECs. The Tonks-Girardeau gas]
           {Weakly- and strongly-interacting 1D Bose gases. The Tonks-Girardeau gas}

In the previous subsections we discussed the case of ultracold {\it weakly-interacting} quasi-1D BECs, 
which, in the absence of thermal or quantum fluctuations, are described by an effectively 1D GP equation 
[cf.~Eq.~(\ref{chap01:dimgpe})]. On the other hand, and in the same context of 1D Bose gases, 
there exists the opposite limit of {\it strong interatomic coupling} \cite{ldbec1a,ldbec1b,dunjko}. In this case, the 
collisional properties of the bosonic atoms are significantly modified, with the interacting bosonic gas 
behaving like a system of free fermions; such, so-called, {\it Tonks-Girardeau} gases of impenetrable bosons  
\cite{chap01:tgA,chap01:tgB} have recently been observed experimentally \cite{chap01:blA,chap01:blB}. 
The transition between the weakly and strongly interacting regimes is usually characterized by
a single parameter $\gamma = 2/ (\rhon a_{1D})$ \cite{ldbec1a,ldbec1b}, with $a_{1D} \equiv a_{r}^{2}/a_{3D}$ and $a_{3D}$  
being the effective 1D and the usual 3D scattering lengths, respectively ($a_r$ is the transverse 
harmonic oscillator length) \cite{book2}. This parameter quantifies the ratio of the average interaction energy to the 
kinetic energy calculated with mean-field theory. Notice that $\gamma$ varies smoothly 
as the interatomic coupling is increased from values $\gamma \ll 1$ for a weakly interacting 1D Bose gas, 
to $\gamma \gg 1$ for the strongly-interacting Tonks-Girardeau gas, with an approximate 
``crossover regime'' being around $\gamma \sim {\cal O}(1)$, attained experimentally as well \cite{ool1,ool2}. 

The above mentioned weakly- and strongly-interacting 1D Bose gases can effectively be described by 
a generalized 1D NLS of the form of Eq.~(\ref{gengpe}). In such a case, 
while the functional dependence of $f(\rhon)$ on $\gamma$ (and its analytical asymptotics) are known \cite{LL}, 
its precise values in the crossover regime can only be evaluated numerically. 
Such intermediate values have been tabulated in Ref.~\cite{dunjko}, 
and subsequently discussed by various authors in the framework of the local density approximation 
\cite{santos,jb}. Note that following the methodology of Ref.~\cite{chap01:npse}, 
Salasnich {\it et al.} have proposed a model different from the NPSE, but still of 
the form of Eq.~(\ref{gengpe}), with the nonlinearity function depending on the density $|\psi|^2$ 
and the transverse width $\sigma$ of the gas, to describe the weakly- and the strongly-interacting regimes, 
as well as the crossover regime \cite{sal1d}. Finally, it is noted that a much simpler approximate 
model (with $f(\rhon)$ being an explicit function of the density),  
which also refers to these three regimes, was recently proposed in Ref.~\cite{ourpla}.


Coming back to the case of the Tonks-Girardeau gas, it has been suggested that an effective 
mean-field description of this limiting case can be based on a 1D {\it quintic} NLS equation, 
i.e., an equation of the form (\ref{gengpe}) with a nonlinearity function \cite{kolom1}: 
\begin{equation}
f(\rhon)= \frac{\pi^{2}\hbar^{2}}{2m} \rhon^2.
\label{nltg}
\end{equation}
The quintic NLS equation was originally derived by Kolomeisky {\it et al.} \cite{kolom2} 
from a renormalization group approach, and then by other groups, using different techniques \cite{tanatar,lee0,kim}; 
it is also worth noticing that 
its time-independent version has been rigorously derived from the many-body Schr{\"o}dinger equation 
\cite{liebprl}. 
Although the applicability of the quintic NLS equation has been criticized 
(as in certain regimes it fails to predict correctly the coherence properties of 
the strongly-interacting 1D Bose gases \cite{girardeaubad}),  
the corresponding hydrodynamic equations for the density $\rhon$ and the phase $S$ arising from 
this equation 
are well-documented in the context of the local density approximation \cite{dunjko,santos}. 
In fact, this equation is expected to be valid as long as the number of atoms exceeds a 
certain minimum value (typically much larger than 10), for which oscillations in the density 
profiles become essentially suppressed \cite{kolom1,lee0,anna}; in other words, the 
density variations should occur on a length scale which is larger than the Fermi 
healing length $\xi_{\rm F} \equiv 1/(\pi \rhon_{\rm p})$ (where $\rhon_{\rm p}$ is the peak density of the trapped gas). 

The quintic NLS model has been used in various studies \cite{kolom1,fpk,kavoulakis,bkp}, basically 
connected to the dynamics of dark solitons in the Tonks-Girardeau gas, 
and in the aforementioned crossover regime of $\gamma \sim {\cal O}(1)$ \cite{ourpla}. In this connection, 
it is relevant to note that in the above works it was found that, towards the strongly-interacting regime,  
the dark soliton oscillation frequency is up-shifted, which may be used as a possible diagnostic tool 
of the system being in a particular interaction regime.


\subsection{Reduced mean-field models for BECs in optical lattices}

Useful reduced mean-field models can also be derived in the case where the BECs are confined 
in periodic (optical lattice) potentials. Here, we will discuss both continuous and discrete 
variants of such models, focusing ---as in the previous subsections--- on the 1D case (generalizations 
to higher-dimensional settings will be discussed as well). 
We start our exposition upon considering that the external potential in 
Eq.~(\ref{chap01:dimgpe}) is of the form $V(z) = V_0 \sin^2(kz)$, i.e., 
an optical lattice of periodicity $L=\pi/k$. Then, measuring length, energy and time in units of  
$a_{L}=L/\pi$, $E_{L}= 2E_{\rm rec} = \hbar^{2}/ma_{L}^{2}$ 
(where the recoil energy $E_{\rm rec}$ is the kinetic energy gained by an atom when it absorbs a photon from the optical lattice), 
and $\omega_{L}^{-1}=\hbar/E_{L}$, respectively, we express Eq.~(\ref{chap01:dimgpe}) in the dimensionless form 
of Eq.~(\ref{chap01:adimgpe}) with $V(z)=\sin^2 (z)$.

Generally, the stationary states of Eq.~(\ref{chap01:adimgpe}) can be found upon employing the 
usual ansatz, $\psi(z,t)= F(z) \exp(-i\mu t)$, where $\mu$ is the dimensionless chemical potential. 
In the limiting case $g \rightarrow 0$ (i.e., for a noninteracting condensate) where the Bloch-Floquet theory is relevant, 
the function $F(x)$ can be expressed as \cite{jordan} $F(z)= u_{k,\alpha}(z)\exp(ikz)$, where the functions $u_{k,\alpha}(z)$ share the 
periodicity of the optical lattice, i.e., $u_{k,\alpha}(z) = u_{k,\alpha}(z +nL)$ where $n$ is an integer. If the 
Floquet exponent (also called ``quasimomentum'' in the physics context) $k$ is real, the wavefunction $\psi$ 
has the form of an infinitely extended wave, known as a {\it Bloch wave}. 
Such waves exist in {\it bands} (which are labeled by the index $\alpha$ introduced above), 
while they do not exist in {\it gaps}, which are spectral regions characterized by Im$(k) \ne 0$. 

The concept of Bloch waves can also be extended in the nonlinear case ($g \ne 0$) \cite{wudi,brcarr,nbw1,diakonov,nbw2,pearl,nicolin,seaman}.
In particular, when the coupling constant is small, the nonlinear band-gap spectrum and the nonlinear Bloch waves are similar to 
the ones in the linear case \cite{diakonov}. However, when the coupling constant is increased (or, physically speaking, the 
local BEC density grows), the chemical potential of the nonlinear Bloch wave is increased (decreased) for $g>0$ ($g<0$) 
and, thus, the nonlinearity effectively ``shifts'' the edges  of the linear band. In this respect, it is relevant to note 
that for a sufficiently strong nonlinearity, ``swallowtails'' (or loops) appear in the band-gap structure, 
both at the boundary of the Brillouin zone and at the zone center. This  
was effectively explained in Ref.~\cite{seaman}, 
where an adiabatic tuning of a second lattice with half period was considered. 
Swallowtails in the spectrum are related to several interesting effects, such as a non-zero Landau-Zener tunneling probability \cite{wudi}, 
the existence of two nonlinear complex Bloch waves (which are complex conjugate of each other) at the edge of the Brillouin zone \cite{nbw1,diakonov}, 
as well as the existence of period-doubled states (in the case of sufficiently strong optical lattices -- see Sec.~3.6.2), also related to 
periodic trains of solitons \cite{nicolin} in this setting. We finally note that experimentally it is possible 
to load a BEC into the ground or excited Bloch state with an unprecedented control over both the lattice and the atoms \cite{chap01:ol1C}.

\subsubsection{Weak optical lattices.} 

Let us first consider the case of weak optical lattices, with $V_0 \ll \mu$. In this case, and in connection to the above discussion, 
a quite relevant issue is the possibility of nonlinear localization of matter-waves in the gaps of the linear spectrum, i.e., the 
formation of fundamental nonlinear structures in the form of {\it gap solitons}, as observed in the experiment \cite{chap01:gap}. 
The underlying mechanism can effectively be described by means of the so-called Bloch-wave envelope approximation near the band edge, 
first formulated in the context of optics \cite{sterke}, and then  used in the BEC context as well \cite{steel,chap01:VVK2A,bks} 
(see also Ref.~\cite{sakmal0} and the review \cite{brkon}). In particular, a multiscale asymptotic method was used to show that 
the BEC wave function can effectively be described as $\psi(z,t)=U(z,t)u_{k,0}(z)\exp(ikz)$, where $u_{k,0}(z)\exp(ikz)$ represents 
the Bloch state in the lowest band $\alpha=0$ (at the corresponding central quasimomentum $k$), while the envelope function 
$U(z,t)$ is governed by the following dimensionless NLS equation:
\begin{equation}
i \frac{\partial U}{\partial t} = - \frac{1}{2m_{\rm eff}} \frac{\partial^{2} U}{\partial z^{2}}
+ g'_{1D} |U|^{2} U,
\label{efnls}
\end{equation}
where $g'_{1D}$ is a renormalized (due to the presence of the optical lattice) effectively 1D coupling constant, 
and $m_{\rm eff}$ is the effective mass. Importantly, the latter is proportional to the inverse effective diffraction 
coefficient $\partial^2 \mu / \partial k^2$ whose sign may change, as it is actually determined by the curvature of the band structure of the 
linear Bloch waves. Obviously, the NLS Eq.~(\ref{efnls}) directly highlights the abovementioned nonlinear localization 
and soliton formation, which occurs for $m_{\rm eff} g'_{1D} <0$. Note that the above envelope can be extended in higher-dimensional 
(2D and 3D) settings \cite{baks}, in which the effective mass becomes a tensor. In such a case, and for repulsive BECs ($g<0$), 
all components of the tensor have to be negative for the formation of multi-dimensional gap solitons \cite{ost,gapvort2}. 

{\it Coupled-mode theory}, originally used in the optics context \cite{sterke,christo}, has also been used to describe 
BECs in optical lattices \cite{gs1,gs2,gs3,masdep,abdg}. According to this approach, and in the same case of weak optical 
lattice strengths, the wavefunction is decomposed into 
forward and backward propagating waves, $A(z,t)$ and $B(z,t)$, 
with momenta $k=+1$ and $k=-1$, respectively, namely $$\psi(z,t)=[A(z,t)\exp(ix) + B(z,t)\exp(-ix)]\exp(-i\mu t).$$ This way, 
Eq.~(\ref{chap01:adimgpe}) can be reduced to the following system of coupled-mode equations (see also the recent work 
\cite{agudep} for a rigorous derivation),
\begin{eqnarray}
i\left(\frac{\partial A}{\partial t} + 2\frac{\partial A}{\partial x} \right) &=& V_{0} B + g (|A|^2 +2|B|^2)A, 
\label{cme1}
\\
i\left(\frac{\partial B}{\partial t} - 2\frac{\partial B}{\partial x} \right) &=& V_{0} A + g (2|A|^2 +|B|^2)B.
\label{cme2}
\end{eqnarray}
Notice that Eqs.~(\ref{cme1})--(\ref{cme2}) are valid at the edge of the first Brillouin zone, 
i.e., in the first spectral gap of the underlying linear problem (with $g=0$). There, the 
assumption of weak localization of the wave function 
(which is written as a superposition of just two momentum components) is quite relevant: for example, 
a gap soliton near the edge of a gap can indeed be approximated by a modulated Bloch wave, which itself 
is a superposition of a forward and backward propagating waves \cite{sterke}. Thus, coupled-mode theory was 
successfully used to describe matter-wave gap solitons in Refs.~\cite{gs1,gs2,gs3,masdep,abdg}. Note that the 
coupled-mode Eqs.~(\ref{cme1})--(\ref{cme2}) can directly be connected by a NLS equation of the form (\ref{efnls}) 
by means of a formal asymptotic method \cite{masdep} that uses the distance from the band edges as a small parameter.
We finally mention that the coupled-mode equations can formally be extended in higher dimensions \cite{agudep}.

\subsubsection[Strong optical lattices and the discrete NLS equation]
{Strong optical lattices and the discrete nonlinear Schr\"{o}dinger equation.
\label{Sec:DNLS}}

Another useful reduction, which is relevant to deep optical lattice potentials with $V_0 \gg \mu $,  
is the one of the GP equation to a genuinely discrete model, the so-called {\it discrete NLS} (DNLS) equation \cite{chap01:IJMPB}. 
Such a reduction has been introduced in the context of arrays of BEC droplets confined in the wells of an optical lattice
in Refs.~\cite{chap01:tromb,chap01:konot} and further elaborated in Ref.~\cite{chap01:wann1}; we will follow the latter below.

When the optical lattice is very deep, the strongly spatially localized wave functions at the wells of the optical lattice 
can be approximated by Wannier functions, 
%
%
%
%
%
i.e., the Fourier transform of the Bloch functions.
%
%
Due to the completeness of the Wannier basis, any solution of Eq.~(\ref{chap01:adimgpe}) can be expressed as 
%
$\psi (z,t)=\sum_{n, \alpha }c_{n,\alpha }(t)w_{n,\alpha }(z)$, 
%
where 
$n$ and $\alpha$ label wells and bands, respectively.  
Substituting the above expression into Eq.~(\ref{chap01:adimgpe}),
and using the orthonormality of the Wannier basis, we obtain a
set of differential equations for the coefficients. 
%
%
%
%
%
Upon suitable decay of the Fourier coefficients and
the Wannier functions' prefactors (which can be systematically checked for given potential parameters),
the model can be reduced to
\begin{eqnarray}
i {\frac{dc_{n, \alpha}}{dt}}&=&\hat{\omega}_{0,\alpha}c_{n,\alpha}
+ \hat{\omega}_{1,\alpha }\left(c_{n-1,\alpha} +c_{n+1,\alpha}\right)
\nonumber \\
&+& g \sum_{\alpha_1, \alpha_2, \alpha_3}
 W^{nnnn}_{\alpha \alpha_1 \alpha_2 \alpha_3 }  c_{n,\alpha_1}^{\ast}
 c_{n,\alpha_2} c_{n,\alpha_3},
\label{chap01:i-ii}
\end{eqnarray}
where $W^{n n_1 n_2 n_3}_{\alpha \alpha_1 \alpha_2
\alpha_3} = \int_{-\infty}^{\infty}
w_{n,\alpha}w_{n_1,\alpha_1}w_{n_2,\alpha_2}w_{n_3,\alpha_3} dx$.
The latter equation degenerates into the so-called tight-binding model \cite{chap01:tromb,chap01:konot},
\begin{eqnarray}
\hskip-1cm
i \frac{dc_{n,\alpha}}{dt} =\hat{\omega}_{0,\alpha}c_{n,\alpha}
+ \hat{\omega}_{1,\alpha}\left(c_{n-1,\alpha}
+c_{n+1,\alpha}\right)
+g W^{nnnn}_{1111}  |c_{n,\alpha}|^2 c_{n,\alpha},
\label{chap01:TB}
\end{eqnarray}
if one restricts consideration only to the first band. Equation 
(\ref{chap01:TB}) is precisely the reduction of the GP equation to its
discrete counterpart, namely, the 
DNLS equation. Higher-dimensional versions of the latter are of course physically relevant models 
and have, therefore, been used in various studies concerning quasi-2D and 3D BECs confined 
in strong optical lattices (see subsequent sections). 



\section{Some mathematical tools for the analysis of BECs \label{Sec:math}}

Our aim in this section will be to present an overview 
of the wide array of mathematical techniques that have emerged 
in the study of BECs. Rather naturally, one can envision multiple 
possible partitions of the relevant methods,  
e.g., 
based on the type of the nonlinearity (repulsive or attractive, depending on the sign of the scattering length),  
or based on the type of the external potential  
(periodic, decaying or confining) involved in the problem. However, in the present
review, we will 
classify the mathematical methods 
based on the mathematical nature of the underlying considerations.
We will focus, in particular, on four categories of methods.
The first one will concern ``direct'' methods which analyze 
the mean-field model 
directly, without initiating the analysis at
some appropriate, mathematically tractable limit. 
Such approaches include, e.g.,
the method of moments, self-similarity and rescaling methods, 
or the variational techniques among others.
The second one will concern methods that make detailed use of the
understanding of the linear limit of the problem (in a parabolic,
or a periodic potential or in combinations thereof including, e.g.,
a double-well potential). The third will entail perturbations from
the nonlinear limit of the system (such as, e.g., the integrable
NLS equation), while the fourth one will concern discrete systems 
where perturbation methods from the so-called anti-continuum limit
of uncoupled sites are extremely helpful. 

\subsection{Direct methods}

Perhaps one of the most commonly used direct methods in BEC is
the so-called variational approximation 
(see Ref.~\cite{malomedprogopt} for a detailed review). It consists
of using an appropriate ansatz, often a solitonic one or a Gaussian
one (for reasons of tractability of the ensuing integrations) in
the Lagrangian or the Hamiltonian of the model at hand, with some
temporally dependent variational parameters. Often these parameters
are the amplitude and/or the width of the BEC wavefunction.
Then, subsequent derivation of the Euler-Lagrange equations leads
to ordinary differential equations (ODEs) for 
such quantities which can be studied either analytically
or numerically shedding light on the detailed dynamics of the
BEC system. Such methods have been extensively used in
examining very diverse features of BECs including collective 
excitations \cite{perezgarcia}, studying the dynamics of BECs
in 1D optical lattices \cite{chap01:tromb}, offering insights 
on the collapse or absence thereof in higher dimensional settings
and potentials ---see, e.g., Ref.~\cite{baizakovepl} and references therein---,
or on the behavior of BECs on space- or time-dependent nonlinearity
settings ---see e.g., Ref.~\cite{ueda03} as an example---. However, both 
due to the very widespread use of the method and due to the fact that detailed
reviews of it already exist \cite{malomedprogopt,borisbook},
we will not focus on reviewing it here. Instead, we direct the
interested reader to the above works and references therein.
We also note in passing one of the concerns about the validity of
the variational method, which consists of its strong restriction of
the infinite-dimensional GP dynamics to a small finite dimensional
subspace (freezing the remaining directions by virtue of the selected
ansatz). This restriction is well-known to potentially lead to
invalid results \cite{kauptaras}; it is worthwhile to note, however,
that there are efforts underway to systematically compute corrections
to the variational approximation \cite{kaupnew}, thereby increasing
the accuracy of the method.

Another very useful tool for analyzing BEC dynamics 
is the so-called moment method \cite{victormoment}, whereby appropriate moments
of the wavefunction 
$\psi=\sqrt{\rho} \exp(i \phi)$ (where $\rho=|\psi|^2$ and $\phi$ are 
the BEC density and phase, respectively) 
are defined such as $N=\int \rho\, d{\bf r}$ (the number of atoms), 
$X_i=\int x_i \rho\, d{\bf r}$ (the center of mass location),
$\tilde{V}_i=\int \rho \partial \phi/\partial x_i\, d{\bf r}$ 
(the center speed), 
$W_i=\int x_i^2 \rho\, d{\bf r}$ (the width of the wavefunction), 
$B_i=2 \int x_i \rho \partial \phi/\partial x_i\, d{\bf r}$ (the growth speed),
$K_i=-(1/2) \int \psi^{\ast} \partial^2 \phi/\partial x_i^2\, d{\bf r}$
(the kinetic energy) and $J=\int G(\rho)\, d{\bf r}$ (the interaction energy).
Notice that in the above expressions 
the subscript $i$ denotes the $i$-th direction.
Then for the rather general 
GP-type mean-field model 
of the form:
\begin{eqnarray}
i \frac{\partial \psi}{\partial t}=-\frac{1}{2} \Delta \psi
+ V({\bf r}) \psi + g(|\psi|^2,t) \psi - i \sigma(|\psi|^2,t) \psi,
\label{s5_1}
\end{eqnarray}
with a generalized nonlinearity $g(\rho)=\partial G/\partial \rho$, the
generalized dissipation $\sigma$ and, say, the typical parabolic potential 
of the form $V({\bf r})=\sum_k (1/2) \omega_k^2 x_k^2$, the moment equations
read \cite{victormoment}:
\begin{eqnarray}
\frac{dN}{dt} &=& -2 \int \sigma \rho\, d{\bf r}, 
\label{s5_2a}
\\[1.0ex]
\frac{d X_i}{dt} &=& \tilde{V}_i - 2 \int \sigma x_i \rho\, d{\bf r},
\label{s5_2b}
\\[1.0ex]
\frac{d \tilde{V}_i}{dt} &=& - \omega_i X_i - 2 \int \sigma \frac{\partial
\phi}{\partial x_i} \rho\, d{\bf r},
\label{s5_2c}
\\[1.0ex]
\frac{d W_i}{dt} &=& B_i - 2 \int \sigma x_i^2 \rho\, d{\bf r},
\label{s5_2d}
\\[1.0ex]
\frac{d B_i}{dt} &=& 4 K_i -2 \omega_i^2 W_i- 2 \int \delta G\, d{\bf r}
-4 \int \sigma \rho \partial \phi/\partial x_i\,  d{\bf r},
\label{s5_2e}
\\[1.0ex]
\frac{d K_i}{dt} &=& -\frac{1}{2} \omega_i^2 B_i - 
\int \delta G \frac{\partial^2 \phi}{\partial x_i^2}\, d{\bf r} \nonumber \\
&+& \int \sigma \left[\sqrt{\rho} \frac{\partial^2 \sqrt{\rho}}{\partial x_i^2}
-\rho \left(\frac{\partial \phi}{\partial x_i}\right)^2 \right]\, d{\bf r},
\label{s5_2f}
\\[1.0ex]
\frac{dJ}{dt} &=&
\sum_i \int \delta G  \frac{\partial^2 \phi}{\partial x_i^2}\, d{\bf r}
-2 \int g \sigma \rho d{\bf r}  + \int \frac{\partial G}{\partial t}\, d{\bf r},
\label{s5_2i}
\end{eqnarray}
where $\delta G \equiv G(\rho)-\rho g(\rho)$.
One can then extract, for parabolic potentials, a closed-form exact
ODE describing the motion of the center of mass (assuming that the 
dissipation does not dependent on $\rho$) of the form:
\begin{eqnarray}
\frac{d^2 X_i}{d t^2}=-\omega^{2}(t) X_i - 2 \sigma(t) \frac{d X_i}{d t}
-2 \sigma(t) X_i.
\label{oscillation}
\end{eqnarray}
In the absence of dissipation (and for constant in time 
magnetic trap frequencies), this yields a simple harmonic oscillator
for the center of mass of the condensate. This is the so-called Kohn mode
\cite{kohn}, which has been observed experimentally (see, e.g., 
Refs.~\cite{book1,book2}). In fact, more generally, for conservative potentials
one obtains a general Newtonian equation of the form \cite{victorpra99}:
\begin{eqnarray}
\frac{d^2 X_i}{d t^2}=-\int \frac{\partial V}{\partial x_i} \rho\, d{\bf r}
\label{s5_3}
\end{eqnarray}
which is the analog of the linear quantum-mechanical Ehrenfest theorem.

There are some simple dissipationless (i.e., with $\sigma=0$) 
cases for which the Eqs.~(\ref{s5_2a})--(\ref{s5_2i}) close.
For example, if the potential is spherically symmetric ($\omega_i(t)=\omega(t)$),
one can close the equations for $R=\sqrt{W}$, together with 
the equation for $K$ into the so-called Ermakov-Pinney (EP) equation 
\cite{epreview} of the form:
\begin{eqnarray}
\frac{d^2 R}{d t^2}=-\omega(t) R + \frac{M}{R^3},
\label{s5_4}
\end{eqnarray}
where $M$ is a constant depending only on the initial data and the
interaction strength $U$ (the equations close only for the two-dimensional
case and for $G=U \rho^2$). One of the remarkable features of such
EP equations \cite{epreview} is that they can be solved analytically provided
that the underlying linear Schr{\"o}dinger problem 
${d^2 R}/{d t^2}=-\omega(t) R$ is explicitly solvable with 
linearly independent solutions
$R_1(t)$ and $R_2(t)$. In that case, the EP equation has a general
solution of the form $(A R_1^2 + B R_2^2 + 2 C R_1 R_2)^{1/2}$, 
where the constants satisfy $A B -C^2 =M/w^2$ and $w$ is the
Wronskian of the solutions $R_1$ and $R_2$.
Such EP equations have also been used to examine
the presence of parametric resonances for 
time-dependent frequencies (such that the linear
Schr{\"o}dinger problem has parametric resonances) 
in Refs.~\cite{victor99,krb00}. Another place where such EP approach
has been quite relevant is in the examination of BECs with
temporal variation of the scattering length; 
the role of the latter in preventing 
collapse has been studied in the EP framework in Ref.~\cite{montesinos}.
It has also been studied in the context of producing exact solutions
of the second moment of the wavefunction, which is associated with
the width of the BEC, 
which are either oscillatory (breathing
condensates) or decreasing in time (collapsing condensates)
or increasing in time (dispersing BECs) in Ref.~\cite{herring}. It should
be mentioned that when the scattering length is time-dependent the EP equation
(\ref{s5_4}) is no longer exact, but rather an approximate equation,
relying on the assumption of a quadratic spatial dependence of
the condensate phase. 
Another case where exact results can be obtained for the moment
equations is when the nonlinearity $g$ is time-independent and
the phase satisfies Laplace's equation
$\Delta \phi=0$, in which case vortex-line solutions can be found
for the wavefunction $\psi$ \cite{victormoment}. 

It should also be mentioned in passing that such moment methods
are also used in deriving rigorous conditions for avoiding 
collapse in NLS equations more generally, where this class
of methods is known under the general frame of variance
identities (see, e.g., the detailed discussion of 
Sec.~5.1 in Ref.~\cite{chap01:sulem}). 

Another comment regarding the above discussion is that 
Newtonian dynamical equations of the form of Eq.~(\ref{s5_3}) are
more generally desirable in describing the dynamics not only of
the full wavefunction but also of localized modes (nonlinear waves),
such as bright solitons. This approach can be rigorously developed
for small potentials $V(x)=\epsilon W(x)$ or wide potentials
$V(x)=W(\epsilon x)$ in comparison to the length scale of the 
soliton. In such settings, it can be proved \cite{frolich} that
the motion of the soliton is governed by an equation of the form
\begin{eqnarray}
m_{\rm eff} \frac{d^2 s}{d t^2} = - \nabla U(s),
\label{s5_5}
\end{eqnarray}
where $m_{\rm eff}$ is an effective mass (found to be $1/2$ independently
of dimension in Ref.~\cite{frolich}) and the effective potential is given by
\begin{eqnarray}
U(s)=\frac{\int V({\bf r}) \psi_0^2({\bf r}-s)\, d {\bf r}}{\int 
\psi_0^2({\bf r})\, d {\bf r}}.
\label{s5_6}
\end{eqnarray}
In 1D, Eq.~(\ref{s5_6}) can be used to characterize not only
the dynamics of the solitary wave, but also its stability around
stationary points such that $U'(s_0)=0$. It is natural to expect
that its motion will comprise of stable oscillations if $U''(s_0)>0$,
while it will be unstable for $U''(s_0)<0$. In the case of multiple
such fixed points, the equation provides global information on the stability
of each equilibrium configuration and local dynamics in the 
neighborhood of all equilibria. In higher dimensions, an approach
such as the one yielding Eq.~(\ref{s5_5}) is not applicable due to 
the instability of the corresponding multi-dimensional (bright) solitary
waves to collapse \cite{chap01:sulem}. This type of dynamical equations
was originally developed formally using asymptotic multi-scale 
expansions as, e.g., in Refs.~\cite{abdul,gorshk}. We will return to a more
detailed discussion of such techniques characterizing the
dynamics of the nonlinear wave in Secs.~\ref{nlwd} and \ref{Sec:math}.3.3.

Another class of methods that can be used to obtain reduced ODE 
information from the original GP partial differential equation (PDE) concerns scaling 
transformations, such as the so-called lens transformation \cite{chap01:sulem}.
An example of this sort with
\begin{eqnarray}
\psi(x,t)=\frac{1}{l(t)} \exp(i f(t) r^2)\,\, u\left(\frac{x}{L(t)},\tau(t)\right),
\label{s5_7}
\end{eqnarray}
has been used in Ref.~\cite{victkon} to convert the more general
(with time dependent coefficients) form of the GP equation
\begin{eqnarray}
i \frac{\partial \psi}{\partial t}=-\frac{\alpha(t)}{2} \Delta \psi
+ \frac{1}{2} \Omega(t) r^2 \psi + g(t) \psi - i \sigma(t) \psi,
\label{s5_8a}
\end{eqnarray}
into the simpler form with time {\em independent} coefficients:
\begin{eqnarray}
i \frac{\partial u}{\partial \tau}=-\frac{1}{2} \Delta_{\eta} u
+ s |u|^2 u,
\label{s5_8}
\end{eqnarray}
where $\eta=x/L(t)$ and $s=\pm 1$. This happens if the temporally
dependent functions $l(t), f(t), L(t)$ and $\tau(t)$ satisfy the similarity
conditions:
\begin{eqnarray}
\frac{d l}{dt} &=& \alpha(t) d l + \sigma(t) l,
\label{s5_9a}
\\
\frac{d f}{dt} &=& -2 \alpha(t) f^2 - \frac{1}{2} \Omega(t),
\label{s5_9b}
\\[1.0ex]
\frac{d L}{dt} &=& 2 \alpha(t) f L,
\label{s5_9c}
\\[1.0ex]
\frac{d \tau}{dt} &=& \frac{\alpha(t)}{L^2},
\label{s5_9d}
\\[1.0ex]
\frac{\alpha(t)}{L^2} &=& \sigma \frac{g(t)}{l^2}.
\label{s5_9e}
\end{eqnarray}
Some of these ODEs can be immediately solved, e.g.,
\begin{eqnarray}
L(t) &=& \exp \left(2 \int_0^t \alpha(t') f(t') dt' \right),
\label{s5_10b}
\\[1.0ex]
l(t) &=& \Gamma(t) \exp\left(d \int_0^t \alpha(t') f(t') dt' \right)
= \Gamma(t) L(t)^{\frac{d}{2}},
\label{s5_10a}
\\[1.0ex]
g(t) &=& s \alpha(t) \Gamma(t) L(t)^{d-2},
\label{s5_10c}
\end{eqnarray}
where $\Gamma(t)=\exp(\int_0^t \sigma(t') dt')$. Notice that
this indicates that $\alpha(t), g(t)$ and $\sigma(t)$ are 
inter-dependent through Eq.~(\ref{s5_10c}). While, unfortunately,
Eq.~(\ref{s5_9b}) cannot be solved in general, it can be integrated
in special cases, such as, e.g., $\Omega(t)=0$. Notice that in that
case, periodic $\alpha(t)$ 
with zero average, i.e., $\int_0^T \alpha(t') dt'=0$
implies that $L(t), l(t)$ and $f(t)$ will also be periodic. In other
settings the above equations can be used to construct collapsing
solutions or to produce pulsating two-dimensional profiles, as is
the case, e.g., for $\Omega(t)=m (1-2 {\rm sn}^2(t,m))$ in the form:
\begin{eqnarray}
\hskip-1.5cm
\psi &=& \frac{1}{{\rm dn}(t,m)} U\left(\frac{x}{{{\rm dn}(t,m),\tau(t)}}
\right)
\exp \left(i \tau(t)-
i r^2 \frac{m {\rm cn}(t,m) {\rm sn}(t,m)}{2 {{\rm dn}(t,m)}} \right),
\label{s5_11}
\end{eqnarray}
where $U(\eta)$ is the well-known 2D Townes soliton (see also
Sec.~\ref{sec:BS_collapse}) profile 
and $\tau(t)= \int_0^t {\rm dn}(t',m)^{-2} dt'$. Similar types of
lens transformations were used to study collapse type phenomena 
\cite{ghosh,rybin} and to examine modulational instabilities
in the presence of parabolic potentials \cite{chap01:ourpraA}.

In the same as the above class of transformation methods one
can classify also the scaling methods that arise from the
consideration of Lie group theory and canonical transformations
\cite{victorlie1}
of nonlinear Schr{\"o}dinger equations with spatially
inhomogeneous nonlinearities of the form (for the stationary problem)
\begin{eqnarray}
-\psi_{xx} + V(x) \psi + g(x) \psi^3=\mu \psi.
\label{s5_12}
\end{eqnarray}
Considering the generator of translational invariance motivates
the scaling of the form
$U(x)=b(x)^{-1/2} \psi$ and $X=\int_0^x[1/b(s)] ds$ with
$g(x)=g_0/b(x)^3$. Then, $U$ satisfies the regular 
1D NLS equation (whose solutions 
are known from the inverse scattering method 
\cite{segur}) and $b$ satisfies:
\begin{eqnarray}
b'''(x)-2 b(x) V'(x) + 4 b'(x) \mu-4 b'(x) V(x)=0,
\label{s5_13}
\end{eqnarray}
which can remarkably be converted to an EP equation, upon the scaling
$\tilde b(x)=b^{1/2}(x)$. Then, combining the knowledge of solvable
EP cases (as per the discussion above) 
with that of the spatial profiles of the various (plane
wave, solitary wave and elliptic function) solutions of the standard
NLS equation for $U$, we can obtain special cases of $g(x)$ for which
explicit analytical solutions are available \cite{victorlie1}.

\subsection{Methods from the linear limit}
\label{lin_lim}

When considering the GP equation as a perturbation problem, one way
to do so is to consider the underlying {\it linear} Schr{\"o}dinger
problem, obtain its eigenvalues and eigenfunctions; subsequently one
can consider the cubic nonlinear term within the realm of Lyapunov-Schmidt
(LS) theory (see Chap.~7 in Ref.~\cite{golub}), to identify the nonlinear
solutions bifurcating from the linear limit. 

We will discuss 
this approach in a 
general 1D problem, with 
both a magnetic trap and an optical lattice potential, 
\begin{eqnarray}
V(x)=V_{\rm MT}(x) + V_{\rm OL}(x) \equiv \frac{1}{2} \Omega^2 x^2 
+ V_0 \cos(2 x),
\label{s5_14}
\end{eqnarray}
following the approach of Ref.~\cite{toddchaos}. 
Considering the linear problem of the GP equation, using
$\psi(x,t)=\exp(-i E t) u(x)$ (where $E$ is the linear eigenvalue)
and rescaling spatial variables by $\Omega^{1/2}$ one obtains
\begin{eqnarray}
{\cal L} u= -\frac{1}{2} \frac{d^2 u}{dx^2} + \frac{1}{2} x^2 u
+ \frac{V_0}{\Omega} \cos \left(2 \frac{x}{\Omega^{1/2}}\right) u = \frac{E}{\Omega} u.
\label{s5_15}
\end{eqnarray}
If we work in the physically relevant regime of $0<\Omega \ll 1$
and for $V_0/\Omega={\cal O}(1)$, then one can use $\mu=\Omega^{1/2}$
as a small parameter and develop methods of multiple scales and
homogenization techniques \cite{toddchaos} in order to obtain analytical
predictions for the linear spectrum. In particular, the one-dimensional
eigenvalue problem using as fast variable $X=x/\mu$ and
setting $\lambda=E_1/\Omega$ becomes:
\begin{eqnarray}
\left[\mu^2 {\cal L}_{\rm MT} -\mu \frac{\partial^2}{\partial x \partial X} + 
{\cal L}_{\rm OL}\right] u= \mu^2 \lambda u, 
\label{eqn5}
\end{eqnarray}
where 
\begin{eqnarray}
{\cal L}_{\rm MT} &=& -\frac{1}{2} \frac{\partial^2}{\partial x^2} +\frac{1}{2} x^2,  
\label{eqn6}
\\[1.0ex]
{\cal L}_{\rm OL} &=& -\frac{1}{2} \frac{\partial^2}{\partial X^2}
+ V_0 \cos(2 X).
\label{eqn7}
\end{eqnarray}
One can then use a formal series expansion (in $\mu$) for $u$ and $\lambda$
\begin{eqnarray}
u &=& u_0+ \mu u_1 + \mu^2 u_2 + \dots\,\,,
\label{eqn8}
\\[1.0ex]
\lambda &=& \frac{\lambda_{-2}}{\mu^2} + \frac{\lambda_{-1}}{\mu} + \lambda_0 
+ \dots\,\, .
\label{eqn9}
\end{eqnarray}
Substitution of this expansion in the eigenvalue problem of Eq.~(\ref{eqn5}),  
and after tedious but straightforward algebraic manipulations and use of
solvability conditions for the first three orders of the expansion 
(${\cal O}(1)$, ${\cal O}(\mu)$ and ${\cal O}(\mu^2)$), yields the following
result for the eigenvalue and the corresponding eigenfunction of 
the eigenvalue problem of the original operator. The relevant
eigenvalue of the $n$-th mode is 
approximated by:
\begin{eqnarray}
E^{(n)} =-\frac{1}{4} V_0^2 + \left(1-\frac{1}{4} V_0^2 \right) 
\Omega \left(n+\frac{1}{2}\right),
\label{eqn10}
\end{eqnarray}
and the corresponding eigenfunction is given by:
\begin{equation}
\hskip-1.5cm
u^{(n)}(x) = c_n H_n \left(\frac{x}{\sqrt{1-\frac{V_0^2}{4}}}\right)
             e^{-\frac{x^2}{2-\frac{V_0^2}{2}}} 
\times \frac{1}{\sqrt{\pi}} \left[1-\frac{V_0}{2} \cos \left(\frac{2 x}{\Omega^{1/2}}
\right) \right],
\label{eqn11}
\end{equation}
where $c_n=(2^n n! \sqrt{\pi})^{-(1/2)}$ is the normalization factor and 
$H_n(x)=e^{-x^2} (-1)^n (d^n/dx^n) e^{x^2}$ are the Hermite polynomials.

\begin{figure}[t]
\center
\includegraphics[width=13cm]{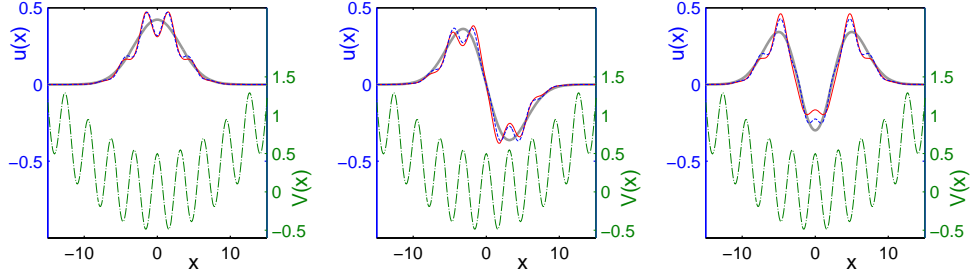}
\caption{
(Color online)
The first three (left to right) eigenfunctions of the linear 
Schr{\"o}dinger equation with the potential of Eq.~(\ref{s5_14}).
The thick gray solid line corresponds to the eigenfunction for
the case of $V_{\rm MT}(x)$, while the thin red solid and blue
dashed lines correspond to the eigenfunction for $V_{\rm MT}(x) + V_{\rm OL}(x)$, 
as computed numerically and theoretically [see Eq.~(\ref{eqn11})], respectively.
The full potential $V_{\rm MT}(x) + V_{\rm OL}(x)$ 
(shifted for visibility, see scale on the right axis) is 
illustrated by the green dash-doted line. The parameters used
in this example correspond to $V_0 = 0.5$ and $\Omega=0.1$.}
\label{rev_fig1}
\end{figure}

Considering now the nonlinear problem ${\cal L} u=-s u^3$ through
LS theory, we obtain the bifurcation function 
\begin{eqnarray}
G(\mu,\Delta E)= -\Delta E \mu + s\, \left\langle (u^{(n)})^2,(u^{(n)})^2 \right\rangle\,  \mu^3,
\label{s5_16}
\end{eqnarray}
for bifurcating solutions $U_n=\mu u^{(n)}$, which bifurcate from
the linear limit of $E=E^{(n)}$, with $\Delta E=E-E^{(n)}$. 
The notation $ \langle f,g \rangle =\int f(x) g(x) dx $ will be used to denote the inner
product of $f$ and $g$.
This calculation 
shows that a nontrivial solution exists only
if $s \Delta E>0$ (i.e., the branches bend to the left
for attractive nonlinearities with $s=-1$ and to the right for repulsive
ones with $s=1$) and the nonlinear solutions are created via a pitchfork
bifurcation (given the symmetry $u \rightarrow -u$) from the linear solution.
The bifurcation is subcritical for $s=-1$ and supercritical for $s=1$.
Typical examples of the relevant solutions of the linear problem 
(from which the nonlinear states
bifurcate) are shown in Fig.~\ref{rev_fig1}.

Notice that for $V_0=0$ the problem becomes the linear quantum harmonic
oscillator (parabolic potential) 
whose eigenvalues and eigenfunctions are known explicitly
and are a special case of those of Eqs.~(\ref{eqn10})--(\ref{eqn11})
for $V_0=0$. This perspective has been used in numerous studies
as a starting point for numerical computation, e.g., in 1D 
\cite{chap01:gp1dB,rodrigues,alfimov}, as well as in higher dimensions
\cite{tristram2}.

It is natural to subsequently examine the stability of the
ensuing nonlinear states, stemming from the linear problem. This can
be done through the linearization of the problem around the
nonlinear continuation of the
solutions $u^{(n)}$, with a perturbation $\tilde{u}=w+i v$. Then,
the ensuing linearized equations for the eigenvalue 
$\lambda$ and the eigenfunction $\tilde{u}$ can be written in the standard
(for NLS equations) $L_+, L_-$ form:
\begin{eqnarray}
L_+ w &=& \left[{\cal L} +3 s \left(u^{(n)}\right)^2 \right] w = - \lambda v
\label{s5_17}
\\[1.0ex]
L_- v &=& \left[{\cal L} + s \left(u^{(n)}\right)^2 \right] v = \lambda u.
\label{s5_18}
\end{eqnarray}
Then, define $n(L)$ and $z(L)$ the number of negative
and zero eigenvalues respectively of operator $L$, $k_{r}$, $k_i^{-}$
and $k_c$ the number of eigenvalues with, respectively,
real positive, imaginary with positive imaginary
part and negative Krein sign (see below) and complex with positive
real and imaginary part. The Krein signature of an eigenvalue 
$\lambda$ is the ${\rm sign}( \langle w,L_+ w \rangle )$. One can then use the 
recently proven theorem for general Hamiltonian systems of the
NLS type of 
Ref.~\cite{kks} based on the earlier work of Refs.~\cite{grillakis1a,grillakis1b,grillakis3a,grillakis3b}
(see also Ref.~\cite{pelinovsky}) according to which:
\begin{eqnarray}
k_r + 2 k_i^{-} + 2 k_c = n(L_+) + n(L_-) - n (D),
\label{s5_19}
\end{eqnarray}
where $D=dN/dE$ ($N$ is the number of atoms of the state of interest). 
Then from Sturm-Liouville theory and given that $u^{(n)}$ is the
only eigenfunction of $L_-$ with eigenvalue 0 (by construction),
we obtain $n(L_-)=j$ and $z(L_-)=1$ for the eigenstate $U=\mu u^{(j)}$
(since it possesses $j$ nodes).
Similarly, using the nature of the bifurcation, one obtains 
$n(L_+)=j$ and $n(D)=0$ if $s=1$, while $n(L_+)=j+1$ and
$n(L_-)=1$ if $s=-1$. Combining these results one has 
that $k_r + 2 k_i^{-} + 2 k_c = 2 j$. More detailed considerations
in the vicinity of the linear limit \cite{toddchaos} in fact show that the
resulting eigenvalues have to be simple and purely imaginary i.e.,
$k_r=k_c=0$ and, hence, each of the waves bifurcating from the linear
limit is {\it spectrally stable} close to that limit. However, the
nonlinear wave bifurcating from the $j$-th linear eigenstate has $j$
eigenvalues with negative Krein signature which may result in
instability if these collide with other eigenvalues (of opposite sign).
That is to say, the state $u^{(j)}$ has $j$ potentially unstable eigendirections.

While the above results give a detailed handle on the stability
of the structures near the linear limit, it is important to also 
quantify the bifurcations that may occur (which may also, in turn,
alter the stability of the nonlinear states), as well as to 
examine the dynamics of the relevant waves further away from
that limit. A reduction approach 
that may be used to address both of these issues is that of 
projecting the dynamics to a full basis of eigenmodes of the
underlying linear operator. Notice that we have seen this method
before in the reduction of the GP equation 
in the presence of a strong OL to the DNLS equation. Considering the problem 
$i u_t={\cal L} u+ s |u|^2 u - \omega u$, we can use the decomposition
$u(x)=\sum_{j=0}^M c_j(t) q_j(x)$, where $q_j(x)$ are the orthonormalized
eigenstates of the linear operator ${\cal L}$.
Setting $a_{klm}^{j}= \langle q_k q_l q_m,q_j \rangle $ and following Ref.~\cite{toddnon}
straightforwardly yields 
\begin{eqnarray}
i \dot{c}_j= (\mu_j-\omega) c_j + s \sum_{k,l,m=0}^M a_{klm}^{j} c_k c_l
c_m^{\ast},
\label{s5_20}
\end{eqnarray}
where $\mu_j$ are the corresponding eigenvalues of the eigenstates
$q_j(x)$.
It is interesting to note that this system with $M \rightarrow \infty$
is equivalent to the original dynamical system, but is practically
considered for finite $M$, constituting a Galerkin truncation of the
original GP PDE. This system preserves both the Hamiltonian structure
of the original equation, as well as additional conservation laws such
as the $L^2$ norm $||u||_{L^2}^2=\sum_{j=0}^M |c_j|^2$.
 
\begin{figure}[t]
\center
\includegraphics[width=6.4cm,height=4.5cm]{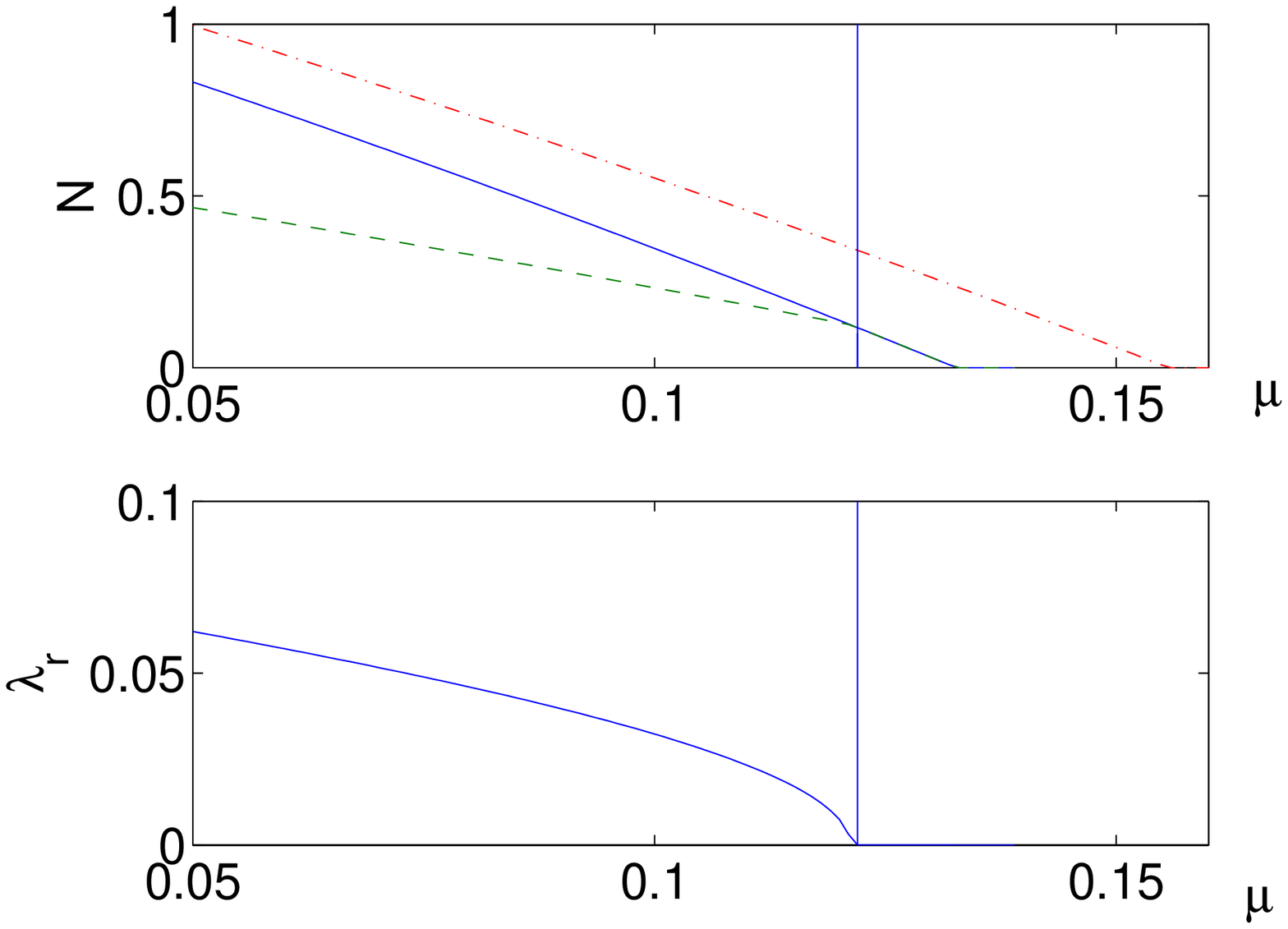}~~~
\includegraphics[width=6.4cm,height=4.5cm]{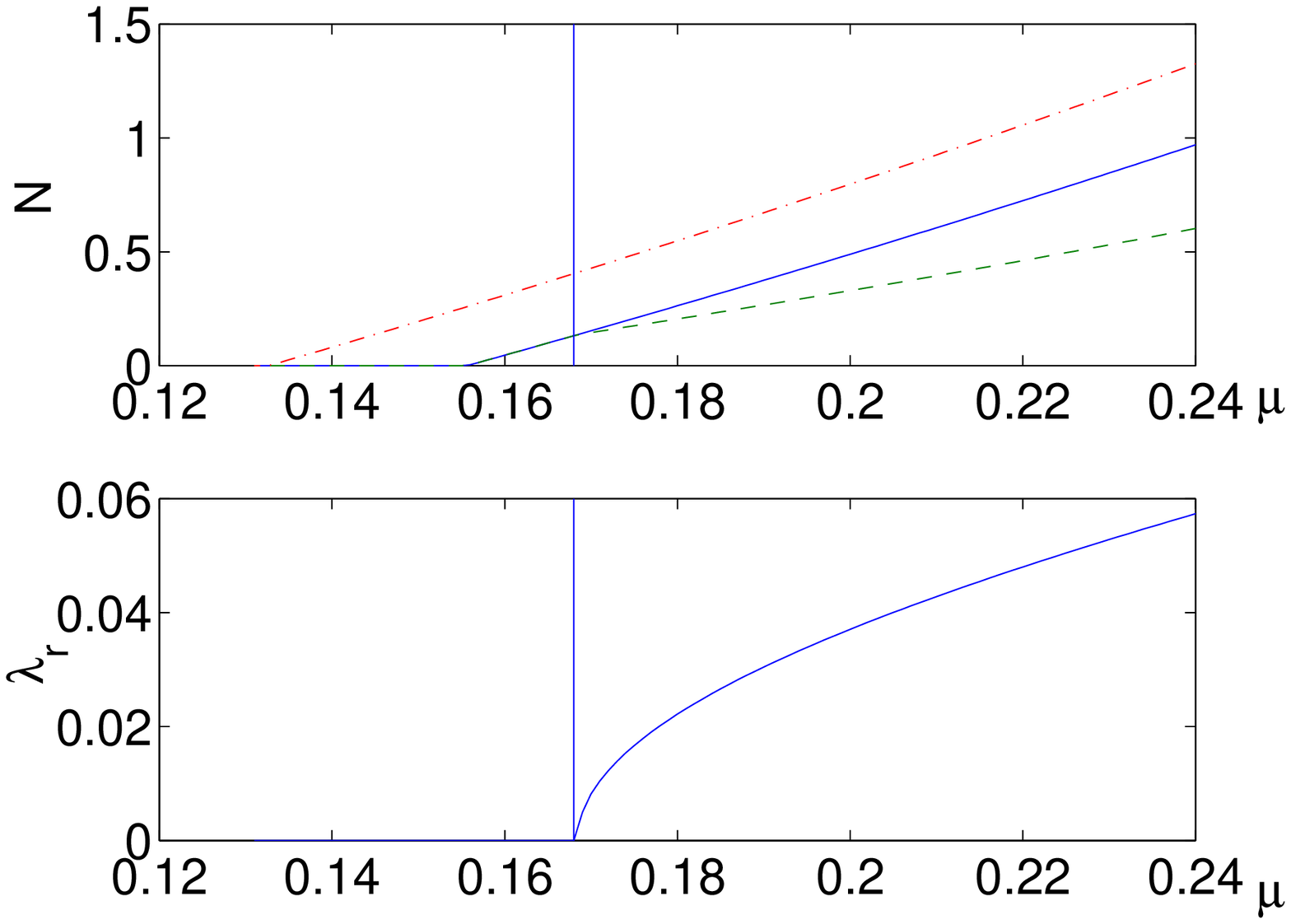}\\
\includegraphics[width=6.5cm,height=4.2cm]{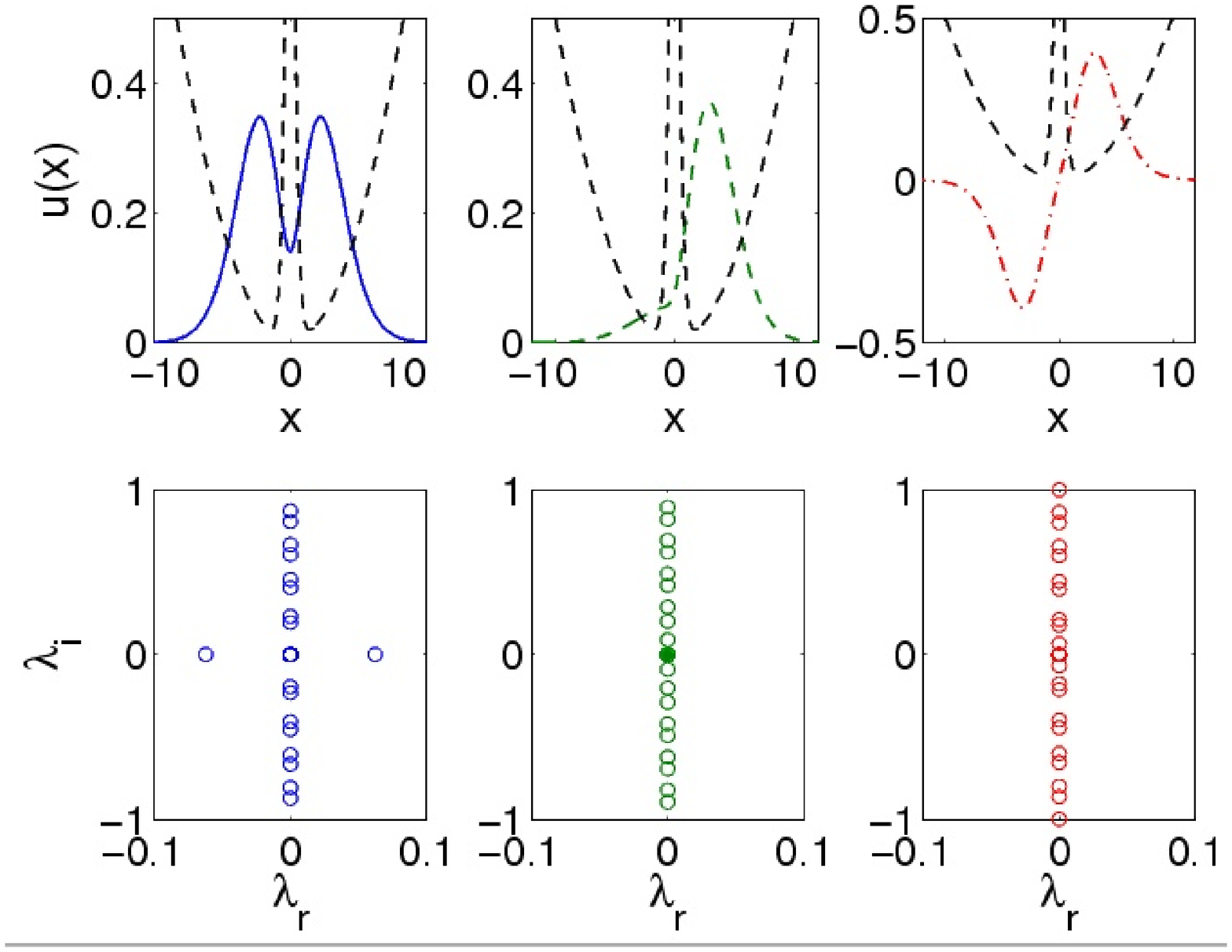}~
\includegraphics[width=6.5cm,height=4.2cm]{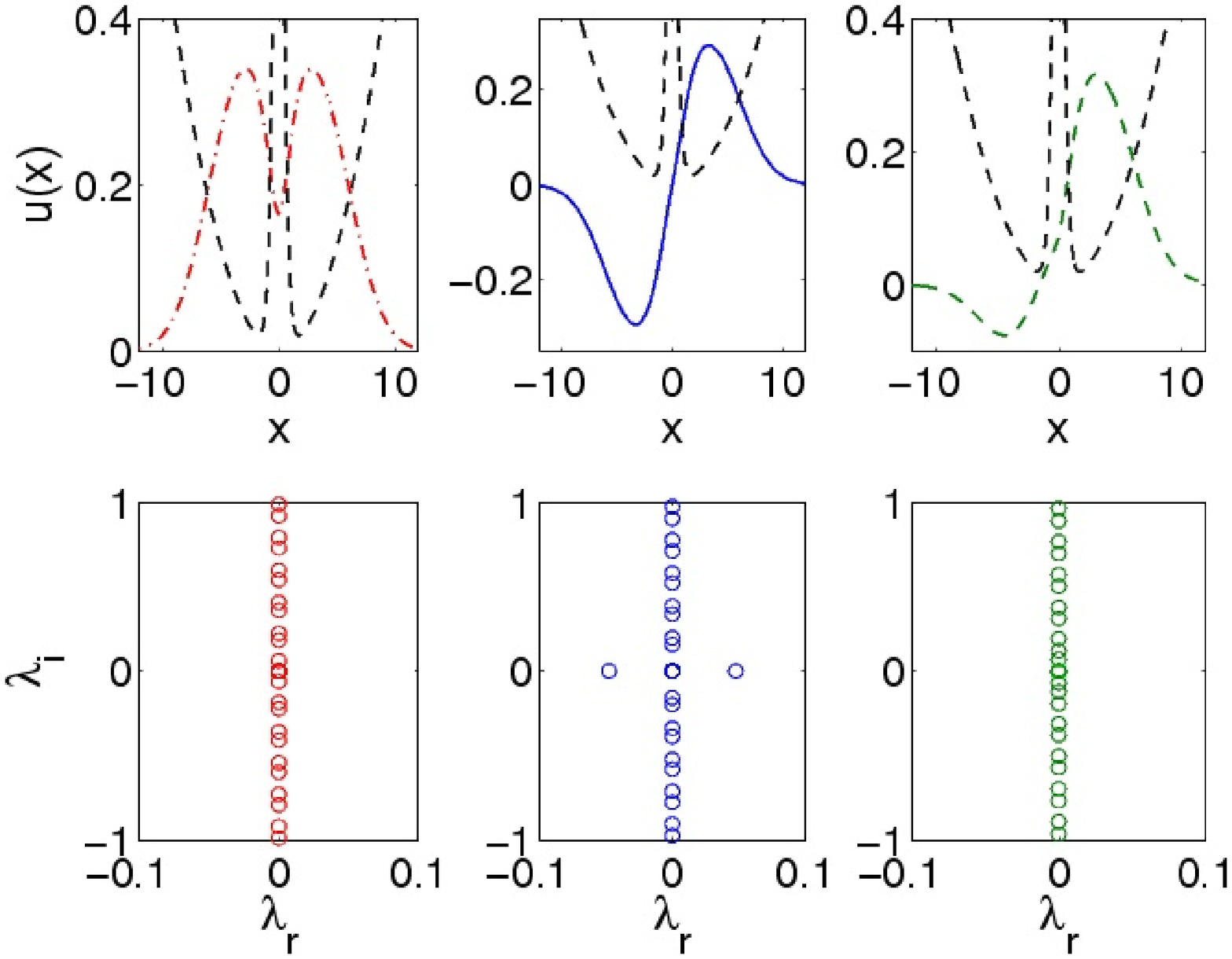}
\caption{
(Color online)
The left panels illustrate a typical double well problem
in the focusing case, while the right ones in the defocusing case.
The top row shows the squared $L^2$ norm of the solutions ($N$) 
as a function of the
eigenvalue parameter $\mu$ (illustrating in each case the bifurcation
of a new asymmetric branch). The second row shows the instability of
the solid (blue) branch (symmetric in the left and antisymmetric in the right)
past the critical point through the appearance of a real eigenvalue. 
The third row shows a particular example of the profiles
for each branch and the fourth
row shows the spectral plane of the linearization around them (absence
of a real eigenvalue indicates stability).}
\label{rev_fig2}
\end{figure}

A relevant question is then how many modes one
should consider to obtain a useful/interesting/faithful description
of the original infinite dimensional dynamical system. 
The answer, naturally, depends on the form of the potential.
The above reduction has been extremely successful in tackling
double well potentials in BECs \cite{toddnon,giorgossb,miwkirr},
as well as in optical systems \cite{zhigangssb}. In this simplest
case, a two-mode description is sufficient to extract the prototypical
dynamics of the system with $M=2$. Then the relevant dynamical equations
become:
\begin{eqnarray}
i \dot{c}_0 &=& (\mu_0-\omega) c_0 + s a_{000}^0 |c_0|^2 c_0 + s a_{110}^0
(2 |c_1|^2 c_0+c_1^2 c_0^{\ast}), 
\label{s5_21}
\\[1.0ex]
i \dot{c}_1 &=& (\mu_1-\omega) c_1 + s a_{111}^1 |c_1|^2 c_1
+ sa_{110}^0 (2 |c_0|^2 c_1 + c_0^2 c_1^{\ast}),
\label{s5_22}
\end{eqnarray}
where we have assumed a symmetric double well potential so that
terms such as $a_{000}^1$ or $a_{111}^0$ disappear 
and 
$a_{001}^1=a_{110}^0=a_{010}^1=\dots=\int q_0^2 q_1^2 dx$.
None of these assumptions is binding and the general cases have in fact
been treated in Refs.~\cite{toddnon,giorgossb}. An angle-action variable
decomposition $c_j=\rho_j e^{i \phi_j}$ leads to 
\begin{eqnarray}
\dot{\rho}_0 &=& s a_{110}^0 \rho_1^2 \rho_0 \sin(2 \delta \phi),
\label{s5_23}
\\[1.0ex]
\dot{\delta \phi} &=& -\Delta \mu + s (a_{000}^0 \rho_0^2-a_{111}^1 \rho_1^2)
\nonumber
\\
&& -s a_{110}^0 \left(2 + 2 \cos(\delta \phi)\right) (\rho_0^2-\rho_1^2),
\label{s5_24}
\end{eqnarray} 
where $\delta \phi$ is the relative phase between the modes. 
Straightforwardly analyzing the ensuing equations [particularly
Eq.~(\ref{s5_23})], we observe that the nonlinear problem can support
states with $\rho_1=0$ and $\rho_0 \neq 0$ (symmetric ones, respecting
the symmetry of the ground state of the double well potential).
It can also support ones with $\rho_0=0$ and $\rho_1 \neq 0$ (antisymmetric
ones); these states are not a surprise since they did exist even at the
linear limit. However, in addition to these two, the nonlinear problem
can support states with $\rho_0 \neq 0$ {\it and} $\rho_1 \neq 0$, provided
that $\sin(2 \delta \phi)=0$. These are {\it asymmetric} states that
have to bifurcate because of the presence of nonlinearity. A more
detailed study of the second equation shows that, typically,
the bifurcation occurs
for $||u||_{L^2}^2>N_c=\Delta \mu/(3 a_{110}^0-a_{000}^0)$ for $s=-1$ (attractive case) 
and is a bifurcation
from the symmetric ground state branch, while it happens for
$||u||_{L^2}^2>N_c=\Delta \mu/( 3 a_{110}^0-a_{111}^1)$ 
in the $s=1$ (repulsive case)
and is a bifurcation from the antisymmetric 
first excited state (see also Fig.~\ref{rev_fig2}
which shows the relevant states and bifurcation diagram). 
Notice that this is a pitchfork bifurcation (since there are two
asymmetric states born at the critical point, each having principally
support over each of the two wells). It should be mentioned that 
although this bifurcation is established for the two-mode reduction,
it has been systematically confirmed by numerical analysis of the
GP PDE in Refs.~\cite{toddnon,giorgossb} for the case of a magnetic
trap and an optical lattice or a magnetic trap and a defect respectively
and it has been rigorously proved for a decaying at infinity double
well potential in Ref.~\cite{miwkirr} (in the latter the corrections
to the above mentioned $N_c$ were estimated). Based on the nature of
the bifurcation (but this can also be proved within the two-mode reduction
and rigorously from the GP equation), we expect the ensuing asymmetric
solutions to be stable, destabilizing the branch from which they are
stemming, as is confirmed in the numerical linear stability results
of Fig.~\ref{rev_fig2}. It is also worthwhile to point out that such
predictions (e.g., the stabilization of an asymmetric state
beyond a critical power) have been directly confirmed in optical experiments
\cite{zhigangssb} in photorefractive crystals, and also have a direct
bearing on relevant BEC experiments analyzed in Ref.~\cite{chap01:albiez}.
We note in passing that in the physics literature, the problem is often
tackled using wavefunctions that are localized in each of the wells of
the double well potential (as linear combinations of the states
$q_1$ and $q_2$ used herein) \cite{rag,ostr,an}, especially to study
Josephson tunneling (but also to examine self-trapping) \cite{chap01:albiez}.
We refer the interested reader to these publications for more details.

Such a few-mode approximation has also been successfully used in
the case of three wells (in connection to applications in experiments
in photorefractive crystals) in Ref.~\cite{toddzhigang}. Naturally in that
case, one uses three modes for the relevant decomposition and the
corresponding dynamical equations grow in complexity very rapidly
(as numerous additional overlap terms become relevant and the
analysis becomes almost intractable). Similarly, such Galerkin approaches
can also be used in the case where there is only a magnetic trap, in
which case the underlying basis of expansion becomes that of the
Hermite-Gauss polynomials \cite{pelioscil}. In that case, in addition
to the persistence of the linear states and a detailed quantitative analysis of
their linearization spectrum  that becomes available
near the linear limit, one can importantly predict the formation of 
new types of solutions. An example of this type consists of the
space-localized, time-periodic (i.e., breathing) solutions in the 
neighborhood of, e.g., the first excited state (which has the form of a dark soliton) of the 
harmonically confined linear problem 
\cite{pelioscil}.

\begin{figure}[t]
\center
\hskip2.0cm
\includegraphics[width=4cm,height=3cm]{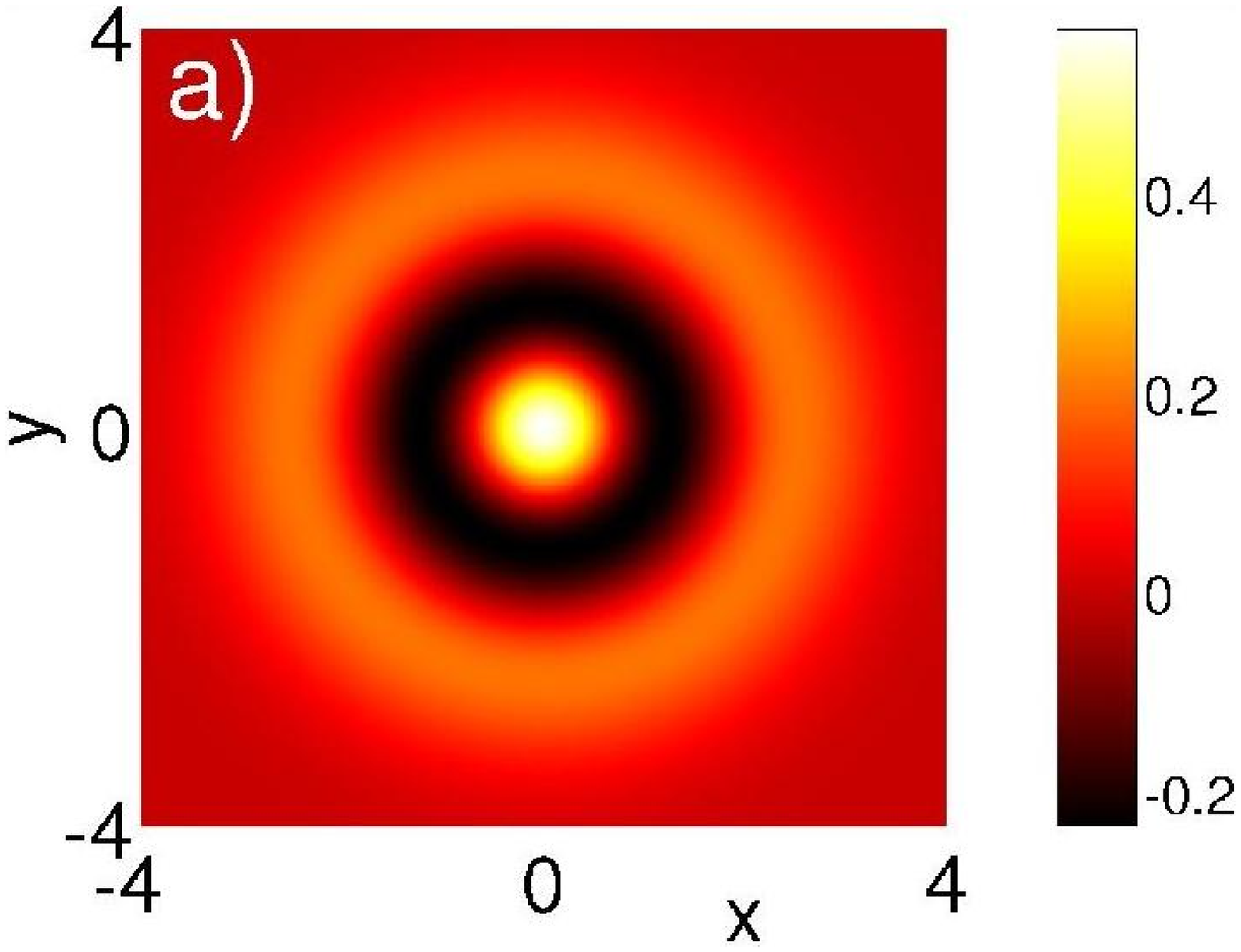}
\includegraphics[width=4cm,height=3cm]{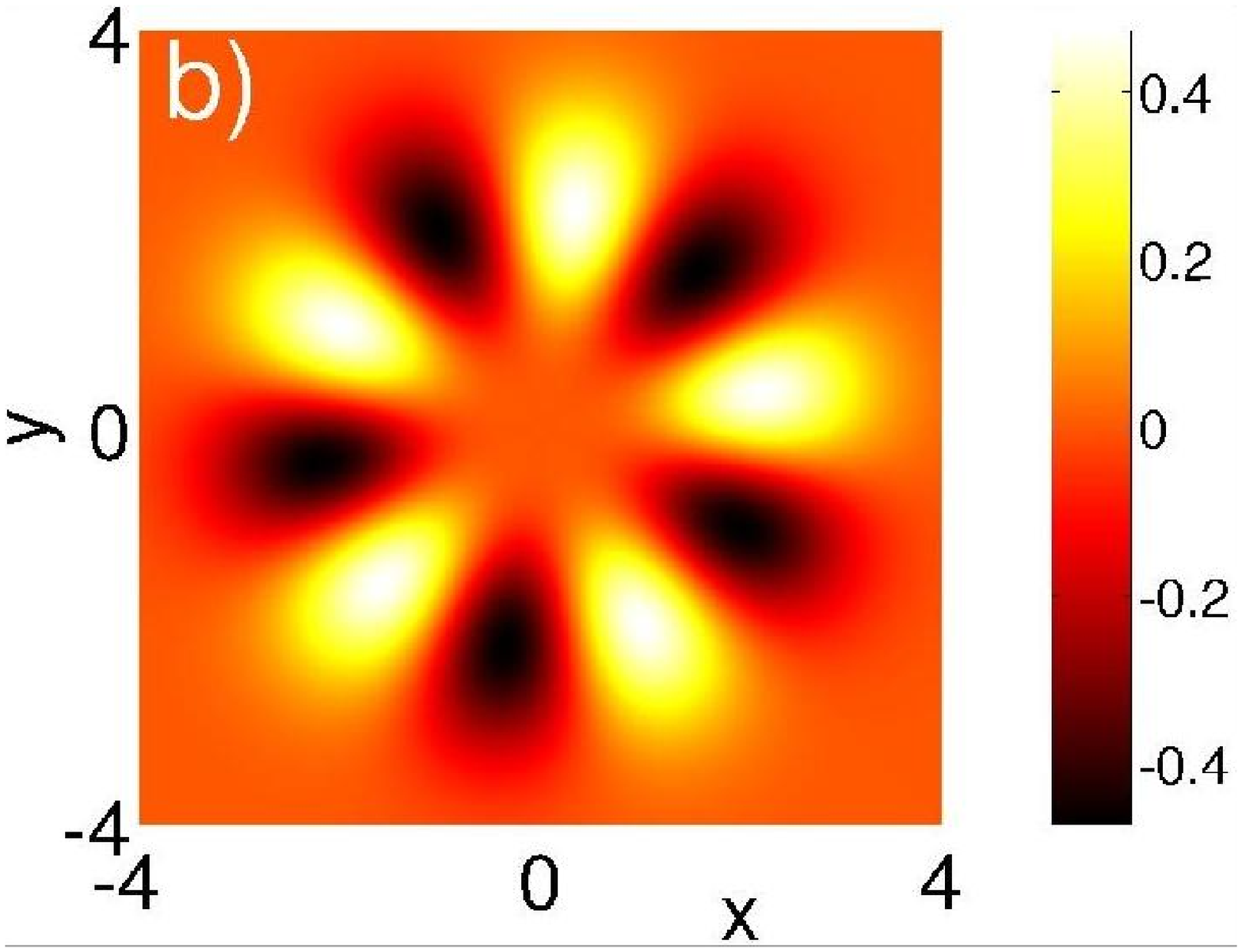}
\\[1.0ex]
\hskip2.0cm
\includegraphics[width=4cm,height=3cm]{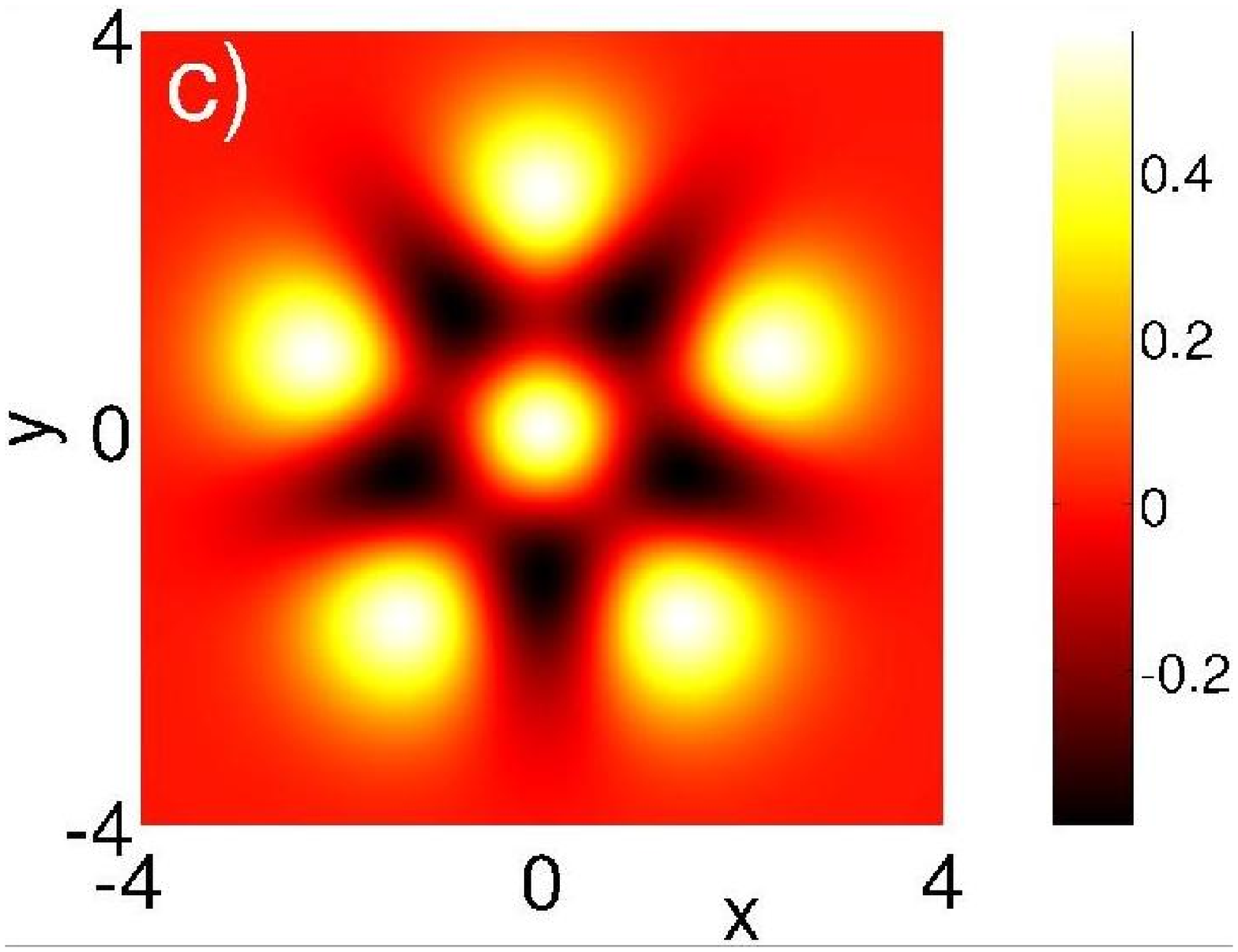}
\includegraphics[width=4cm,height=3cm]{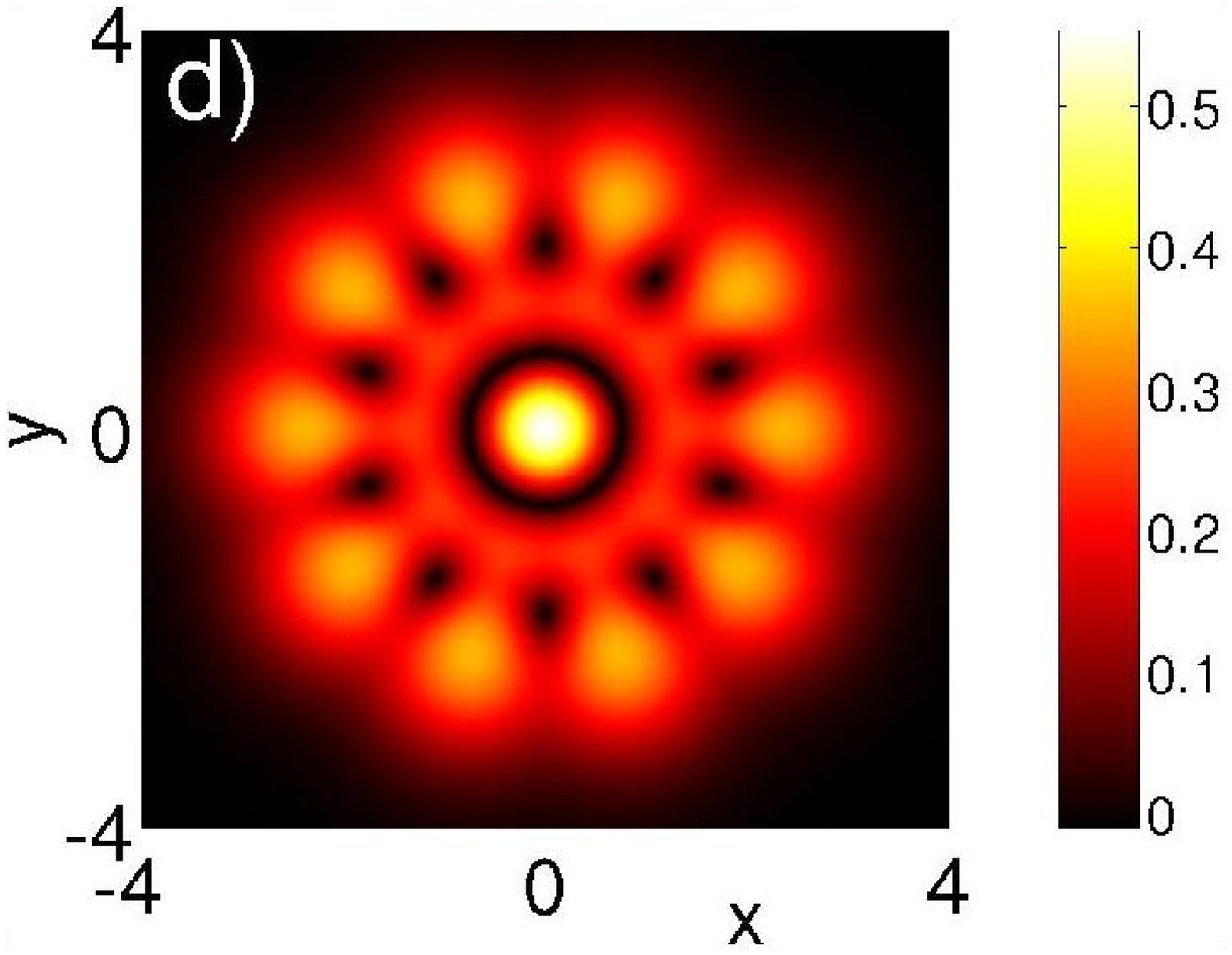}
\caption{
(Color online)
A typical ring solution (top left), multipole solution
(top right), soliton necklace (bottom left) and vortex necklace (bottom
right) that can be obtained from the near-linear analysis of the 2D problem
through Eqs.~(\ref{s5_27})--(\ref{s5_29}).
Reprinted from Ref.~\cite{toddricardo} with permission.}
\label{rev_fig3}
\end{figure}

Finally, we indicate that such methods from the linear limit can
equally straightforwardly be applied to higher dimensional settings
and be used to extract complex nonlinear states. For instance, 
considering the problem
\begin{eqnarray}
-\frac{\partial^2 u}{\partial r^2} - \frac{1}{r} \frac{\partial u}{\partial r}
-\frac{1}{r^2} \frac{\partial^2 u}{\partial \theta^2} + i \Omega \frac{\partial u}{\partial \theta} + r^2 u + s |u|^2 u= \omega u,
\label{s5_25}
\end{eqnarray}
where also a rotational stirring term (frequent to condensate
experiments \cite{book1,book2}) is included, one can use a
decomposition into the linear states $q_{m,l}(r) \exp(i l \theta)$
\cite{toddricardo}, 
where $m$ is the number of nodes of $q_{m,l}$. Then the underlying
linear problem has eigenvalues $\lambda_{m,l}=2(|l|+1)+4 m + l \Omega$
and e.g., for solutions bifurcating from $\lambda=6$, one can write
\begin{eqnarray}
u=\left(x_1 q_{1,0}(r) + y_1 q_{0,l'} \cos(l'\theta) + i y_2 q_{0,l'} 
\sin(l' \theta) \right) \epsilon^{1/2}, 
\label{s5_26}
\end{eqnarray}
together with $\omega=6 + \epsilon \delta \omega$, and derive algebraic
equations for $x_1, y_1$ and $y_2$:
\begin{eqnarray}
0 &=& x_1 \left[\mu+2 x_1^2 + g_1 (2 |y_1|^2+y_1^2+y_2^2) \right],
\label{s5_27}
\\
0 &=& c_g \mu y_1 + g_2 x_1^2 (2 y_1+y_1^{\ast}) + \frac{3}{4} |y_1|^2 y_1
+\frac{1}{4} y_2^2 (2 y_1-y_1^{\ast}),
\label{s5_28}
\\
0 &=& y_2 \left[c_g \mu + g_2 x_1^2 + 
\frac{1}{4} (2 |y_1|^2-y_1^2) + \frac{3}{4} y_2^2 \right], 
\label{s5_29}
\end{eqnarray}
where $\mu=-s \delta \omega/(g_0 \pi)$, $g_1=g_{0,l'}/g_0$,
$g_2=g_{0,l'}/g_{l'}$ and $c_g=g_2/g_1$ and $g_1=\int r q_{1,0}^4 dr$,
$g_{l'}=\int r q_{0,l'}^4 dr$, $g_{0,l'}=\int r q_{1,0}^2 q_{0,l'}^2 dr$.
From these equations one can find real solutions containing
only $x_1$ (ring solutions), only $y_1$ (multipole solutions),
both $x_1$ and $y_1$ (soliton necklaces), as well as complex
solutions also involving $y_2 \neq 0$ such as vortices and vortex
necklaces. A sampler of these solutions is illustrated in 
Fig.~\ref{rev_fig3}; more details can be found in Ref.~\cite{toddricardo},
where the stability of such states is also analyzed,
leading to the conclusion that the most robust among them
are the (soliton and vortex) necklace and the vortex states.

\subsection{Methods from the nonlinear limit}

We partition our consideration of such methods to ones that
tackle the stationary problem (in connection to the existence
and the stability of the solutions) and to ones that address the
dynamics of the perturbed solitary waves.

\subsubsection{Existence and stability methods.}
\label{esm_non}

Consider a general Hamiltonian system of the form:
\begin{eqnarray}
\frac{d v}{dt}=J E'(v),
\label{s5_30}
\end{eqnarray}
where the $J$ matrix has the standard symplectic structure ($J^2=-I$)
and $E=\int (1/2) [|v_x|^2 + s |v|^4] dx$ for the case of the GP equation
without external potential (although the formalism presented below 
is very general \cite{kks,kkjpa,kapitulapd}). Given that the above model in the case of the GP
equation has certain invariances (e.g., with respect to translation
and phase shift), one can use the generator $T_{\omega}$ of the 
corresponding semigroup $T(\exp(\omega t))$
of the relevant symmetry to make a change of variables
$v(t)=T(\exp(\omega t)) u(t)$, which in turn leads to 
$du/dt=J E_0'(u;\omega)$, where $E_0'(u;\omega)=E'(u)-J^{-1} T_{\omega} u$.
Defining then the appropriate conserved functional $Q_{\omega}=(1/2)
 \langle J^{-1}T_{\omega} u,u \rangle $, we note that relative equilibria 
satisfying $E_0'(u;\omega)=0$ will be critical points of $E_0(u;\omega)=
E(u)-Q_{\omega}(u)$. Then the linearization problem around such a
stationary wave $u_0$ reads:
\begin{eqnarray}
J E_0''(u_0;\omega) w = \lambda w.
\label{s5_31}
\end{eqnarray}
Given the symmetries of the problem, this linearization operator has
a non-vanishing kernel since:
\begin{eqnarray}
J E_0''(u_0;\omega) T_{\omega_i} u_0 &=& 0,
\label{s5_32}
\\[1.0ex]
J E_0''(u_0;\omega) \partial_{\omega_i} u_0 &=& T_{\omega_i} u_0,
\label{s5_33}
\end{eqnarray}
where each $i$ corresponds to one of the relevant symmetries and
the latter equation provides the generalized eigenvectors of the
operator.

The consideration of the perturbed Hamiltonian problem with a
Hamiltonian perturbation such that the perturbed energy is
$E_0(u)+\epsilon E_1(u)$ was considered in 
Refs.~\cite{kks,pelinovsky,kkjpa,kapitulapd} and a number of 
conclusions were reached regarding the existence and stability
of the ensuing solitary waves. Firstly, a necessary condition
for the persistence of the wave is:
\begin{eqnarray}
 \langle E_1'(u_0;\omega),T_{\omega_i} u_0 \rangle =0,
\label{s5_34}
\end{eqnarray}
for all $i$ pertaining to the original symmetries. 
This is a rather natural condition intuitively since it implies
that the perturbed wave is a stationary solution if it is a critical
point of the perturbation energy functional.
The condition
is also sufficient if the number of zero eigenvalues $z(M)$ of
the matrix 
$M_{ij}= \langle T_{\omega_i} u_0,E_1''(u_0;\omega) T_{\omega_j} u_0 \rangle $
is given by $n-k_s$, where $n$ is the multiplicity of the original
symmetries and $k_s$ the number of symmetries broken by the perturbation.

As a result of the perturbation, $2 k_s$ eigenvalues (corresponding
to the $k_s$ broken symmetries) will leave the origin, and can be tracked 
by the following result proved by means of 
LS reductions
in Ref.~\cite{kks}. The eigenvalues will be $\lambda=\sqrt{\epsilon} \lambda_1
+ {\cal O}(\epsilon)$, where the correction $\lambda_1$ is given by
the generalized eigenvalue problem:
\begin{eqnarray}
(D_0 \lambda_1^2  + M_1)\, v = 0,
\label{s5_35}
\end{eqnarray}
where the matrix of symmetries $(D_0)_{ij}= \langle \partial_{\omega_i} u_0,
E_0''(u_0;\omega) \partial_{\omega_j} u_0 \rangle $.

In addition to this perturbative result on the eigenvalues, one
can obtain a general count on the number of unstable eigendirections
of a Hamiltonian system \cite{kks}, 
using the functional analytic framework of 
Refs.~\cite{grillakis1a,grillakis1b,grillakis3a,grillakis3b} (see also Ref.~\cite{pelinovsky} for
a different approach). In particular, for a linearization operator
${\cal L}_{\omega}=E''(u)-J^{-1} T_{\omega}$ and a symmetry matrix
$D_{ij}= \langle \partial_{\omega_i} u, {\cal L}_{\omega} \partial_{\omega_j} u \rangle $,
\begin{eqnarray}
k_r + 2 k_i^{-} + 2 k_c = n({\cal L}_\omega) - n(D) -z(D),
\label{s5_36}
\end{eqnarray}
where the relevant symbolism has been introduced in Sec.~\ref{lin_lim}.
In fact, the latter subsection constitutes a special case example
of this formula, in the case of the form of 
\begin{eqnarray}
{\cal L}_\omega=
\left( \begin{array}{cc}
L_+ & 0 \\
0  & L_- \end{array} \right).
\label{call}
\end{eqnarray}

We now give a special case example of the application of the theory
in the presence of a linear and a nonlinear lattice potential of the
form \cite{rkkj}:
\begin{eqnarray}
i u_t =-\frac{1}{2} u_{xx} - \left(1+\epsilon n_1(x) \right) |u|^2 u 
+\epsilon n_2(x)u.
\label{pnls}
\end{eqnarray}
Then the problem can be rephrased in the above formalism
with 
\begin{eqnarray}
E_1(u)=\int_{-\infty}^{+\infty} \left(n_2(x)|u|^2- \frac{1}{2} n_1(x)|u|^4\right) dx.
\end{eqnarray}
Therefore, as indicated above, the persistence of the stationary bright solitary 
wave of the form $u_0=\sqrt{\mu}\, {\rm sech} [\sqrt{\mu} (x-\xi)] e^{i \delta}$
(with $\delta=\mu/2$) is tantamount to: $\nabla_{\xi} E_1(u)=0$. This implies
that the wave is going to
persist only if centered at the parameter-selected 
extrema of the energy (which are now going to form, at best, a countably
infinite set of solutions, as opposed to the one-parameter infinity of 
solutions previously allowed by the translational invariance).

\begin{figure}[t]
\begin{center}
\hskip1.0cm
\includegraphics[width=5.9cm]{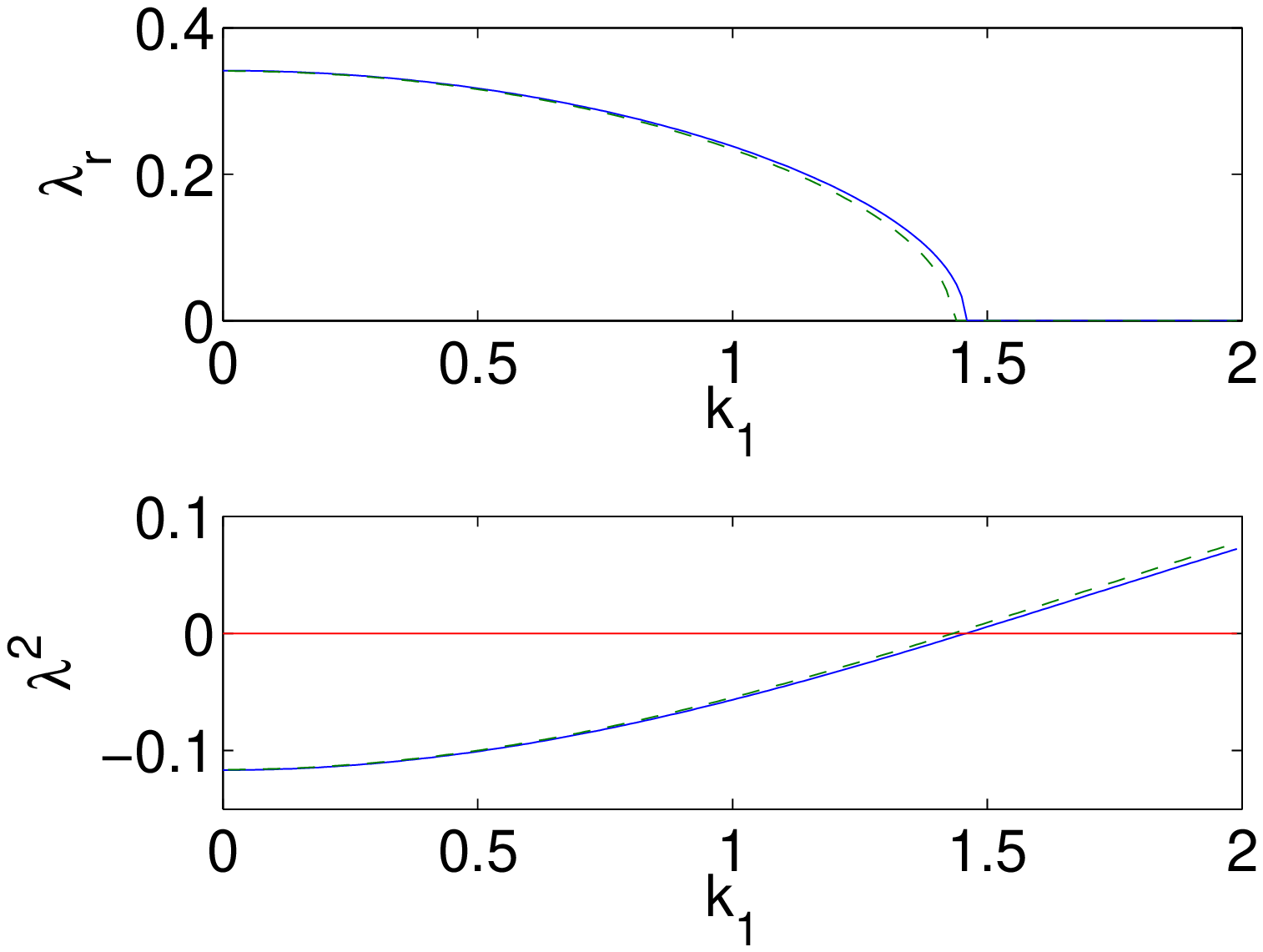}
\includegraphics[width=5.9cm]{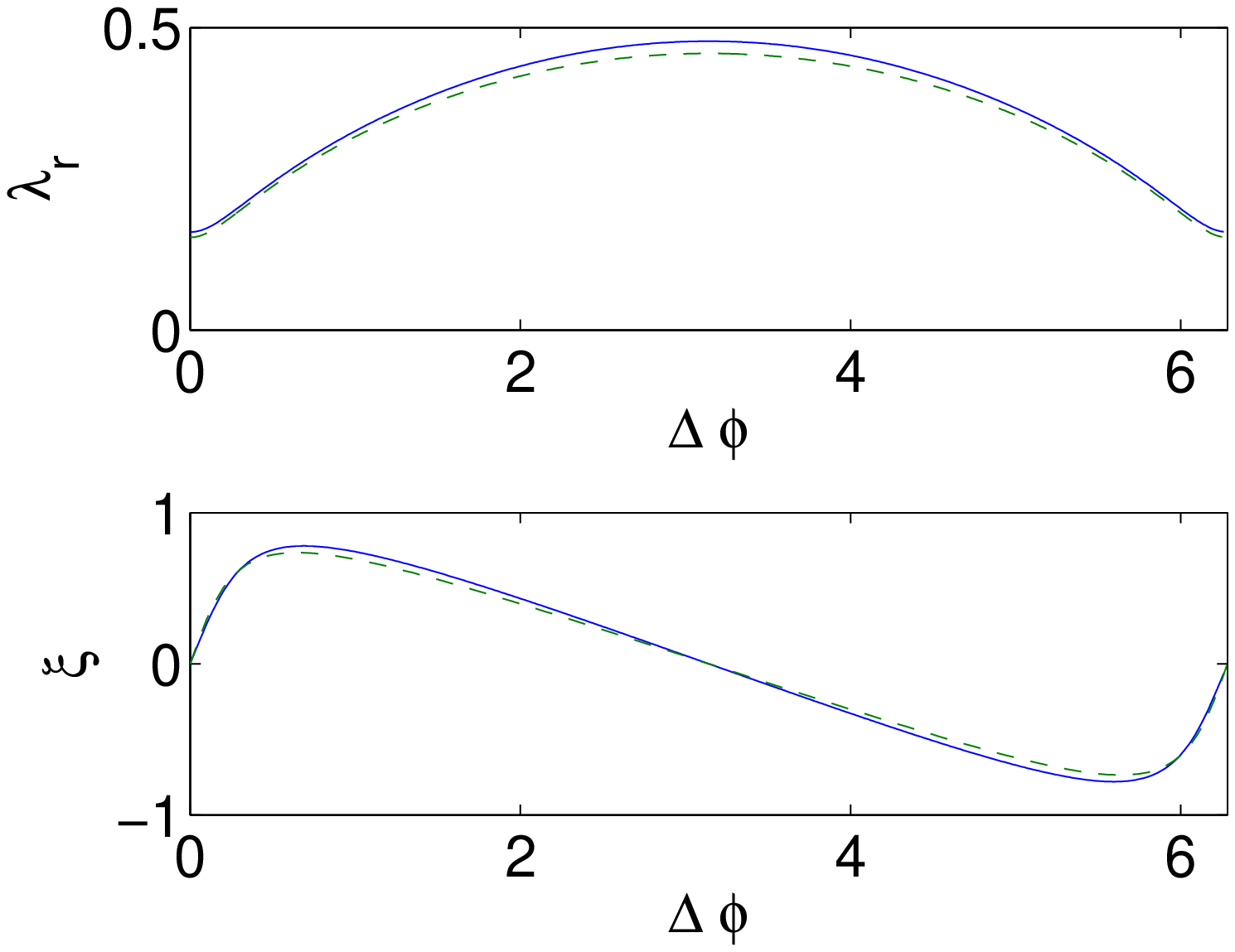}
\end{center}
\vskip-0.4cm
\caption{
(Color online)
Typical examples of the translational eigenvalue as obtained
numerically (solid/blue line) versus the analytical prediction (dashed/green line).
The linear and nonlinear potentials are: $n_1(x)=A \cos(k_1 x)$
and $n_2(x)=B \cos(k_2 x +\Delta \phi)$. In the left panels we assume 
that $A=B=1$, and fix $k_2=2 \pi/5$ and $\Delta \phi=0$ and examine
the relevant translational eigenvalue (its real part and its square)
as a function of $k_1$. Notice that there is a transition from
instability to stability as $k_1$ is increased. In the right panels we 
select $A=B=1$ and $k_1=k_2=2 \pi/5$ and vary $\Delta \phi \in [0,2\pi]$.
Notice that in the latter case the (critical point associated with the 
stationary) 
location of the solitary wave also
changes with $\Delta \phi$ and its theoretical and numerical values 
are also given (again in dashed and solid lines, respectively). Notice
in all the cases the accuracy of the theoretical prediction.}
\label{rev_fig4}
\end{figure}

Furthermore, the stability of the perturbed wave is determined by
the location of the eigenvalues associated with the translational
invariance; previously, the relevant 
eigenvalue pair was located at the origin $\lambda=0$
of the spectral plane of eigenvalues $\lambda=\lambda_r + i \lambda_i$.
On the other hand, we expect the eigenvalues associated with the
U(1) invariance (i.e., the phase invariance associated with the
$L^2$ conservation) to remain at the origin, given the preservation
of the latter symmetry under the perturbations considered herein.
Adapting the framework of Ref.~\cite{kks}, we have that the matrices that
arise in Eq.~(\ref{s5_35}) are given by:
\begin{eqnarray}
D_0=
\left( \begin{array}{cc}
(\partial_x u_0,-x u_0) & 0 \\
0  & 2 (u_0,\partial_{\mu} u_0) \end{array} \right)
= \left( \begin{array}{cc}
\mu^{1/2} & 0 \\
0  & -\mu^{-1/2} \end{array} \right),
\label{new2}
\end{eqnarray}
and
\begin{eqnarray}
M_1 &=&
\left( \begin{array}{cc}
\frac{\partial}{\partial \xi}(\frac{\partial E_1}{\partial u_0^{\ast}}, \partial_{\xi} u_0) & 0 \\
0  & 0 \end{array} \right)
\nonumber
\\
&=& \left( \begin{array}{cc}
\int \left( \frac{1}{2} \frac{d^2 n_2}{d x^2} (u^0)^2 - \frac{1}{4}  
\frac{d^2 n_1}{d x^2} (u_0)^4 \right) dx & 0 \\
0 & 0 
\end{array} \right).
\end{eqnarray}
One can then use the above along with Eq.~(\ref{s5_35}) to obtain the
relevant translational eigenvalue as:
\begin{eqnarray}
\lambda^2=- \frac{\epsilon}{\mu^{1/2}} 
\int \left( \frac{1}{2} \frac{d^2 n_2}{d x^2} (u_0)^2 - \frac{1}{4}  
\frac{d^2 n_1}{d x^2} (u_0)^4 \right) dx.
\label{extra2}
\end{eqnarray}
Based on this expression, the corresponding eigenvalue can be
directly evaluated, provided that the extrema of the effective energy 
landscape 
$E_1$ are evaluated first.
This effective energy landscape $V_{\rm eff}(\xi)=\epsilon E_1$ is a function
of the solitary wave location $\xi$. 
The physical intuition of the above results is that the 
stability or instability of the configuration will be associated
with the convexity or concavity, respectively, 
of this effective energy landscape. Some examples of the accuracy
of such a prediction are provided in Fig.~\ref{rev_fig4}, for specific
forms of $n_1(x)$ and $n_2(x)$.

This class of techniques has been applied to different problems with
spatial variation of the linear \cite{superl} 
or nonlinear \cite{yiota}
potential. They can also be 
applied to multi-component problems \cite{kkjpa} or to problems
with different nonlinearity exponents \cite{miwpd} or higher dimensions
\cite{miwprl}. We note in passing that in addition to these
methods, for periodic variations of the potential, and for 
appropriate regimes (for details see Refs.~\cite{miwpd,miwprl,wang}),
one can develop multiple-scale techniques exploiting the disparity
in spatial scales between the solution and the potential. We refer
the interested reader to the above references for further technical
details. 
This type of averaging techniques is popular not
only when the linear or nonlinear potential presents spatial variations
of a characteristic scale, but also similarly when these variations
are temporal \cite{kkp1,kkpz,peza}, especially because it is more
straightforward in the averaged equations to extract conclusions
about the possible existence or potential collapse or dispersion of
the solutions \cite{miwpd,miwprl,stefanov}.

\begin{figure}[t]
\begin{center}
\hskip1.0cm
\includegraphics[width=5.9cm,height=4cm]{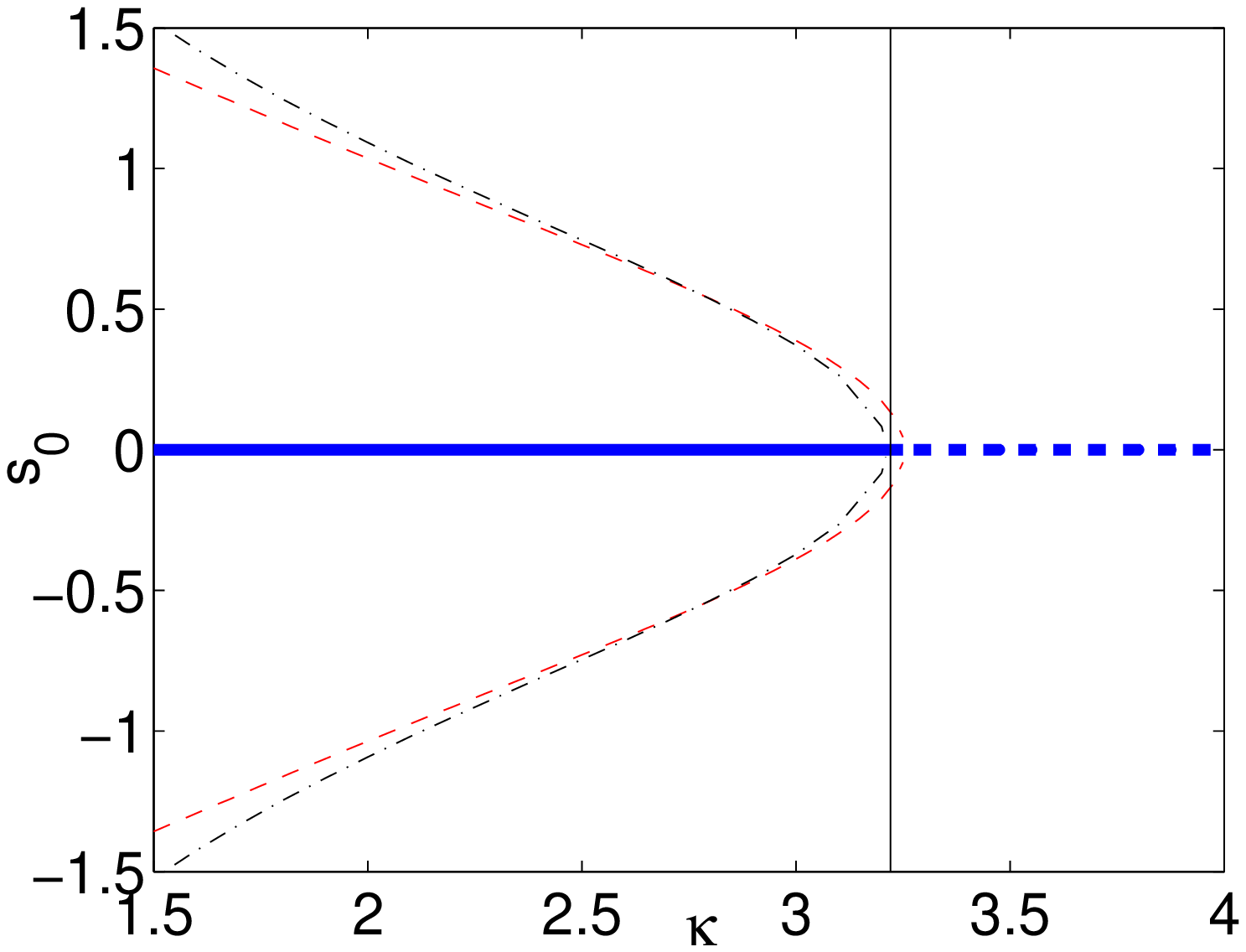}
\includegraphics[width=5.9cm,height=4cm]{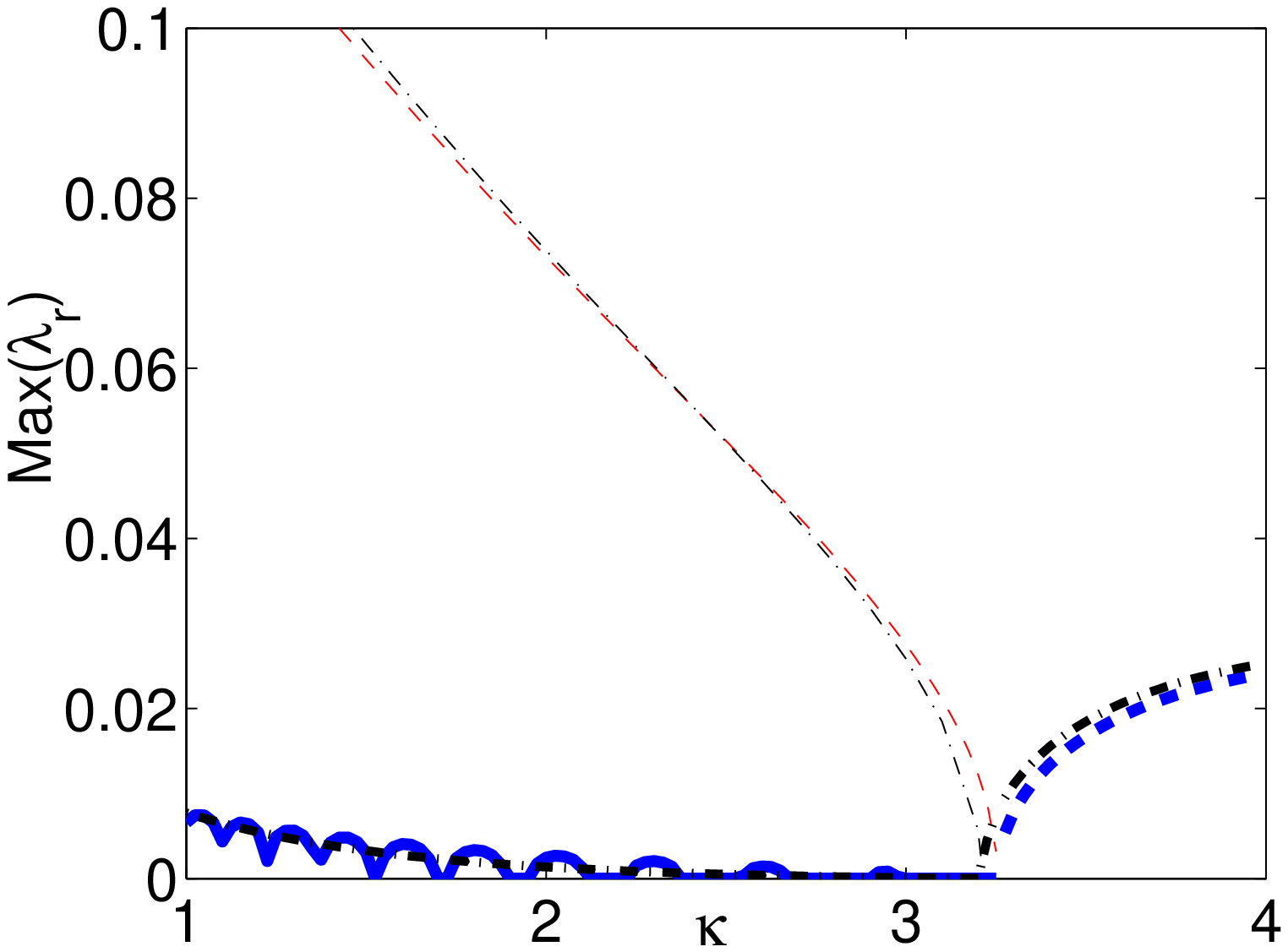}
\end{center}
\vskip-0.4cm
\caption{
(Color online)
An example of a subcritical pitchfork bifurcation in the parameter
$\kappa$ of the potential $V(x)= x^2 \exp(-\kappa |x|)$
for fixed $\epsilon = 0.2$ in Eq.~(\ref{s5_41}).
The left panel shows the center of mass $s_0 \equiv \xi$ of the 
dark soliton kink modes
($s_0 \neq 0$ by dashed line, $s_0 = 0$ by thick solid and dashed
lines). The theoretical predictions of $s_0$ based on
Eq.~(\ref{s5_42}) are shown by
dash-dotted line. The vertical line gives the theoretical
prediction for the bifurcation point $\kappa = \kappa_0 =3.21$. The
right panel shows the real part of the unstable eigenvalues for
the corresponding solutions, using the same symbolism as the left
panel.  The theoretical predictions of eigenvalues are shown by
thick and thin dash-dotted lines, respectively for the branches
with $s_0=0$ and $s_0 \neq 0$. Notice that for the quartet of eigenvalues
of the branch centered at the origin, the small jumps are due to the
finite size of the computational domain (see Ref.~\cite{kevpeli} for
details).}
\label{rev_fig5}
\end{figure}

A similar approach can be used in the case of dark solitons
in examining the persistence and stability of the waves in the presence
of external potentials; however in the latter case, it is a much
harder task to control the linearization spectrum of the problem since
it encompasses the origin. This complication has allowed this problem
to be tackled only recently at the nonlinear limit \cite{dimenza}
and perturbatively away from that limit \cite{kevpeli}. The main
result of Ref.~\cite{dimenza} is that by using the limit
\begin{eqnarray}
\lim_{\lambda \rightarrow 0} g(\lambda)= \lim_{\lambda \rightarrow 0}
 \langle (L_-\lambda)^{-1} u',u' \rangle ,
\label{s5_40}
\end{eqnarray}
one can infer the stability of the black soliton, since if this
quantity is positive the soliton will have a real eigenvalue
and will be unstable, while if non-positive, it will be stable.
However, one of the problems with this expression is that
even when the soliton is analytically available, it is relatively
hard to evaluate (see, e.g., the example of the integrable cubic
case worked out in Ref.~\cite{dimenza}). On the other hand, although
there exist
results on the orbital stability of dark solitons \cite{bar1}
and of other structures such as bubbles (black solitons with zero
phase shift) \cite{bouard} (see Ref.~\cite{kevpeli} for a more detailed
discussion of earlier works), the work of Ref.~\cite{kevpeli} was the
first one to establish detailed estimates on the relevant eigenvalues,
using once again the technique of Lyapunov-Schmidt reductions
in the limit of small potential perturbations. The main results
can be summarized as follows. For the equation:
\begin{eqnarray}
i u_t=-\frac{1}{2} u_{xx} + f(|u|^2) u + \epsilon V(x) u
\label{s5_41}
\end{eqnarray} 
\begin{enumerate}
\item The analogous condition to the persistence condition
(\ref{s5_34}) is now:
\begin{eqnarray}
M'(\xi)=\int V'(x) [q_0-u_0^2(x-\xi)] dx =0,
\label{s5_42}
\end{eqnarray}
where the unperturbed solution $u_0$ asymptotes to $\pm \sqrt{q_0}$
at $\pm \infty$; i.e., the background of the solution is now 
appropriately incorporated in Eq.~(\ref{s5_42}) [in comparison to
Eq.~(\ref{s5_34})].
\item The dark soliton will be spectrally unstable in the GP
equation with exactly one real eigenvalue (for small $\epsilon$)
in the case where 
\begin{eqnarray}
M''(\xi)=\int V''(x) [q_0-u_0^2(x-\xi)] dx <0,
\label{s5_43}
\end{eqnarray}
while it will be unstable due to two complex-conjugate eigenvalues
with positive real part if $M''(\xi)>0$. This is the analogous condition
to the curvature of the effective potential; however, notice the
disparity of this condition from what would be expected intuitively
based on the notion of convexity/concavity. The latter result is
a direct byproduct of the nature of the essential spectrum (encompassing
the origin in this defocusing case), which upon bifurcation of the 
translational eigenvalue along the imaginary axis makes it directly
complex (case of $M''(\xi)>0$). The location of the relevant
eigenvalue in the GP case of cubic nonlinearity is given to leading
order by:
\begin{eqnarray}
\lambda^2+\frac{\epsilon}{4} M''(\xi) \left(1-\frac{\lambda}{2} \right) =0,
\label{s5_44}
\end{eqnarray}
which is consonant with the above result. A form of this expression
for general nonlinearities was also obtained in Ref.~\cite{kevpeli}.
\end{enumerate}
A case example of the possible dark soliton solutions for
a potential of the form $V(x)=x^2 \exp(-\kappa |x|)$ is shown in
Fig.~\ref{rev_fig5}, illustrating the quantitative accuracy of 
Eqs.~(\ref{s5_42})--(\ref{s5_44}). In this case, the formalism
elucidates a subcritical pitchfork bifurcation whereby three
dark soliton solutions (one centered at $\xi=0$ and two symmetrically
at $\xi \neq 0$) eventually merge into a single unstable kink
centered at $\xi=0$.

One of the fundamental limitations of this result is its being dependent
upon the decaying nature of the potential at $\pm \infty$. The fundamentally
different nature of the spectrum in the presence of parabolic or periodic
potentials (or both) makes it much harder to provide such considerations
in the latter cases. While this can be done in some special limits 
(such as the Thomas-Fermi, large chemical potential limit of
the appendix of Ref.~\cite{pelioscil}), in that setting it is generally
easier to use the linear limit approach of Sec.~\ref{lin_lim}.

\subsubsection{Perturbation theory for solitons.}
\label{nlwd}


Dynamics of either bright or dark matter-wave solitons can be studied by means of the perturbation theory for solitons 
\cite{pert1,pert2,pert3,pert4} (see also Ref.~\cite{pertkm} for a review). Here, we will briefly discuss an application of 
this approach upon considering the example of soliton dynamics in BECs confined in an external potential, say $V(x)$, which is  
smooth and slowly-varying on the soliton scale. This means that in the case, e.g., of the conventional parabolic trap 
$V(x)=(1/2)\Omega^{2}x^{2}$, the effective trap strength is taken to be $\Omega \sim \epsilon$, where $\epsilon \ll 1$ 
is a formal small (perturbation) parameter. Taking into regard the above, we consider the following perturbed NLS equation 
\begin{eqnarray} 
i \frac{\partial u}{\partial t}+\frac{1}{2} \frac{\partial^{2} u}{\partial x^{2}} - g |u|^{2} u = R(u), 
\label{gpe1d_u} 
\end{eqnarray} 
%
with the perturbation being the potential term $R(u) \equiv V(x) u$,  
and $g=\pm 1$ corresponding to repulsive and attractive interactions. Soliton dynamics in the framework of Eq.~(\ref{gpe1d_u}) 
can then be treated perturbatively, assuming that a perturbed soliton solution can be expressed as
\begin{equation}
u(x,t)=u_{s}(x,t)+\epsilon u_{d}(x,t)+\epsilon u_{r}(x,t).
\label{pes}
\end{equation}
Here, $u_{s}(x,t)$ has the same functional form as the soliton solutions (\ref{chap01:bs}) 
for $g=-1$ and (\ref{chap01:ds}) for $g=+1$, but with the soliton parameters depending on time.  
On the other hand, $u_{d}$ is a function localized near the soliton, 
describing the deformation of the soliton (i.e., the change of the soliton shape) and $u_{r}$ is the radiation 
(in the form of sound waves) emitted by the soliton. In fact, the effect described by 
$u_{d}$ is not significant, as the small change in the soliton shape does not modify its motion, while the 
emission of radiation may be neglected for sufficiently weak perturbations.  
Thus, here we will consider solely the first term in Eq.~(\ref{pes}), 
which corresponds to the so-called {\it adiabatic approximation} of the perturbation theory for solitons \cite{pertkm}.

First we discuss the dynamics of bright solitons in external potentials. 
Taking into regard  that for $g=-1$ and $R(u)=0$, Eq.~(\ref{gpe1d_u}) has a bright soliton solution 
of the form given in Eq.~(\ref{chap01:bs}), we assume that in the perturbed case with $R(u) \ne 0$  
a soliton solution can be expressed as 
\begin{eqnarray} 
u(x,t)=\eta\, {\rm sech}[\eta(x-x_{0})]\exp[i(kx-\phi(t)]
\label{ansatzbs} 
\end{eqnarray} 
where $x_{0}$ is the soliton center, the parameter $k=dx_{0}/dt$ defines both the soliton wavenumber and velocity, and 
$\phi(t)=(1/2)(k^2-\eta^2)t$ is the soliton phase. In the case under consideration, the soliton 
parameters $\eta$, $k$ and $x_0$ are considered to be unknown, slowly-varying functions of time $t$.
Then, from Eq.~(\ref{gpe1d_u}), it is found that the 
number of atoms  $N$ and the momentum $P$ (which are integrals of motion of the unperturbed system), 
evolve, in the presence of the perturbation, according to the following equations,
\begin{eqnarray} 
\hskip-1.5cm
\frac{dN}{dt} = -2\, {\rm Im} \left[ \int_{-\infty}^{+\infty} R(u)\, u^{\ast} dx \right], \,\,\,\,\,\,\
\frac{dP}{dt} = 2\, {\rm Re} \left[ \int_{-\infty}^{+\infty} R(u)\, \frac{\partial u^{\ast}}{\partial x} dx \right].
\label{evin2}
\end{eqnarray}
We now substitute the ansatz (\ref{ansatzbs}) (but with the soliton parameters being functions of time) 
into Eqs.~(\ref{evin2}) and obtain the evolution equations for $\eta(t)$ and $k(t)$,
\begin{eqnarray} 
\frac{d \eta}{dt}= 0, \,\,\,\, 
\frac{dk}{dt}=-\frac{\partial U}{\partial x_{0}},
\label{bright_par2} 
\end{eqnarray}
where $U(x_0)$ is given by the expression of Eq.~(\ref{s5_6}).
In the case, however, of slow variation of the potential on the scale
of the solitary wave (the case of interest here), 
a simple Taylor expansion yields the same
equation but with $U \equiv V$ i.e., the trapping potential.

Now, recalling that $dx_{0}/dt=k$, we may combine the above Eqs.~(\ref{bright_par2}) to derive 
the following equation of motion for the soliton center:
\begin{eqnarray} 
\frac{d^2x_{0}}{dt^2}=-\frac{\partial V}{\partial x_{0}}, 
\label{eq_motb} 
\end{eqnarray} 
which shows that the bright matter-wave soliton behaves effectively like a Newtonian particle. Note that in the case 
of a parabolic trapping potential, i.e., $V(x)=(1/2)\Omega^{2}x^{2}$, 
Eq.~(\ref{eq_motb}) implies that the frequency of oscillation is $\Omega$; this result is consistent with the Ehrenfest 
theorem of the quantum mechanics, or the so-called Kohn theorem \cite{kohn}, implying that the motion of the center of 
mass of a cloud of particles trapped in a parabolic potential is decoupled from the internal excitations. 
Note that the result of Eq.~(\ref{eq_motb}) can be obtained by other methods, such as the WKB approximation \cite{am}, 
or other perturbative techniques \cite{sb,st1,st2}.

We now turn to the dynamics of dark matter-wave solitons in the framework of Eq.~(\ref{gpe1d_u}) for $g=+1$.  
First, the background wavefunction is sought in the form $u = \Phi(x) \exp(-i\mu t)$ ($\mu$ being the normalized chemical potential),  
where the unknown function $\Phi(x)$ satisfies the following real equation,
\begin{equation}
\mu\Phi+\frac{1}{2}\frac{d^{2} \Phi}{dx^2}-\Phi^{3}=V(x)\Phi.
\label{ub11}
\end{equation}
Then, following the analysis of Ref.~\cite{fr1}, we seek for a 
dark soliton solution of Eq.~(\ref{gpe1d_u}) on top of the inhomogeneous background satisfying Eq.~(\ref{ub11}), namely, 
$
u = \Phi(x) \exp(-i\mu t)\upsilon (x,t), 
$
where the unknown wavefunction $\upsilon (x,t)$ represents a dark soliton. 
This way, employing Eq.~(\ref{ub11}), the following evolution equation for the dark soliton wave function is readily obtained:
\begin{equation}
i\frac{\partial \upsilon}{\partial t} +\frac{1}{2} \frac{\partial^{2} \upsilon}{\partial x^2}
- \Phi^{2}(|\upsilon|^{2}-1)\upsilon = -\frac{d}{dx}\ln(\Phi) \frac{\partial \upsilon}{\partial x}.
\label{upsa}
\end{equation}
Taking into regard the fact that in the framework of the Thomas-Fermi approximation
the profile can be simply approximated by Eq.~(\ref{ub11}), 
Eq.~(\ref{upsa}) can be simplified to the following defocusing perturbed NLS equation,
\begin{equation}
i\frac{\partial \upsilon}{\partial t}+\frac{1}{2}\frac{\partial^{2} \upsilon}{\partial x^2}-\mu(|\upsilon |^{2}-1)\upsilon=Q(\upsilon),
\label{pnlsd}
\end{equation}
with the perturbation $Q(\upsilon)$ being of the form, 
\begin{eqnarray}
Q(\upsilon)= \left( 1-|\upsilon |^{2}\right) \upsilon V+ \frac{1}{2(\mu-V)}\frac{dV}{dx} \frac{\partial \upsilon}{\partial x}.  
\label{pertQ}
\end{eqnarray}
In the absence of the perturbation, Eq.~(\ref{pnlsd}) has the dark soliton solution (for $\mu=1$) 
$
\upsilon (x,t)=\cos \varphi \tanh \zeta +i \sin \varphi,
$
where $\zeta \equiv \cos \varphi \left[ x-(\sin \varphi)t \right]$ (recall that $\cos \varphi$ and $\sin \varphi$ are the 
soliton amplitude and velocity respectively and $\varphi$ is the soliton phase angle). 
To treat analytically the effect of the perturbation (\ref{pertQ}) on the dark soliton, we employ the adiabatic 
perturbation theory devised in Ref.~\cite{KY}. Assuming, as in the case of bright solitons, that 
the dark soliton parameters become slowly-varying unknown functions of $t$, the soliton phase angle 
becomes $\varphi \rightarrow\varphi(t)$ and, as a result, the soliton coordinate becomes 
$\zeta \rightarrow \zeta=\cos\varphi(t) \left[x-x_{0}(t) \right]$, where
$x_{0}(t)= \int_{0}^{t}\sin\varphi(t^{\prime })dt^{\prime}$,
is the soliton center. Then, 
the evolution of the parameter $\varphi$ is governed by \cite{KY}, 
\begin{equation}
\frac{d\varphi}{dt}=\frac{1}{2\cos ^{2}\varphi \sin \varphi} {\rm Re}
\left[ \int_{-\infty}^{+\infty}Q(\upsilon)\frac{\partial \upsilon^{\ast}}{\partial t} dx \right],
\label{phi}
\end{equation}
%
which, in turn, yields 
for the perturbation in Eq.~(\ref{pertQ}):
\begin{eqnarray} 
\frac{d \phi}{dt}=- \frac{1}{2} \cos\varphi  \frac{\partial V}{\partial x_{0}}.
\label{dark_par2} 
\end{eqnarray} 
To this end, combining Eq.~(\ref{dark_par2}) with the definition of the dark soliton center, 
we obtain the following equation of motion (for nearly stationary dark solitons with $\cos \varphi \approx 1$), 
\begin{eqnarray} 
\frac{d^2x_{0}}{dt^2}=-\frac{1}{2} \frac{\partial V }{\partial x_{0}} .
\label{eqmd} 
\end{eqnarray} 
Equation (\ref{eqmd}) implies that the dark soliton, similarly to the bright one, behaves like a Newtonian particle. 
However, in an harmonic potential with strength $\Omega$, the dark soliton oscillates with another frequency, namely 
$\Omega/\sqrt{2}$, a result that may be considered as the Ehrenfest theorem for dark solitons. The oscillations of dark solitons in 
trapped BECs has been a subject that has attracted much interest; in fact, many relevant analytical works have been devoted to 
this subject, in which various different perturbative approaches have been employed \cite{fr2,huang1,fr3,fr4,fr5}. It should also 
be mentioned that a more general Newtonian equation of motion, similar to Eq.~(\ref{eqmd}) but also valid for a wider class 
of confining potentials, was recently discussed in Ref.~\cite{kevpeli}.

Perturbation theory for dark solitons may also be applied for dark solitons with radial symmetry, i.e., for ring or spherical dark 
solitons described by a GP equation of the form 
\begin{eqnarray} 
i \frac{\partial \psi}{\partial t}=-\frac{1}{2} \nabla^{2} \psi + |\psi|^{2} \psi + V(r) \psi, 
\label{gpers} 
\end{eqnarray} 
where $\nabla^{2}= \partial^{2}_{r} + (D-1)r^{-1}\partial_{r} $ is the transverse Laplacian, $V(r)=(1/2)\Omega^2 r^2$, 
and $D=2,3$ correspond to the cylindrical and spherical case, respectively. In this case, Eq.~(\ref{gpers}) can be treated 
as a perturbed 1D defocusing NLS equation provided that the potential term and the term $\sim r^{-1}$ can be considered 
as small perturbations; this case is physically relevant for weak trapping potentials with $\Omega \ll 1$, and 
radially symmetric solitons of large radius $r_{0}$. Then, it can be found \cite{rds} (see also Ref.~\cite{fr4}) that the 
radius $r_0$ of the radially symmetric dark solitons is governed by the following Newtonian equation of motion, 
\begin{equation}
\frac{d^{2}r_{0}}{dt^{2}}=- \frac{\partial V_{\rm eff}}{\partial r_{0}}, 
\label{sem}
\end{equation}
where the effective potential is given by $V_{\rm eff}(r_{0})=(1/4)\Omega^{2}r_{0}^{2}- \ln r_{0}^{(D-1)/3}$ 
[note that in the 1D limit of $D=1$, Eq.~(\ref{sem}) is reduced to Eq.~(\ref{eqmd})]. 
Clearly, in this higher-dimensional setting the equation of the soliton motion becomes nonlinear, even for nearly black solitons, 
due to the presence of the repulsive curvature-induced logarithmic potential. We finally note that such radially symmetric solitons 
are generally found to be unstable, as they either decay to radiation (the small-amplitude ones) or are subject to the snaking 
instability (the moderate- and large-amplitude ones), giving rise to the formation of vortex necklaces \cite{rds}. 

\subsubsection{The reductive perturbation method.\label{sec:reductive}} 

Another useful tool in the analysis of the dynamics of matter-wave solitons (and especially the dark ones) is 
the so-called reductive perturbation method (RPM) \cite{jefkaw}. Applying this asymptotic method, one usually introduces proper 
``stretched'' (slow) variables to show that {\it small-amplitude} nonlinear structures governed by a specific nonlinear evolution equation 
can effectively be described by another equation. Such a formal connection between soliton equations was demonstrated for 
integrable systems in Ref.~\cite{zakuz}, and then extended to the reduction of nonintegrable models to integrable ones, first 
in applications in optics (see, e.g., Refs.~\cite{opt1,opt2,opt3}) and later in BECs. 
Here, we will briefly describe this method upon considering, as an example, an inhomogeneous generalized NLS equation 
similar to Eq.~(\ref{s5_41}), namely, 
\begin{equation}
iu_{t}=-\frac{1}{2}u_{xx}+V(X)u+g(\rhon)u,
\label{ggpe}
\end{equation}
which is characterized by a general nonlinearity $g(\rhon)$ (depending on the density $\rhon=|u|^2$), and a 
slowly varying external potential $V(X)$, depending on a slow variable $X \equiv \epsilon^{3/2}x$ (with $\epsilon$ being a 
formal small parameter). Our main purpose is to show that this rather general NLS-type mean-field model can be reduced to 
the much simpler Korteweg-de Vries (KdV) equation with variable coefficients. The latter, has been used in the past 
to describe shallow water-waves over variable depth, or ion-acoustic solitons in inhomogeneous plasmas \cite{asano}, and,  
as we will discuss below, it provides an effective description of the dark soliton dynamics in BECs.

Following the analysis of Ref.~\cite{ourpla}, we first derive from Eq.~(\ref{ggpe}) hydrodynamic equations for the density 
$\rhon$ and the phase $\phi$, arising from the Madelung transformation $u=\sqrt{\rhon}\exp(i\phi)$, and then introduce the 
following asymptotic expansions,
\begin{eqnarray}
\rhon&=&\rhon_{0}(X)+ \epsilon \rhon_{1}(X,T)+\epsilon^{2} \rhon_{2}(X,T)+\cdots, 
\label{ae1} \\[1.0ex]
\phi&=&-\mu_{0}t+ \epsilon^{1/2} \phi_{1}(X,T)+\epsilon^{3/2} 
\phi_{2}(X,T)+\cdots, 
\label{ae2}
\end{eqnarray}
where $\rhon_{0}(X)$ is the ground state of the system determined by the Thomas-Fermi approximation 
$g(\rhon_{0})=\mu_{0}-V(X)$ (with $\mu_0$ being the chemical potential), and 
$T=\epsilon^{1/2}\left( t-\int_{0}^{x} C^{-1}(x')dx'\right)$ is a slow time-variable [where 
$C = \sqrt{ \dot{g}_{0}\rhon_{0}}$ is the local speed of sound and $\dot{g}_{0} \equiv (dg/d\rhon)|_{\rhon=\rhon_{0}}$]. 
This way, in the lowest-order approximation in $\epsilon$ we obtain an equation for the phase, 

\begin{equation}
\phi_{1}(X,T)=-\dot{g}_{0}(X)\int_{0}^{T} \rhon_{1}(X,T')dT', 
\label{r}
\end{equation}
and derive the following KdV equation for the density $\rhon_1$,
\begin{eqnarray}
%
%
\hskip-1.5cm
\rhon_{1X}-\frac{\left(3\dot{g}_{0}+\rhon_{0}\ddot{g}_{0} \right)}{2C^{3}} \rhon_{1}\rhon_{1T}+\frac{1}{8C^{5}}\rhon_{1TTT} 
= -\frac{d}{dX}\left[ \ln \left(|C|\dot{g}_{0} \right)^{1/2} \right] \rhon_{1}, 
\label{kdv1}
\end{eqnarray}
where $\ddot{g}_{0} \equiv (d^{2}g/d\rhon^{2})|_{\rhon=\rhon_{0}}$. Importantly, in a homogeneous gas with 
$\rhon_{0}(X)=\rhon_{\rm p}={\rm const.}$, Eq.~(\ref{kdv1}) is the completely integrable KdV equation; the soliton solution of the latter, 
which is $\sim {\rm sech}^{2}(T)$, is in fact a density notch on the background density $\rhon_{\rm p}$ [see Eq.~(\ref{ae1})], 
with a phase jump across it [see Eq.~(\ref{r}), which implies that $\phi_{1}\sim \tanh(T)$] 
and, thus, it represents an approximate dark soliton solution of Eq.~(\ref{ggpe}). On the other hand, the results obtained earlier for the analysis of 
the KdV equation with variable coefficients \cite{kko,kar2} have been used to analyze shallow soliton dynamics in the BEC context. 
In particular, the KdV Eq.~(\ref{kdv1}) was obtained in the framework of the cubic nonlinear 
version of Eq.~(\ref{ggpe}) (with $g(\rhon) \sim \rhon$), 
and used to study the dynamics and the collisions of shallow dark solitons in BECs \cite{huang1,huang2}.  
Moreover, other versions of Eq.~(\ref{kdv1}) relevant to the Tonks gas (corresponding to $g(\rhon) \sim \rhon^2$) \cite{fpk}, 
as well as the BEC-Tonks crossover regime (corresponding to a generalized nonlinearity) 
\cite{ourpla} were derived and analyzed as well. 

Finally, it is relevant to note that there exist studies in higher-dimensional (disk-shaped) BECs, where the RPM was used 
to predict 2D nonlinear structures, such as ``lumps'' described by an effective Kadomtsev-Petviashvili equation \cite{huang3}, 
and ``dromions'' described by an effective Davey-Steartson equation \cite{huang4,maximo}.

\subsection{Methods for discrete systems}

As our final class of methods, we will present a series of
results that are relevant to systems with periodic potentials
and, in particular, with discrete lattices per the Wannier function
reduction of Sec.~\ref{Sec:dim_reduc}.6.2. 
%
Starting with the prototypical discrete model of the 
1D DNLS equation (see, e.g., Ref.~\cite{chap01:IJMPB} for a review) of the form:
\begin{equation}
i \dot{u}_n=-\epsilon (u_{n+1}+u_{n-1}) - |u_n|^2 u_n,
\label{s5_45}
\end{equation}
we look for standing waves of the form: $u_n=\exp(i \mu t) v_n$
which satisfy the steady state equation
\begin{equation}
(\mu-|v_n|^2) v_n = \epsilon (v_{n+1}+v_{n-1}).
\label{s5_46}
\end{equation}
One of the fundamental
ideas that we exploit in this setting is the so-called anti-continuum
(AC) limit of MacKay and Aubry \cite{mackay} for $\epsilon=0$, where 
Eq.~(\ref{s5_46}) 
is completely solvable $v_n=\{0,\pm \sqrt{\mu} \exp(i \theta_n)\}$,
where $\theta_n$ is a free phase parameter for each site. However, a key
question is which ones of all these possible sequences of $v_n$
will persist as solutions when $\epsilon \neq 0$. A simple way to
see this in the 1D case of Eq.~(\ref{s5_45}) is to multiply 
Eq.~(\ref{s5_46}) by $v_n^{\ast}$ and subtract the complex conjugate
of the resulting equation, which in turn leads to:
\begin{equation} 
v_n^{\ast} v_{n+1}-v_n v_{n+1}^{\ast}= {\rm const.} 
~~\Rightarrow~~ 2 {\rm arg}(v_{n+1})= 2 {\rm arg}(v_n),
\label{s5_47}
\end{equation}
since we are considering solutions vanishing as $n \rightarrow \pm \infty$.
Without loss of generality (using the scaling of the equation), 
we can scale $\mu=1$, in which case the only states that will
persist for finite $\epsilon$ are ones containing sequences with
combinations of $v_n=\pm 1$ and $v_n=0$. A systematic computational
classification of the simplest ones among these sequences and of
their bifurcations is provided in Ref.~\cite{konodisc}. 
Notice that we are tackling
here the focusing case of $s=-1$, however, the defocusing case
of $s=1$ can also be addressed based on the same considerations
and using the so-called staggering transformation $w_n=(-1)^n u_n$
(which converts the defocusing nonlinearity into a focusing one,
with an appropriate frequency rescaling which can be absorbed in a
gauge transformation).

We subsequently consider the issue of stability, using once again
the standard symplectic formalism $J {\cal L} w = \lambda w$, where
${\cal L}$ is given by Eq.~(\ref{call}) and $J$ is the symplectic
matrix. In this case, the $L_+$ and $L_-$ operators are given by: 
\begin{eqnarray}
(1-3 v_n^2) a_n - \epsilon (a_{n+1}+a_{n-1}) &=& {\cal L}_+ a_n = - \lambda b_n
\label{s5_48}
\\[1.0ex]
(1-v_n^2) b_n - \epsilon (b_{n+1}+b_{n-1}) &=& {\cal L}_- b_n = \lambda a_n.
\label{s5_49}
\end{eqnarray}
We again use the AC limit where  we assume a sequence
for $v_n$ with $N$ ``excited'' (i.e., $\neq 0$) sites; then, it is easy
to see that for
$\epsilon=0$ these sites correspond to eigenvalues $\lambda_{L_+}=-2$ for
$L_+$ and to eigenvalues $\lambda_{L_-}=0$ for $L_-$, and they result
in $N$ eigenvalue pairs  with $\lambda^2=0$ for the full Hamiltonian problem. 
Hence, these eigenvalues are potential sources of instability, since for
$\epsilon  \neq 0$, $N-1$ of those will become nonzero (there is
only one symmetry, namely the U(1) invariance, persisting for 
$\epsilon \neq 0$). The key question for stability purposes is to 
identify the location of these $N-1$ small eigenvalue pairs.
One can manipulate Eqs.~(\ref{s5_48})--(\ref{s5_49}) into the form:
\begin{equation}
{\cal L}_- b_n =-\lambda^2  {\cal L}_+^{-1} b_n 
~~\Rightarrow~~ \lambda^2= - 
\frac{(b_n, {\cal L}_- b_n)}{(b_n, {\cal L}_+^{-1} b_n)}
\label{s5_50}
\end{equation}
Near the AC limit, the effect of $L_+$ is a multiplicative
one (by $-2$). Hence:
\begin{equation}
\lim_{\epsilon \rightarrow 0}(b_n, {\cal L}_+^{-1} b_n)=-\frac{1}{2}
~~\Rightarrow~~ \lambda^2 = 2 \gamma =2 (b_n, {\cal L}_- b_n)
\label{s5_51}
\end{equation}
Therefore the problem reverts to the determination of the spectrum
of $L_-$. However, using the fact that $v_n$ is an eigenfunction
of $L_-$ with $\lambda_{L_-}=0$ and the Sturm comparison theorem
for difference operators \cite{levy}, one infers that if the
number of sign changes in the solution at the AC limit is $m$
(i.e., the number of times that adjacent to a $+1$ is a $-1$
and next to a $-1$ is a $+1$), then $n(L_-)=m$ and therefore
from Eq.~(\ref{s5_51}), the number of imaginary eigenvalue
pairs of ${\cal L} $ is $m$, while that of real eigenvalue pairs
is consequently $(N-1)-m$. An immediate conclusion is that
unless $m=N-1$, or practically unless adjacent sites are
out-of-phase with each other, the solution will be immediately
unstable for $\epsilon \neq 0$. Notice that this is also consistent
with the eigenvalue count of Refs.~\cite{kks,pelinovsky} since
$n({\cal L})-n(D)=(N+m)-1=(N+m-1)+2 \times m + 2 \times 0=
k_r + 2 k_i^{-} + 2 k_c$ (it is straightforward to show by the
definition of the Krein signature \cite{pelid1} to show
that these $m$ imaginary pairs have negative Krein signature).

\begin{figure}[t]
\begin{center}
\includegraphics[width=6.4cm,height=5cm]{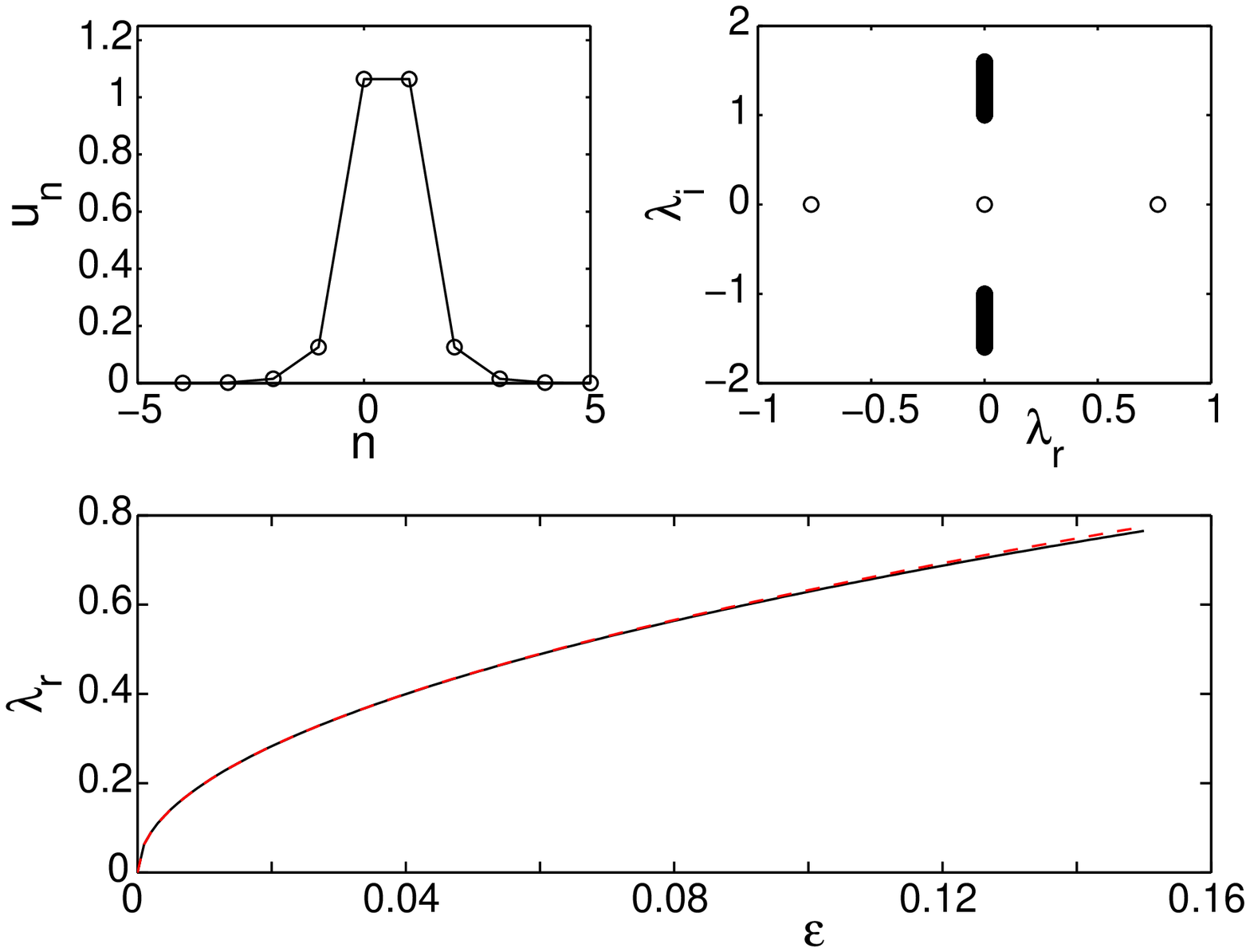}
\includegraphics[width=6.4cm,height=5cm]{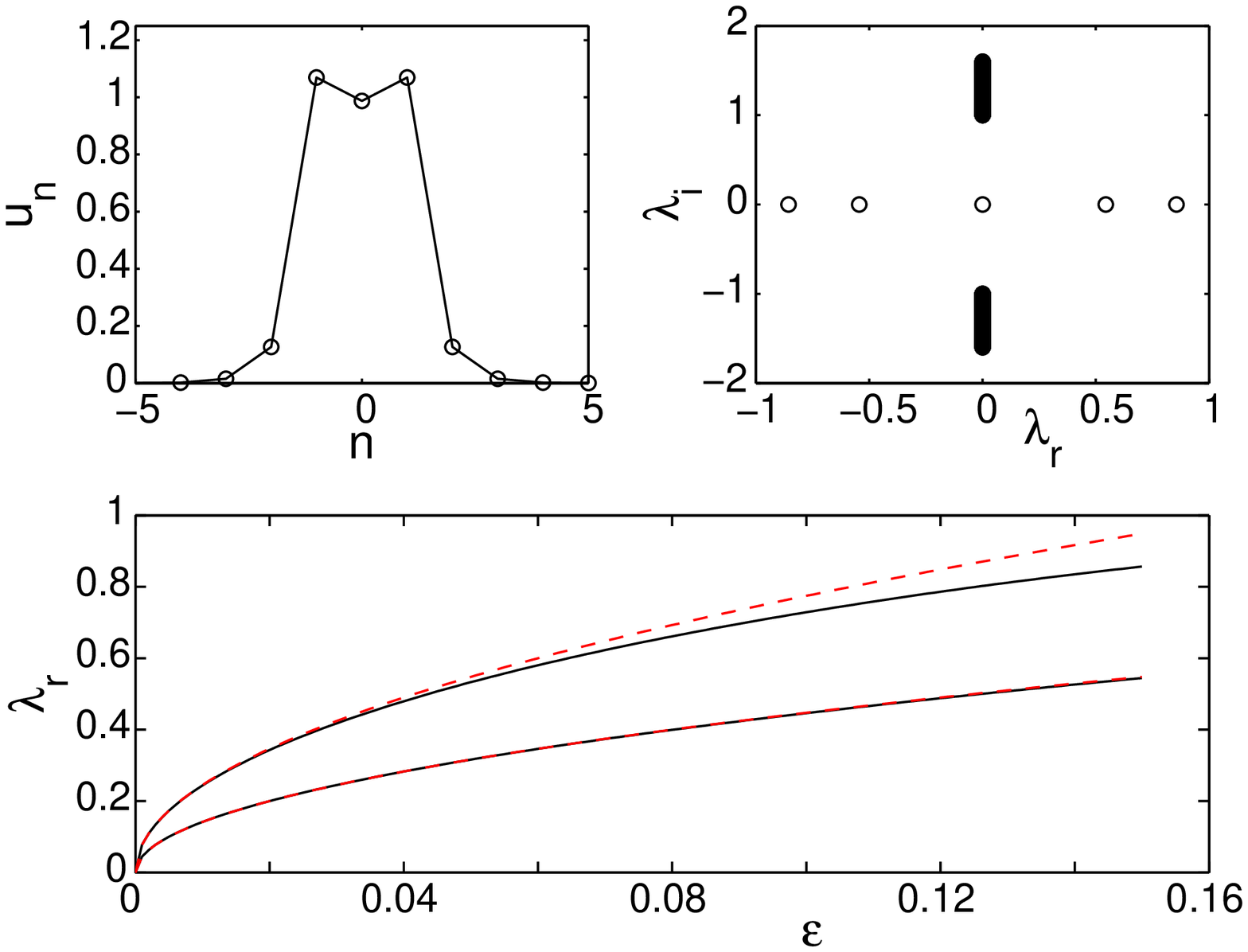}
\end{center}
\vskip-0.4cm
\caption{
(Color online)
A typical example of a two-site configuration is shown in the left
panels of the figure, and a corresponding one of a three-site configuration
in the right panels. The top panels show a solution (for a particular value
of $\epsilon$) and its corresponding spectral plane $(\lambda_r,\lambda_i)$
of eigenvalues $\lambda=\lambda_r+i\lambda_i$, while the bottom ones
show the dependence of the relevant real eigenvalues as a function of
the inter-site coupling $\epsilon$ obtained analytically (dashed/red lines)
and numerically (solid/black lines).}
\label{rev_fig6}
\end{figure}

One can also use Eq.~(\ref{s5_51}) quantitatively to identify
the relevant eigenvalues perturbatively for the full problem by
considering the perturbed (originally zero when $\epsilon=0$)
eigenvalues of $L_-$ in the form:
\begin{equation}
{\cal L}_{-}^{(0)} b_n^{(1)} = \gamma_1 b_n^{(0)} - 
{\cal L}_{-}^{(1)} b_n^{(0)},
\label{s5_52}
\end{equation}
where ${\cal L}_-={\cal L}_-^{(0)} + \epsilon {\cal L}_-^{(1)} + {\cal O}(\epsilon^2)$
and a similar expansion has been used for the eigenvector $b_n$.
Also $\lambda_{L_-}= \epsilon \gamma_1 + {\cal O}(\epsilon^2)$. 
Projecting the above equation
to all the eigenvectors of zero eigenvalue of $L_-^{(0)}$, one
can explicitly convert Eq.~(\ref{s5_52}) into an eigenvalue
problem \cite{pelid1} of the form $M c= \gamma_1 c$, where the
matrix $M$ has off-diagonal entries: 
$M_{n,n+1}=M_{n+1,n}=-\cos(\theta_{n+1}-\theta_n)$ and
diagonal entries $M_{n,n}=\left(\cos(\theta_{n-1}-\theta_n)+
\cos(\theta_{n+1}-\theta_n) \right)$. Then it is straightforward
to compute $\gamma_1$ and subsequently from Eq.~(\ref{s5_51}) to
evaluate the corresponding $\lambda=\pm \sqrt{2 \epsilon \gamma_1}$.
For example, for one-dimensional configurations with two-adjacent
sites with phases $\theta_1$ and $\theta_2$, the matrix $M$ becomes:
\begin{eqnarray}
M=
\left( \begin{array}{cc}
\cos(\theta_1-\theta_2) & -\cos(\theta_1-\theta_2) \\
-\cos(\theta_1-\theta_2) & \cos(\theta_1-\theta_2) \end{array} \right),
\label{s5_53}
\end{eqnarray}
whose straightforward calculation of eigenvalues leads to
$\lambda^2=0$ and $\lambda^2=\pm 2 \sqrt{\epsilon \cos(\theta_1-\theta_2)}$.
Notice that this result is consonant with our qualitative theory above
since for same phase excitations ($\theta_1=\theta_2$), the configuration
is unstable, while the opposite is true if $\theta_1=\theta_2 \pm \pi$.
Similar calculations are possible for 3-site configurations with
phases $\theta_{1,2,3}$, in which case, one of the eigenvalues of
$M$ is again $0$ (as has to generically be the case, due to the 
U(1) invariance), while the other two are given by:
\begin{eqnarray}
\gamma_{1} &=& \cos(\theta_2 - \theta_1) + \cos(\theta_3 - \theta_2)
\nonumber
\\
& \pm & \sqrt{\cos^2(\theta_2 - \theta_1) - \cos(\theta_2 - \theta_1)
\cos(\theta_3 - \theta_2) + \cos^2(\theta_3-\theta_2)}.
\label{s5_54}
\end{eqnarray}
Some of the examples of the accuracy of these theoretical
predictions for some typical two-site and three-site configurations 
(in particular, the in-phase ones, which should have one and
two real eigenvalue pairs respectively) are shown in Fig.~\ref{rev_fig6}.

This approach can be generalized to different settings, such as
in particular higher dimensions \cite{pelid2,pelid3} or
multi-component systems \cite{pelimulti}. Perhaps the fundamental
difference that arises in the higher dimensional settings is
that one can excite sites over a contour and then, for the
$N$ excited sites around the contour the persistence (Lyapunov-Schmidt)
conditions can be obtained as a generalization of Eq.~(\ref{s5_45})
that reads \cite{pelid2,sukhor}:
\begin{eqnarray}
\sin(\theta_1-\theta_2)=\sin(\theta_2-\theta_3)=\dots=\sin(\theta_N-\theta_1),
\label{s5_55}
\end{eqnarray}
which indicates that a key difference of higher dimensional settings
is that not only ``solitary wave'' structures with phases 
$\theta \in \{0,\pi\}$
are possible, but also both symmetric and asymmetric vortex families
\cite{pelid2,sukhor} may, in principle, be possible [although 
Eq.~(\ref{s5_55}) provides the leading order persistence condition and one
would need to also verify 
the corresponding conditions to higher order to confirm
that such solutions persist \cite{pelid2}]. Such vortex solutions had
been predicted numerically earlier \cite{boris1,boris2} and have been
observed experimentally in the optical setting of photorefractive
crystals \cite{yuri1,moti1}. Performing the stability
analysis is possible for these higher dimensional structures, although
the relevant calculations are technically far more involved. However,
the theory can be formulated in an entirely general manner: we give its
outline and some prototypical examples of higher dimensional theory-computation
comparisons below. 

The existence problem can be generally formulated in the multi-dimensional
case as the vanishing of the vector field ${\bf F}_n$ of the form:
\begin{equation}
\label{nonlinear-vector-field} {\bf F}_n(\mbox{\boldmath
$\phi$},\epsilon) = \left[
\begin{array}{c} 
(1 - |\phi_n|^2) \phi_n - \epsilon \Sigma \phi_n \\ [1.0ex]
(1 - |\phi_n|^2) {\phi}_n^{\ast} - \epsilon \Sigma
{\phi}_n^{\ast}  \end{array} \right].
\end{equation}
If we then define the matrix operator:
\begin{eqnarray}
\label{energy} {\cal H}_{n} &=& \left( \begin{array}{cc} 1 - 2
|\phi_n|^2 & - \phi_n^2 \\ - \bar{\phi}_n^2 & 1 - 2 |\phi_n|^2
\end{array} \right) 
\nonumber
\\
&-& \epsilon \left( s_{+e_1} + s_{-e_1} +
s_{+e_2} + s_{-e_2} + s_{+e_3} + s_{-e_3} \right) \left(
\begin{array}{cc} 1 & 0 \\ 0 & 1 \end{array} \right),
\end{eqnarray}
where the $s_{\pm e_i}$ denotes the shift operators along the respective
directions, then the stability problem is given by $\sigma {\cal H} 
\mbox{\boldmath $\psi$} = i
\lambda \mbox{\boldmath $\psi$}$, where the 2-block of $\sigma$ is
the diagonal matrix of $(1,-1)$ at each node $n$.
Furthermore, the existence problem is connected to ${\cal H}$ through:
${\cal H} = D_{\mbox{\boldmath $\phi$}} {\bf
F}(\mbox{\boldmath $\phi$},0)$. At the AC limit of $\epsilon=0$
$$
({\cal H}^{(0)})_n = \left[ \begin{array}{cc} 1 & 0 \\ 0 & 1
\end{array} \right], \; n \in S^{\perp}, \qquad
({\cal H}^{(0)})_n = \left[ \begin{array}{cc} -1 & -e^{2 i \theta_n} \\
-e^{-2 i \theta_n} & -1 \end{array} \right], \; n \in S,
$$
where $S$ is the set of excited sites. Then the eigenvectors
of zero eigenvalue will be of the form:
$$
({\bf e}_n)_k = i \left[ \begin{array}{c} e^{i \theta_n} \\ -
e^{-i \theta_n}\end{array} \right] \; \delta_{k,n}.
$$
Defining the projection operator:
\begin{equation}
({\cal P} {\bf f})_n = \frac{\left( {\bf e}_n,
{\bf f} \right)}{\left( {\bf e}_n, {\bf e}_n \right)} = \frac{1}{2i}
\left( e^{-i \theta_n} ({\bf f}_1)_n - e^{i \theta_n}({\bf f}_2)_n
\right), \qquad n \in S,
\end{equation}
and decomposing the solution as:
\begin{eqnarray}
\mbox{\boldmath $\phi$} = \mbox{\boldmath
$\phi$}^{(0)}(\mbox{\boldmath $\theta$}) 
+ \mbox{\boldmath $\varphi$} \in X,
\end{eqnarray}
one can obtain the Lyapunov-Schmidt persistence conditions as  \cite{pelid3}:
\begin{equation}
\label{g-definition} {\bf g}(\mbox{\boldmath $\theta$},\epsilon) =
{\cal P} {\bf F}(\mbox{\boldmath $\phi$}^{(0)}(\mbox{\boldmath
$\theta$})+\mbox{\boldmath $\varphi$}(\mbox{\boldmath
$\theta$},\epsilon),\epsilon) = 0.
\end{equation}
This leads to the persistence theorem \cite{pelid3}:
The configuration $\mbox{\boldmath
$\phi$}^{(0)}(\mbox{\boldmath $\theta$})$ 
can be continued to the domain $\epsilon \in {\cal O}(0)$
if and
only if there exists a root $\mbox{\boldmath $\theta$}_*$
of the vector field ${\bf g}(\mbox{\boldmath
$\theta$},\epsilon)$. Moreover, if the
root $\mbox{\boldmath $\theta$}_*$ is analytic in $\epsilon \in
{\cal O}(0)$ and $\mbox{\boldmath $\theta$}_* = \mbox{\boldmath
$\theta$}_0 + {\cal O}(\epsilon)$, the solution $\mbox{\boldmath
$\phi$}$ of the difference equation 
is
analytic in $\epsilon \in {\cal O}(0)$, such that
\begin{equation}
\label{Taylor-series-phi} \mbox{\boldmath $\phi$} = \mbox{\boldmath
$\phi$}^{(0)}(\mbox{\boldmath $\theta$}_*) + \mbox{\boldmath
$\varphi$}(\mbox{\boldmath $\theta$}_*,\epsilon) = \mbox{\boldmath
$\phi$}^{(0)}(\mbox{\boldmath $\theta$}_0) + \sum_{k=1}^{\infty}
\epsilon^k \mbox{\boldmath $\phi$}^{(k)}(\mbox{\boldmath
$\theta$}_0).
\end{equation}
One can also formulate on the same footing a general stability theory.
More specifically, let the solution of interest persist for $\epsilon \neq 0$. 
If the 
operator ${\cal H}$ has a small eigenvalue $\mu$ of multiplicity $d$, 
such that $\mu = \epsilon^k \mu_k + {\cal O}(\epsilon^{k+1})$, then
the full Hamiltonian
eigenvalue problem admits $(2D)$ small
eigenvalues $\lambda$. These are such that $\lambda = \epsilon^{k/2}
\lambda_{k/2} + {\cal O}(\epsilon^{k/2+1})$, where non-zero values
$\lambda_{k/2}$ are found from
\begin{eqnarray}
\label{reduced-problem-1} && \mbox{odd $k$:} \quad {\cal M}^{(k)}
\mbox{\boldmath $\alpha$} = \frac{1}{2} \lambda_{k/2}^2 \mbox{\boldmath $\alpha$}, \\[1.0ex]
\label{reduced-problem-2} && \mbox{even $k$:} \quad {\cal M}^{(k)}
\mbox{\boldmath $\alpha$} + \frac{1}{2} \lambda_{k/2} {\cal
L}^{(k)} \mbox{\boldmath $\alpha$} = \frac{1}{2} \lambda_{k/2}^2
\mbox{\boldmath $\alpha$},
\end{eqnarray}
where
\begin{eqnarray}
{\cal M}^{(k)} &=&~ D_{\mbox{\boldmath $\theta$}} {\bf g}^{(k)}(\mbox{\boldmath $\theta$}_0), 
\nonumber
\\[1.0ex]
~{\cal L}^{(k)} &=&~ {\cal P} \left[ {\cal H}^{(1)}
\mbox{\boldmath $\Phi$}^{(k')}(\mbox{\boldmath $\theta$}_0) + ...
+ {\cal H}^{(k'+1)} \mbox{\boldmath $\Phi$}^{(0)}(\mbox{\boldmath
$\theta$}_0) \right],
\nonumber
\end{eqnarray}
and $k'=(k-1)/2$. For more details, we refer the interested
reader to Ref.~\cite{pelid3}.

\begin{figure}[t]
\begin{center}
\includegraphics[width=6.2cm,height=5.4cm]{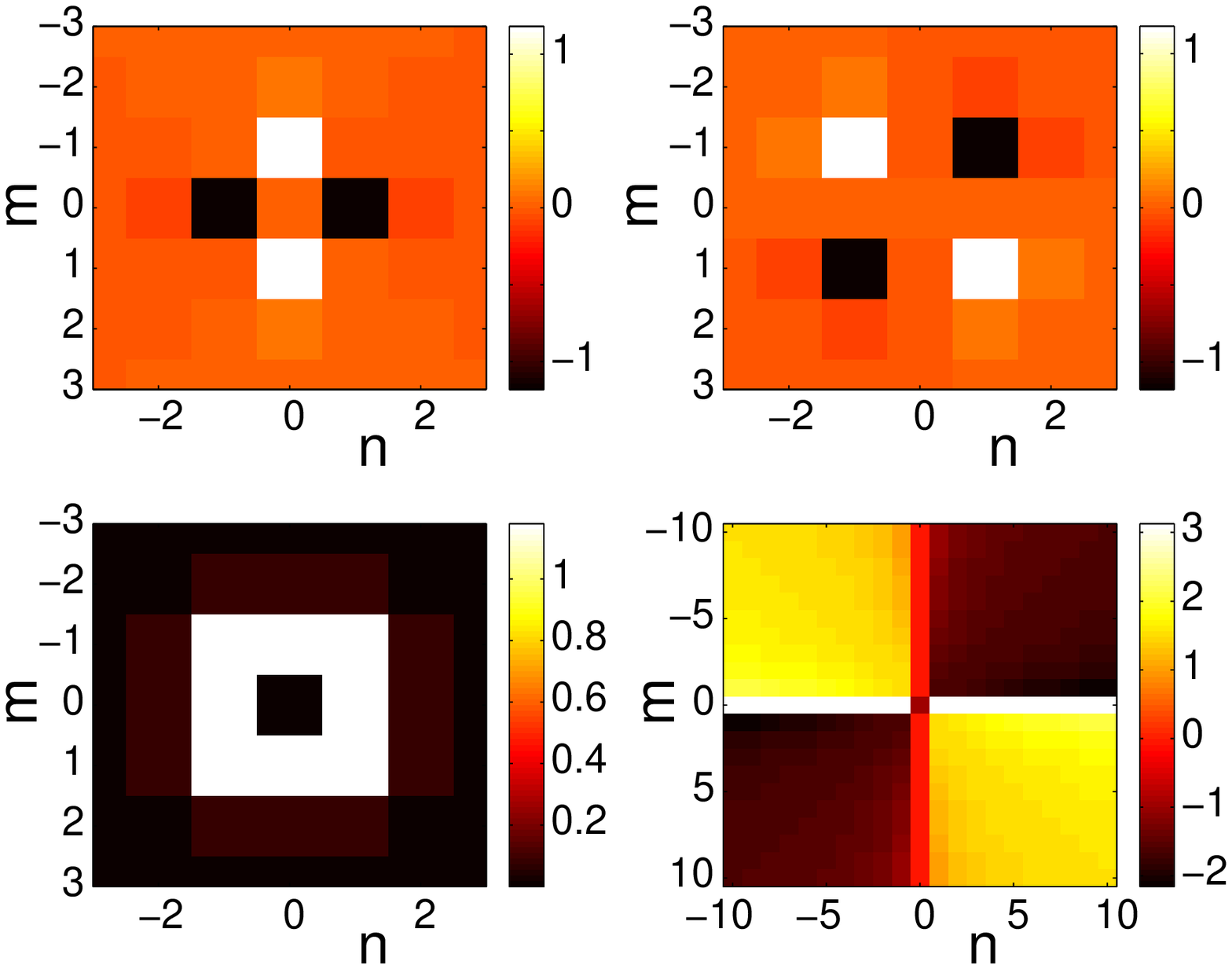}~~
\includegraphics[width=6.2cm,height=5.4cm]{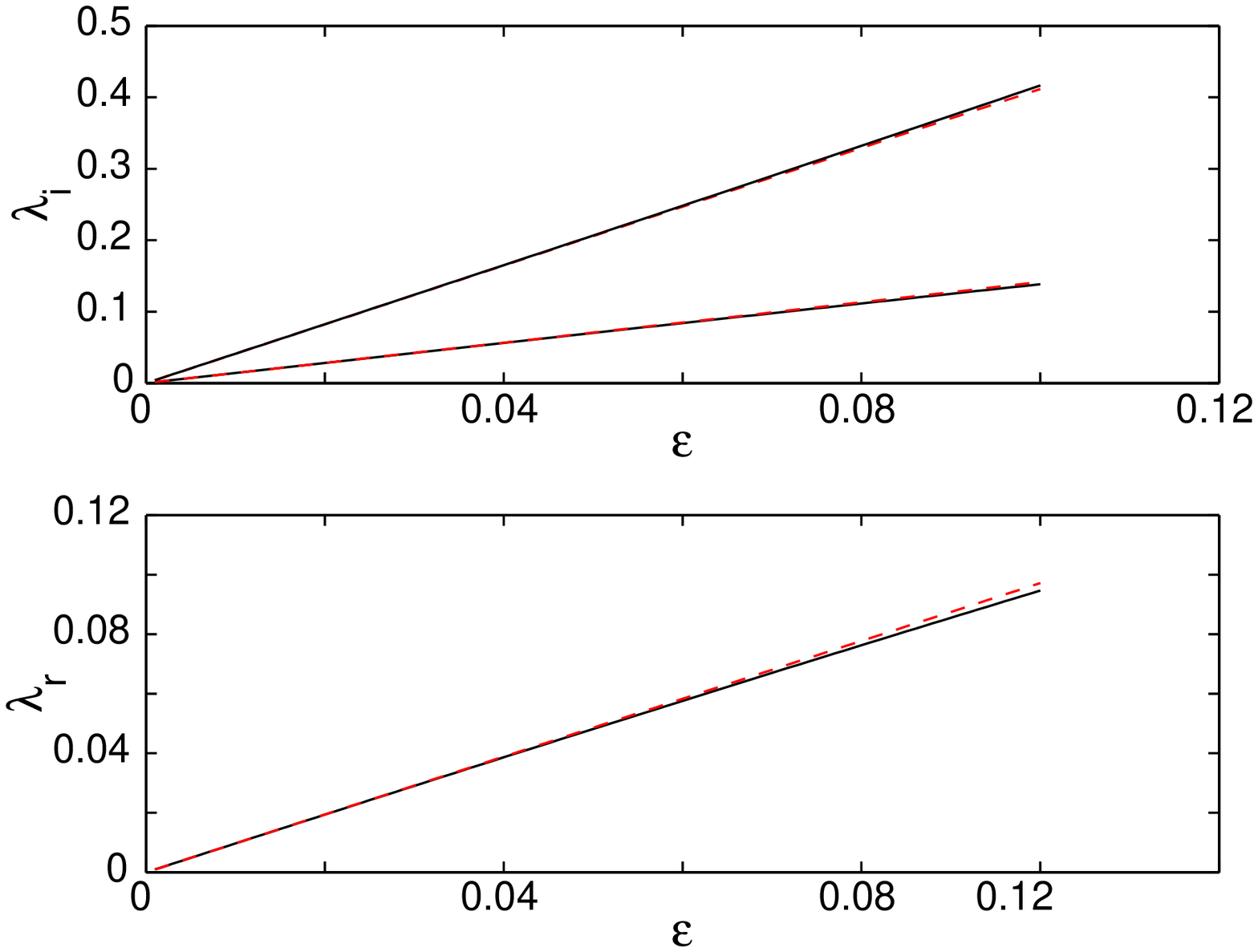}\\
\includegraphics[width=6.2cm,height=4.6cm]{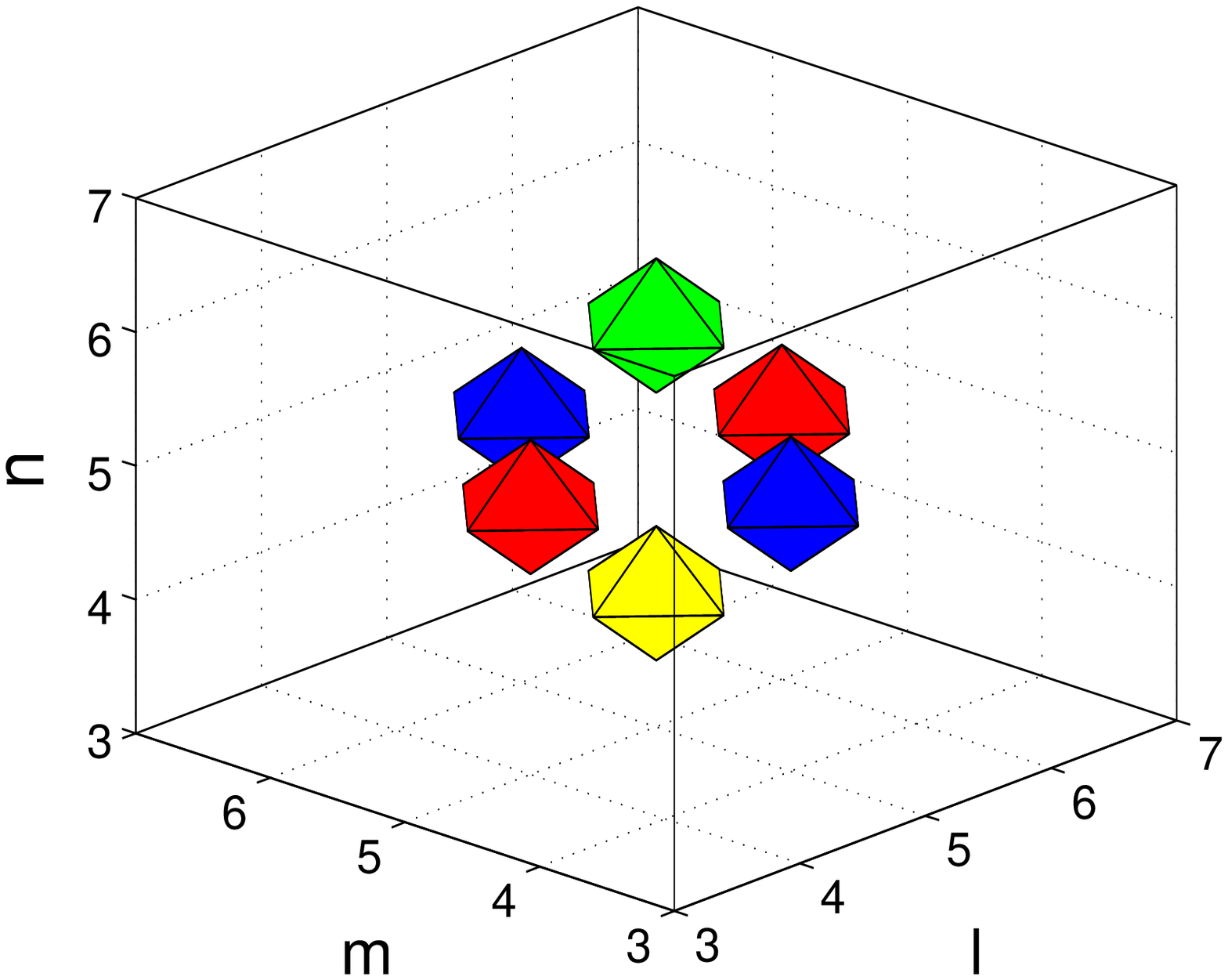}~~
\includegraphics[width=6.5cm,height=4.6cm]{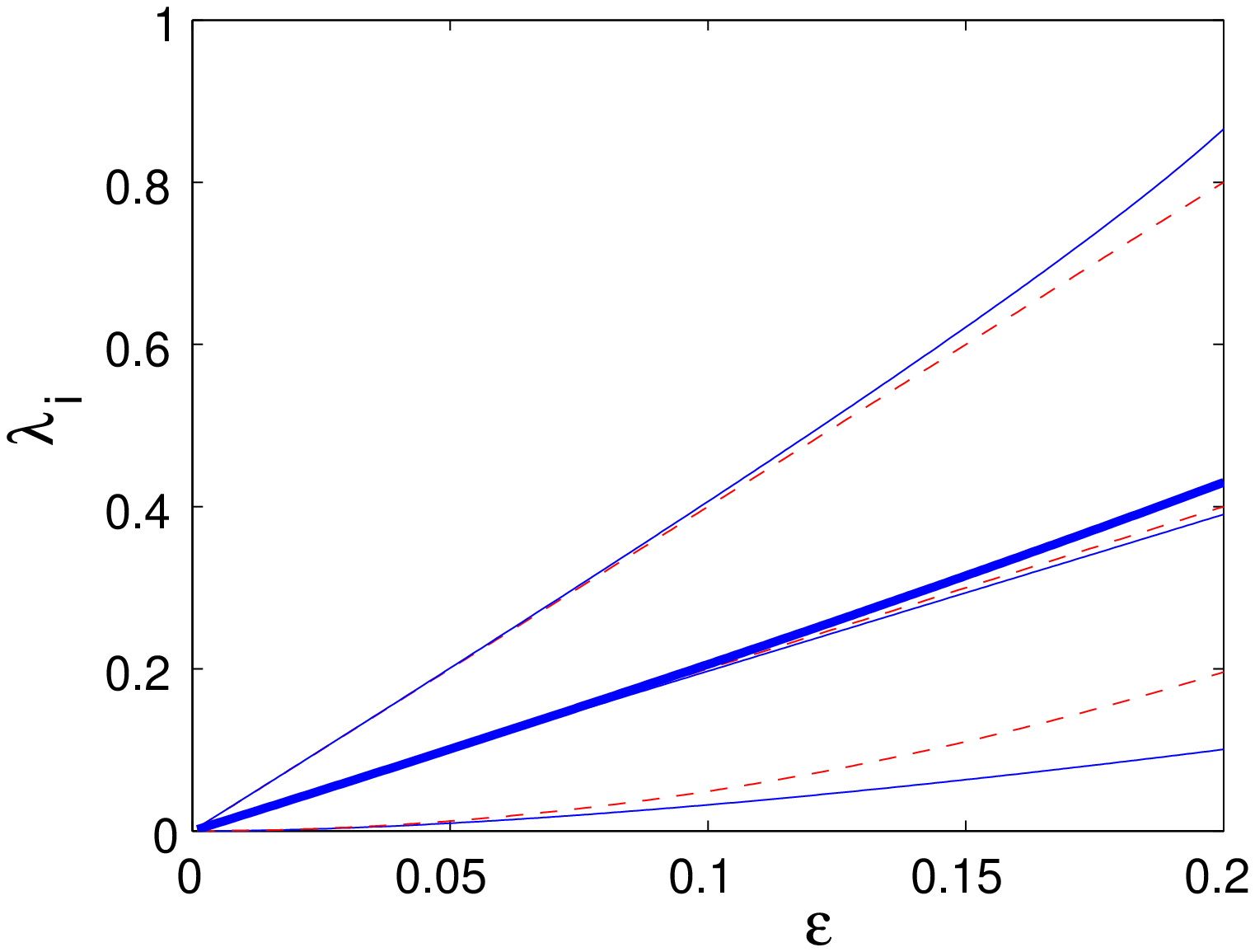}
\end{center}
\vspace{-0.3cm}
\caption{
(Color online)
The top panel shows a 2D $S=2$ vortex configuration
(left panels show its real and imaginary part -top- as well as amplitude
and phase -bottom-) and its linear stability real and imaginary eigenvalues
(right panels). Same thing in the bottom for a diamond 3D configuration
(see the explanation for its phases in the text).
The diamond configuration is shown using iso-level contours of different hues:
blue/red (dark gray/gray in the black-and-white version)
are positive/negative real iso-contours while the green/yellow (light gray/very
light gray in the black-and-white version) correspond to positive/negative 
imaginary iso-contours.
The theoretical 
predictions for the eigenvalues as a function of the coupling are shown
by dashed line, while the full numerical results by solid ones.
Partially reprinted from Ref.~\cite{pelid1} with permission.}
\label{rev_fig7}
\end{figure}

In Fig.~\ref{rev_fig7} we show typical examples of 2D and 
3D configurations that satisfy the
persistence conditions formulated above. The former configuration is a vortex
of topological charge $S=2$ (i.e., its phase changes uniformly by
$\pi/2$ from each site to the next so that it changes from $0$
to $4 \pi$ around the discrete contour of the solution). This structure is
unstable due to a real eigenvalue pair that is theoretically predicted
from Eqs.~(\ref{reduced-problem-1})--(\ref{reduced-problem-2}) to be
$\lambda=\pm \sqrt{\sqrt{80}-8} \epsilon$ (while it also has a 
pair of simple eigenvalues $\lambda = \pm i \epsilon \sqrt{\sqrt{80}+8}$,
a quadruple eigenvalue $\lambda = \pm i \epsilon \sqrt{2}$ and a single
eigenvalue of higher order). The latter configuration is a three-dimensional 
diamond configuration
(a quadrupole in the plane with phases $0, \pi, 0, \pi$ and two 
out-of-plane sites with phases $\pi/2$ and $3 \pi/2$). This is a stable
3D structure with a single eigenvalue $\lambda= \pm 4 i \epsilon$,
a triple eigenvalue $\lambda=\pm 2 i \epsilon$ and an eigenvalue of
higher order according to  
Eqs.~(\ref{reduced-problem-1})--(\ref{reduced-problem-2}). Notice in
both cases the remarkable agreement between the theoretical prediction
for the eigenvalues (dashed lines) and the full numerical results (solid
lines).
 
\section{Special topics of recent physical interest\label{Sec:special}}

In this section we give a very brief overview of some of the more
recent themes of interest in the nonlinear phenomenology emerging in
the realm of Bose-Einstein condensation. We pay special
attention to the physical motivation of the different topics.
Note that an in-depth treatment of the emergent nonlinear
behavior displayed by BECs and the synergy between experiments and theory 
can be found in the recent review~\cite{BECBOOK}.

\subsection{Spinor/Multicomponent condensates}

Advances in trapping techniques for BECs have
opened the possibility to simultaneously confine
atomic clouds in different hyperfine spin states.
The first such experiment, the so-called {\em pseudospinor}
condensate, was achieved in mixtures of two magnetically trapped
hyperfine states of $^{87}$Rb \cite{Myatt1997a}. Subsequently,
experiments in optically trapped $^{23}$Na \cite{Stenger1998a}
were able to produce multicomponent condensates for
different Zeeman sub-levels of the same hyperfine level,
the so-called {\em spinor} condensates. In addition to these
two classes of experiments, mixtures of two different
species of condensates have been created by sympathetic cooling
(i.e., condensing one species and allowing the other one
to condense by taking advantage of the coupling with the
first species) were $^{41}$K atoms were sympathetically
cooled by $^{87}$Rb atoms \cite{Modugno2001a}. 
More exotic mixtures are also being currently explored in degenerate
fermion-boson mixtures in $^{40}$K-$^{87}$Rb \cite{40Kfermion}
and pure degenerate fermion mixtures in $^{40}$K-$^{6}$Li
\cite{Li-fermion1,Li-fermion2,Li-fermion3}.

The mean-field dynamics of such multicomponent condensates
is described by a system of coupled GP equations analogous to
Eq.~(\ref{s5_1}), that for mixtures of $\mathcal N$ bosonic components
reads
\begin{equation}
\hskip-2.3cm
i \frac{\partial \psi_n}{\partial t}=-\frac{1}{2} \Delta \psi_n
+ V_n({\bf r}) \psi_n + \sum_{k=1}^{\mathcal N}\left[ g_{n,k}|\psi_k|^2 \psi_n
-\kappa_{n,k} \psi_k + \Delta\mu_k \psi_n  \right]
- i \sigma_n \psi_n,
\label{2C}
\end{equation}
were $\psi_n$ is the wavefunction of the $n$-th component
($n=1,\dots,{\mathcal N}$), $V_n({\bf r})$ is the potential confining
the $n$-th component, $\Delta\mu_{n,k}$ is the
chemical potential difference between components $n$ and $k$,
and $\sigma_n$ describes the losses of the $n$-th component.
The components $n$ and $k$ are coupled together
via (i) nonlinear coupling with coefficients $g_{n,k}$
and (ii) linear coupling with coefficients $\kappa_{n,k}$;
where, by symmetry, $g_{n,k}=g_{k,n}$ and $\kappa_{n,k}=\kappa_{k,n}$.
The nonlinear coupling results from inter-atomic collisions
while the linear coupling accounts for
spin state interconversion usually
induced by a spin-flipping resonant electromagnetic wave \cite{NZ}.
In the case of fermionic mixtures one needs to replace
the self-interacting nonlinear terms by $g_{n,n}|\psi_n|^{4/3}\psi_n$
\cite{fermionGPE1,fermionGPE2,fermionGPE3}.
In the absence of losses ($\sigma_n=0$), the total number of atoms is
conserved:
\begin{equation}
N\equiv \sum_{k=1}^{\mathcal N} N_{k}=\sum_{k=1}^{\mathcal N}\int |\psi_k|^2 d\mathbf{r}.
\label{norm} \end{equation}
In fact, in the further absence of linear interconversions
($\kappa_{n,k}=0$) each norm $N_{k}$ is conserved separately.

\begin{figure}[t]
\center
\hskip1.0cm
\epsfig{figure=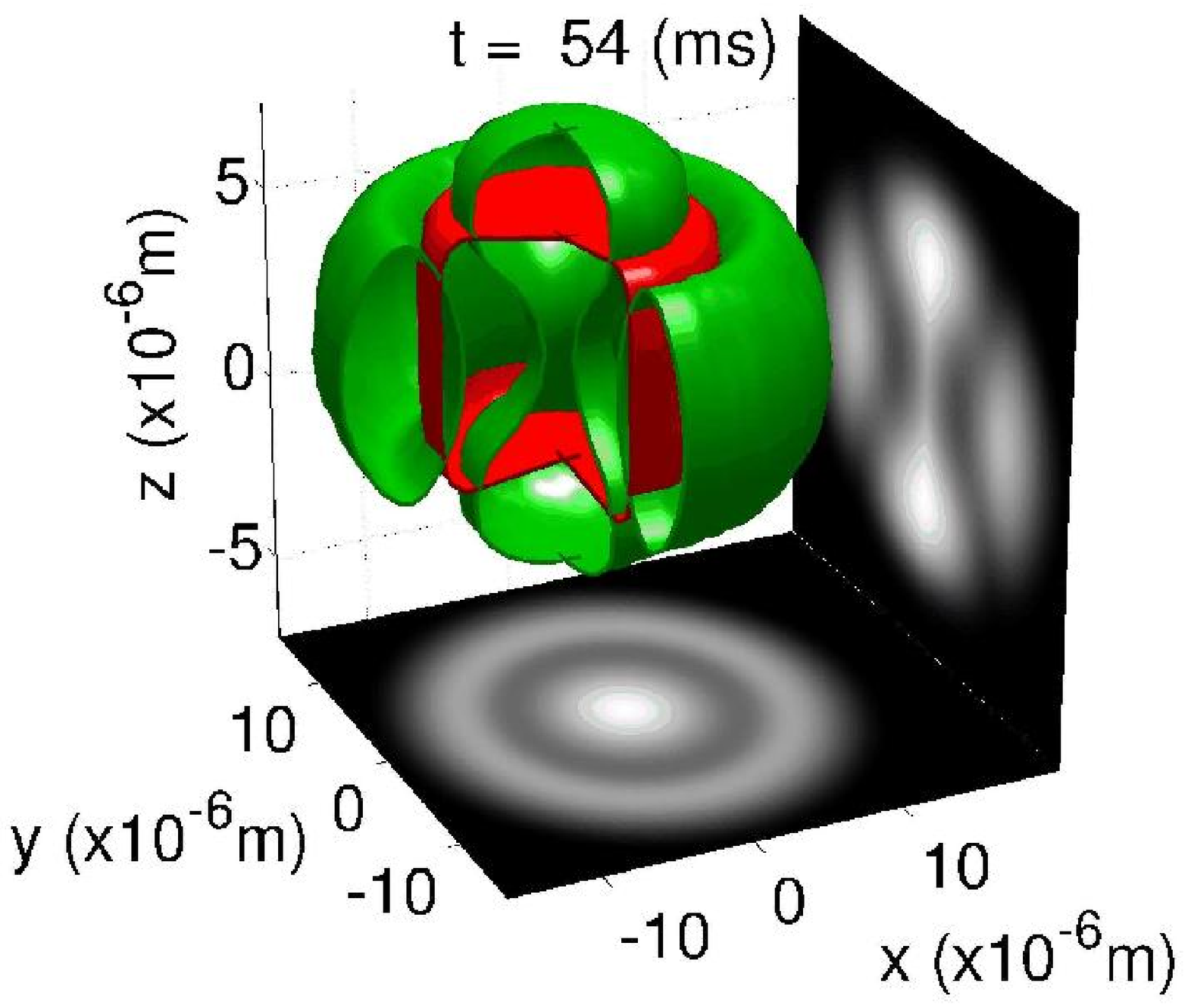,width=3.9cm}
\epsfig{figure=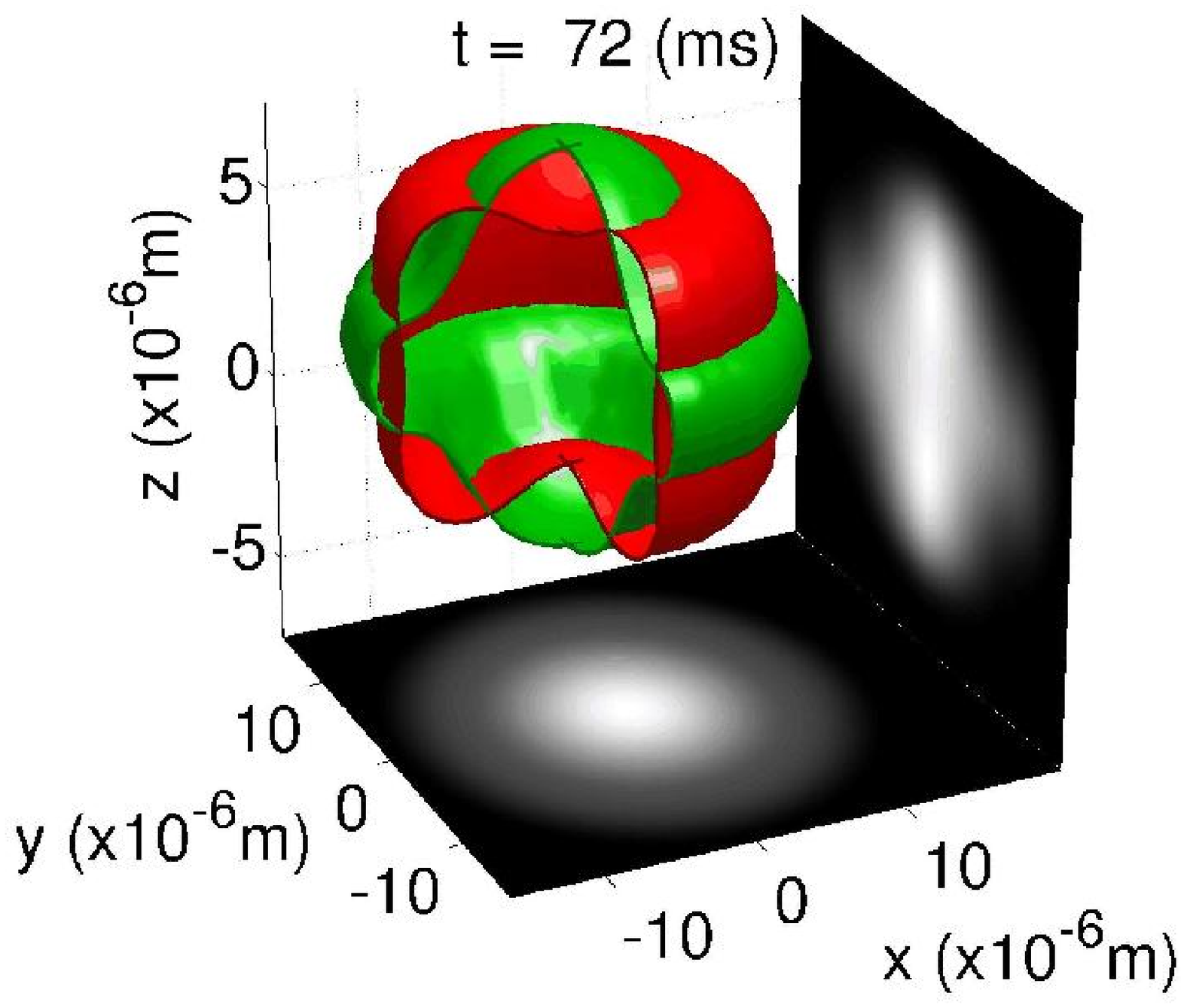,width=3.9cm}
\epsfig{figure=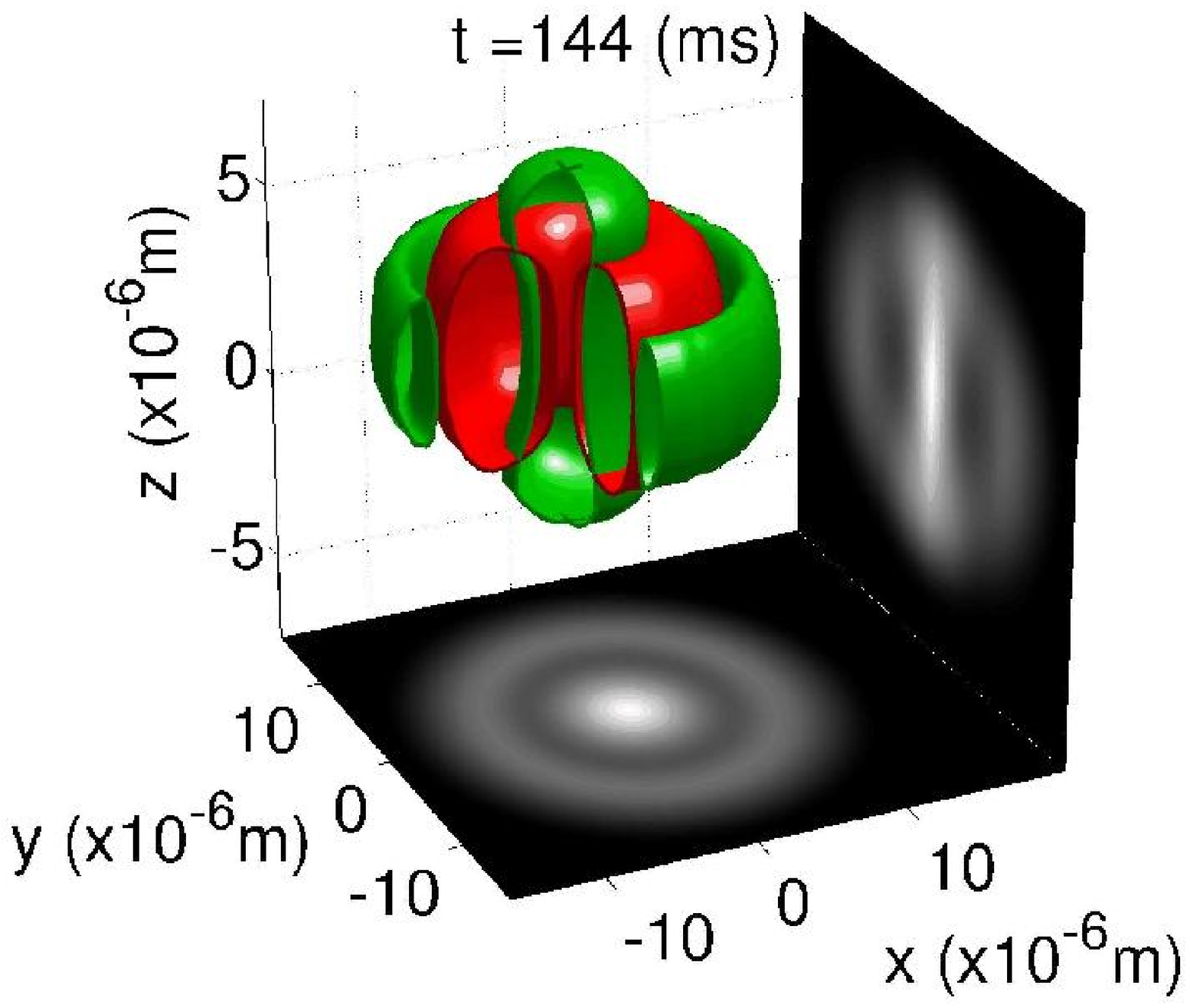,width=3.9cm}
\vspace{-0.1cm}
\caption{
(Color online).
Nonlinear dynamics of a binary 50:50 mixture of two spin states
($|1,-1\rangle $ and $|2,1\rangle $) of
$N=375,000$ Rb atoms.
Each component is depicted by a contour slice at
half of its corresponding maximal density.
The bottom (side) projection corresponds to the $z$- ($x$-) integrated
density for the $|1,-1\rangle $ component as it is observed in the
laboratory experiments \cite{CompSep}.
{Please
visit: http://www.rohan.sdsu.edu/$\sim$rcarrete/
[Publications] to view movies showing the inter-penetrating time evolution
between the two components over a span of 220 ms.}
Reprinted from Ref.~\cite{CompSep} with permission.}
\label{blobs}
\end{figure}

The simplest case of two species (${\mathcal N}=2$)
has been studied extensively.
In particular, if one considers the trapless system ($V_n=0$) in the
absence of linear interconversion, losses, and chemical
potential differences, the two components tend to
segregate if the immiscibility condition
\begin{equation}
\Delta \equiv (g_{12}g_{21}-g_{11}g_{22})/g_{11}^{2} > 0
\label{Delta} \end{equation}
is satisfied \cite{miscibility}. This condition can be interpreted
as if the mutual repulsion between species is stronger than the
combined self-repulsions. In typical experiments, the miscibility
parameter (an adimensional quantity) is rather small:
$\Delta \approx 9\times 10^{-4}$
for $^{87}$Rb \cite{Myatt1997a,proximity} and
$\Delta \approx 0.036$ for $^{23}$Na \cite{chap01:stamp}.
Depending on the various nonlinear coefficients, a vast
array of solutions can be supported by a binary condensate.
These include, ground-state
solutions \cite{pu1,shenoy,esry}, small-amplitude excitations
\cite{pu2,excit1,excit2,excit3},
bound states of
dark-bright \cite{anglin} and dark-dark \cite{obsantos},
dark-gray, bright-gray, bright-antidark and dark-antidark
\cite{epjd} complexes of solitary waves,
vector solitons with embedded domain-walls (DWs) \cite{HS},
spatially periodic states \cite{decon}, and
modulated amplitude waves \cite{mason}.
Extensions of some of these patterns in two dimensions,
namely 
DWs, have
been investigated in
Refs.~\cite{shenoy,esry,Marek,healt,DW2D1,DW2D2,DW2D3,cross1,cross2}.
The non-equilibrium dynamics of a binary condensate
has been shown to support (experimentally and theoretically) long
lived ring excitations whereby each component inter-penetrates
the other one repeatedly \cite{CompSep} (see Fig.~\ref{blobs}).
%
The effects of adding the linear inter-species coupling between the
components has also been studied in some detail
\cite{decon,lincoup1,lincoup2,lincoup3,lincoup4,lincoup5,more-lincoup1,more-lincoup2,more-lincoup3,more-lincoup4,more-lincoup5,more-lincoup6}.
One of the salient features
of adding the linear inter-species coupling is the
suppression or promotion of the transition to miscibility
(cf.~Ref.~\cite{Merhasin} and Chap.~15 in Ref.~\cite{BECBOOK}).
Spinor condensates with three species have also drawn
considerable attention since their experimental creation
\cite{chap01:stamp,cahn}. Such systems
give rise to spin domains \cite{Stenger1998a},
polarized states \cite{spindw},
spin textures \cite{spintext}, and multi-component (vectorial) solitons of
bright \cite{wad1a,wad1b,boris,zh}, dark \cite{wad2}, gap \cite{ofyspin}, and
bright-dark \cite{ourspin} types.
%

\subsection{Vortices and vortex lattices}

\begin{figure}[htb]
\center
\hskip2.6cm
\epsfig{figure=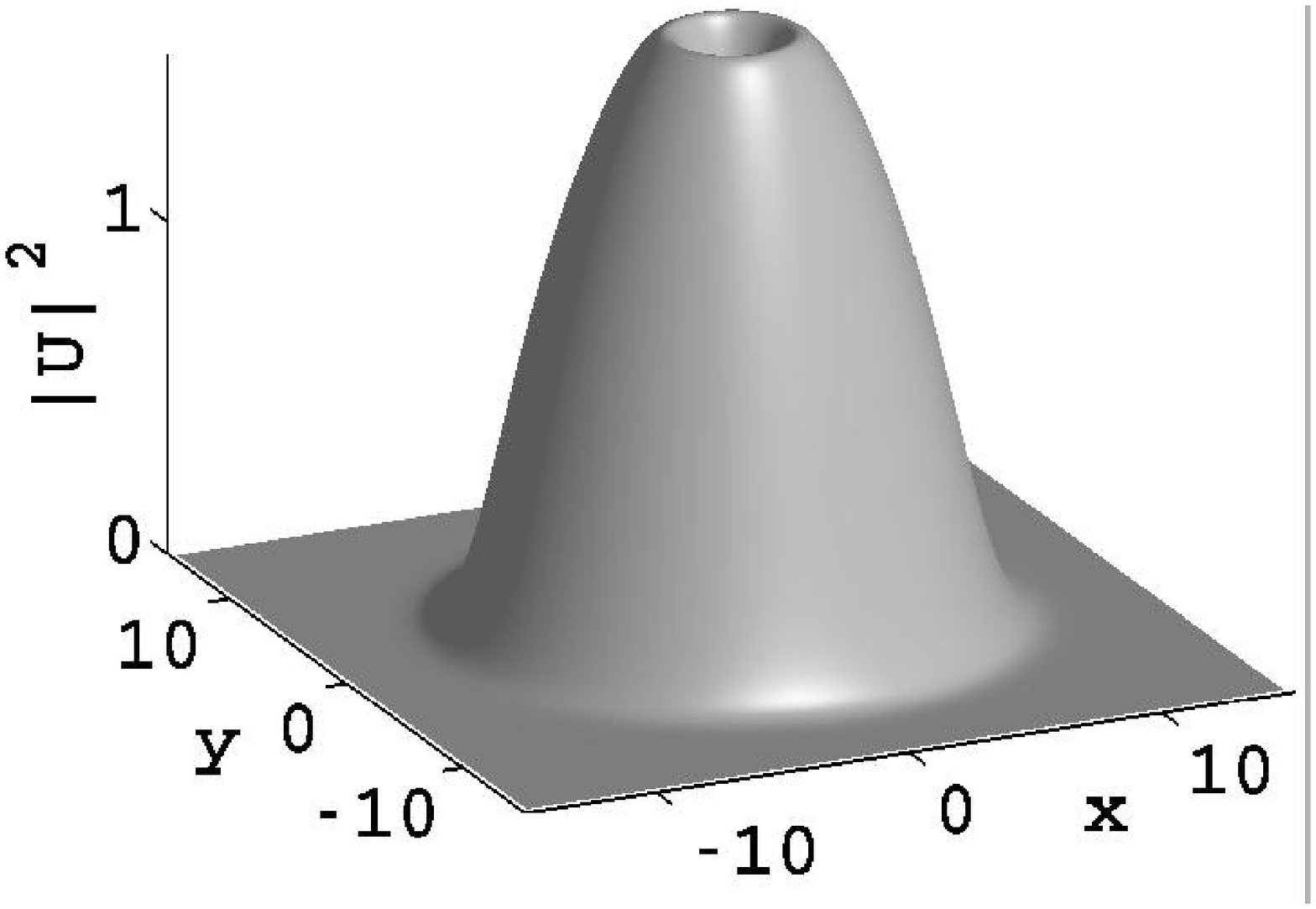,width=4.4cm,height=3.5cm}
~ ~
\epsfig{figure=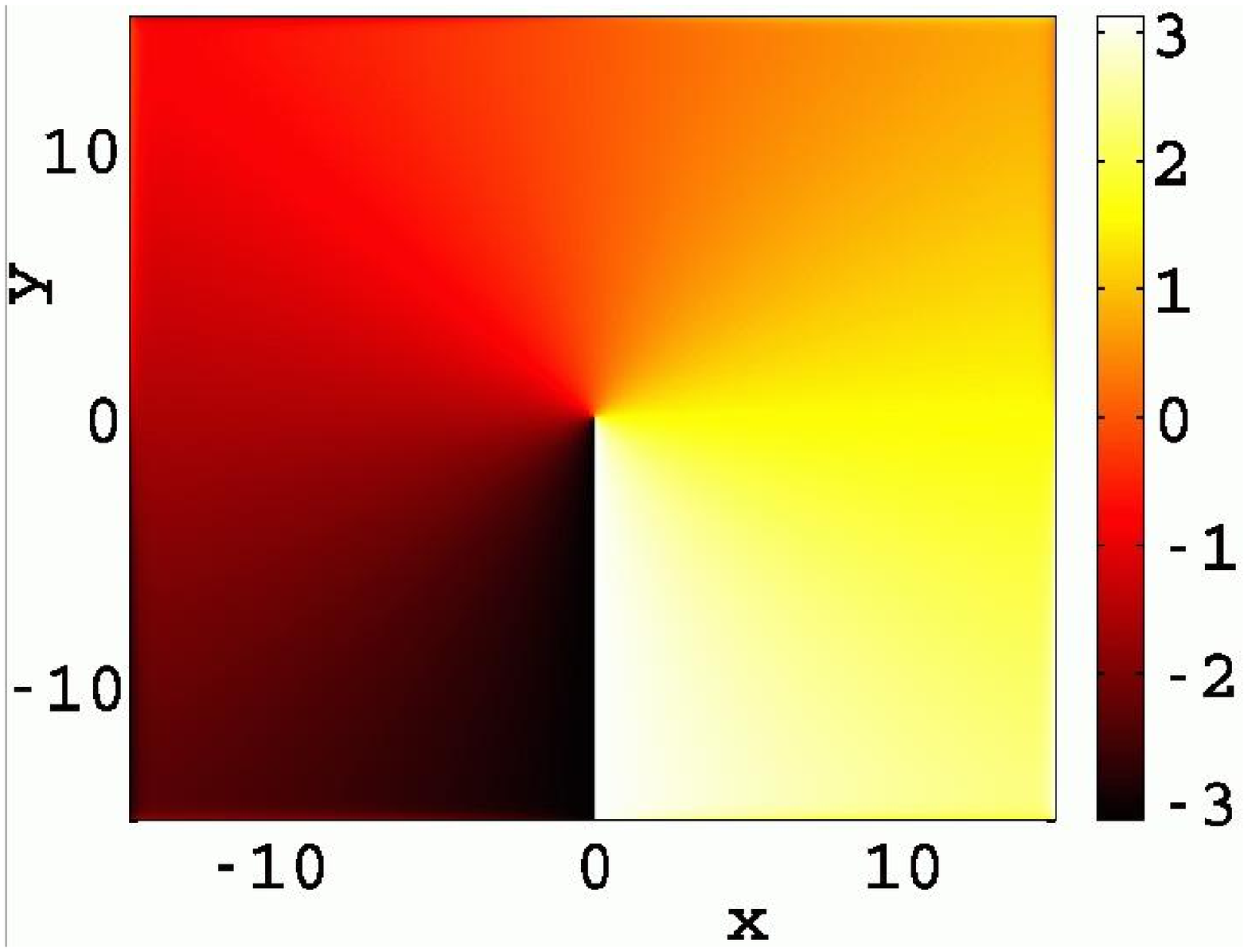,width=4.1cm,height=3.5cm}
\vspace{-0.0cm}
\caption{
(Color online)
Two-dimensional, singly charged ($S=1$), vortex inside a parabolic
magnetic trap $V(z)=\frac{1}{2}\Omega^2 (x^2+y^2)$ with $\Omega=0.2$.
Depicted are the density (left) and
the phase profile (right) for a chemical potential $\mu=2$.
}
\label{chap01:fig2}
\end{figure}

Arguably, one of the most striking nonlinear matter-wave
manifestations in BECs is the possibility of supporting
vortices \cite{fetter,PGK:MPLB:04}. 
Vortices are characterized by their non-zero topological
charge $S$ whereby the phase of the wavefunction has a 
phase jump of $2\pi S$ along a closed contour surrounding the core of
the vortex (cf.~phase profile for a singly charged
vortex, $S=+1$, in the right panel of Fig.~\ref{chap01:fig2}).
Historically, the first observation of vortices in
BECs was achieved \cite{chap01:vort1} 
by phase imprinting
between two hyperfine spin states of Rb \cite{Williams99}.
Nowadays, the standard technique to nucleate vortices in
BECs is based on stirring \cite{chap01:vort2} 
the condensate cloud
above a certain critical angular speed \cite{Recati01,Sinha01,Madison01}.
This technique has proven to be extremely efficient, not only in
creating single vortices, but also, from a few vortices
\cite{Madison01}, to vortex lattices \cite{Raman}.
It is also possible to nucleate vortices by
dragging a moving impurity through the condensate
for speeds above a critical velocity (depending on the
local density and also the shape of the impurity)
\cite{frisch92,Hakim97,Jackson98,Jackson00,adams1a,adams1b,LInew,2Cdark}.
Yet another possibility to nucleate vortices can be
achieved by separating the
condensate in different fragments and allowing them to collide
\cite{BPAPRL,Nate,bpa_us}.

The profile of a vortex in a two dimensional setting
(see left panel of Fig.~\ref{chap01:fig2})
can be obtained by solving for the 
density $U(r)$ when considering a wave function of the
form $\psi(r,\theta) = U(r)\exp(iS\theta-i\mu t)$
that satisfies the 2D GP equation 
with repulsive nonlinearity ($g=+1$), 
where $(r,\theta)$ are the polar coordinates and $\mu$ is the 
chemical potential. 
The equation for  $U$
takes the form
\begin{equation}
\frac{d^2 U}{d r^2}+\frac{1}{r}
\frac{dU}{dr}-\frac{S^2}{r^2}U +\left(\mu-V(r)-U^2\right)U=0,
\label{chap01:ve}
\end{equation}
with boundary conditions $U(0)=0$ and $U(+\infty)=\sqrt{\mu}$
for a confining potential $V(r=+\infty)=+\infty$.
Unfortunately,
the ensuing equation for the vortex profile cannot be
solved exactly (even in the simplest homogeneous case with 
$V(r)\equiv 0$)
and one has to resort to numerical or
approximate methods \cite{Berloff}.
The asymptotic behavior of the vortex profile $U(r)$
can be found from Eq.~(\ref{chap01:ve}), i.e., $U(r) \sim r^{|S|}$ as
$r\rightarrow 0$, and $U(r) \sim \sqrt{\mu}-{S^2}/{(2r^2)}$ as
$r\rightarrow +\infty$.
%
%
%
%
Note that the width of singly charged vortices in BECs is of ${\cal O}(\xi)$ 
[where $\xi$ is the healing length given in Eq.~(\ref{chap01:heal})],  
while higher-charge vortices ($|S|>1$) have cores wider than the healing length and are unstable in the
homogeneous background case ($V(r)\equiv 0$) but might
be rendered stable by external impurities \cite{simula:pra2002}
or by external potentials \cite{lundh:pra2002,Pu99,S2Ket,S2Mottonen}.
When unstable, higher order charge vortices typically split in multiple
single charge vortices.

\begin{figure}[t]
\center
\hskip2.6cm
\hskip0.7cm
\epsfig{figure=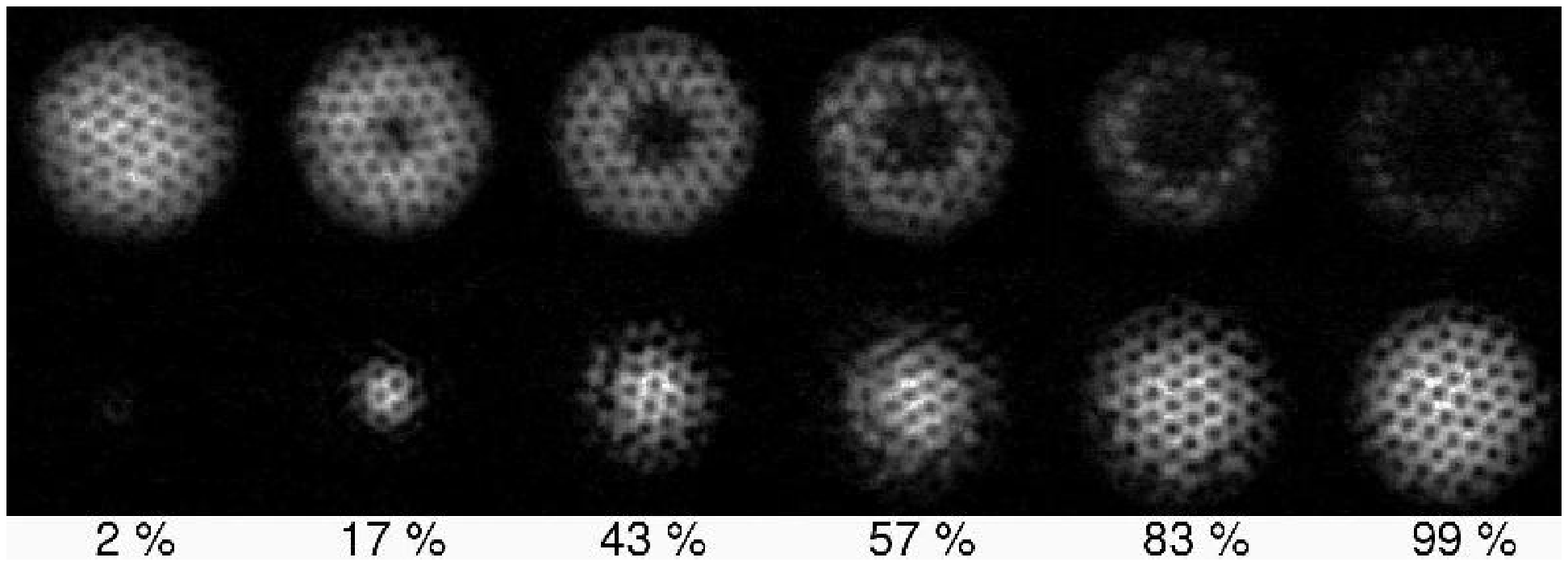,width=9.0cm,height=3cm,angle=0,silent=}\\[2.0ex]
\hskip2.6cm
\epsfig{figure=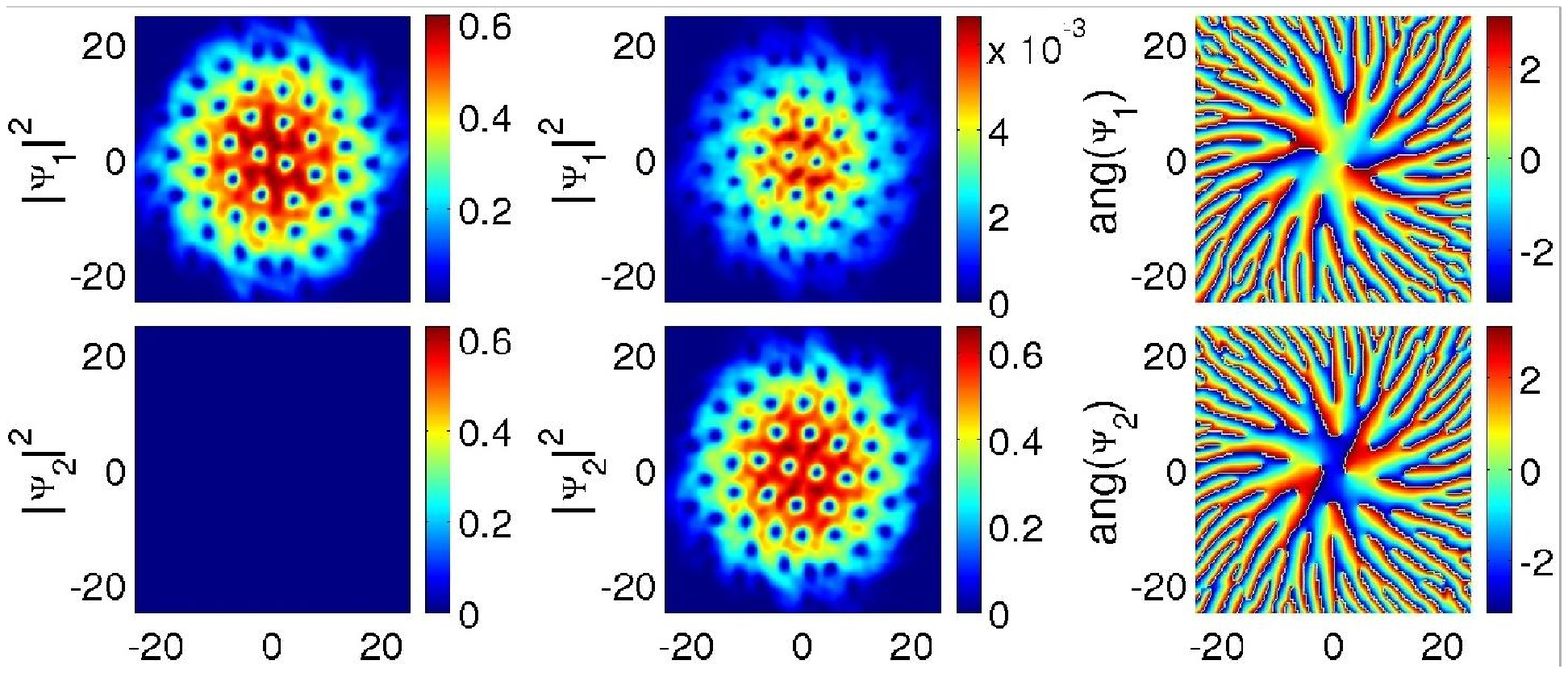,width=10.0cm,height=3.8cm,angle=0,silent=}
\caption{
(Color online).
Top:
Experimental look at component separation in a
rotating BEC of Rb atoms
\cite{DHdata} after 30~ms evolution,
release and 20~ms free expansion from a relatively tight trapping
potential. The percentages quoted are the fraction of atoms
transferred from the $|1,-1\rangle $ state (top row) to the
$|2,1\rangle $ state (bottom row).
%
Bottom:
Numerical vortex lattice (VL) transfer by linear coupling
from the first (top row) to the second component (bottom row). The
initial VL (left column) is successfully transferred between
components (see middle column where the scale in the top panel
clearly indicates that the first component is almost depleted of
atoms after transfer).
More importantly, note that the phase distribution is also
transferred between components (right column).
\label{fig_VL} }
\end{figure}

Single charge vortices are extremely robust due to their inherent topological
charge since continuous transformations/deformation of the
vortex profile cannot eliminate the $2\pi S$ phase jump
---unless that the density is close to zero (this is the reason
why, in the stirring experiments,
vortices are nucleated at the periphery of the condensate
cloud where the density tends to zero for confining potentials
\cite{Gardiner02,Ueda02,Ueda03,Castin04}).
Vortices are prone to motion induced by gradients in both
density and phase of the background \cite{Kivshar98}.
These gradients can be induced by an external potential
or the presence of another vortex. The effect of vortex
precession induced by the external trap has been extensively
studied \cite{precession1,precession2,precession3,precession4,precession5,precession6,precession7,precession8}.
The motion induced on a vortex by another vortex
is equivalent to the one observed in fluid vortices
whereby vortices with same charge travel parallel
to each other at constant speed, while vortices of opposite
charges rotate about each other at constant angular speed.

Another topic that has attracted an enormous deal of attention
in recent years is the ubiquitous existence of robust
vortex lattices in rapidly rotating condensates consisting
of ordered lattices of vortices arranged in
triangular configurations (the so-called Abrikosov lattices
\cite{Abrikosov}). The first experimental observation of
vortex lattices consisted of just a few ($< 15$) vortices
\cite{Raman} but soon experiments were able to maintain
vortex lattices with some 100 vortices \cite{madi00jmo}.
Alternative methods to describe vortex lattice configurations
have been given in terms of Kelvin's variational principle
\cite{Newton1} and through a linear algebra formulation \cite{Newton2}.
Another approach is to treat each vortex as a quasi-particle
and apply ideas borrowed from molecular dynamics to find the
most energetically favorable configurations \cite{rcg:30:vortexMD}.
Some other interesting phenomenology of vortex lattices includes
the excitation of Tkachenko modes \cite{tkach}
via the annihilation of a central chunk of vortex lattice
matter through a localized laser heating \cite{engels2}.

Yet another promising avenue of research that is currently being
explored is the topic of vortex lattices in multicomponent
condensates. For example, starting with a two-component BEC mixture
with only one atomic species containing a vortex lattice,
and subsequently ``activating'' the linear coupling, it is possible
to {\it entirely transfer} the vortex lattice
to the second component (cf.~results in Fig.~\ref{fig_VL}).
This ``Rabi oscillation'' between atomic species \cite{dsh3,decon} is
an extremely useful tool for controllably transferring desirable
fractions of atoms from one state to another and can be
extended to multicomponent, condensates \cite{stamper,uedaspinor}.
Furthermore, it is important to note that the interaction of vortex lattices
in a multicomponent BEC can result in structural changes
in the configurations of the vortex lattices, i.e., resulting
in lattices with different symmetry \cite{UedasquareVL,engelssquareVL}.


\subsection{Shock waves \label{Sec:shocks}}

One of the classical types of nonlinear waves appearing in the context of BECs is shock waves. 
Shock waves were first observed in the experiments reported in Ref.~\cite{chap01:dutton}, 
where a slow-light technique was used to produce density depressions in a sodium BEC. 
More recently, they were observed in rapidly rotating $^{87}$Rb BECs triggered 
by repulsive laser pulses \cite{soundsw}, while their formation was discussed 
in an experiment involving the growth dynamics of a 1D sodium quasi-condensate in a dimple 
microtrap created on top of the harmonic confinement of an atom chip \cite{nppsw}. 
Finally, shock waves were studied experimentally and theoretically in Ref.~\cite{ablsw}; 
in the experiments reported in this work, repulsive laser beams were used (as in Ref.~\cite{soundsw}, 
but with a nonrotating condensate) to push atoms from the BEC center, thus forming ``blast-wave'' patterns. 

On the theoretical side, shock waves in repulsive BECs were mainly studied in the framework of mean-field theory and  
the GP equation for weakly-interacting Bose gases \cite{ablsw,psh,zak1,zak2,damskisw,frmsw,anat,menotti,dekel}, but also 
for strongly interacting ones \cite{damskiswsi} and in the BEC-Tonks crossover \cite{damskiswbt}; additionally, 
the effect of temperature (see Sec.~\ref{Sec:beyond}) on shock wave formation and dynamics and the effect of depleted atoms 
were respectively discussed in Refs.~\cite{nppsw} and \cite{damskiswbt}. 
Many of the above mentioned theoretical studies rely on the hydrodynamic equations that can be obtained 
from the GP equation via the Madelung transformation $\psi=\sqrt{\rhon}\exp(i\phi)$ (with $\rhon$ and  $\phi$ being 
the condensate's density and phase, respectively, see also Sec.~\ref{sec:reductive}). These hydrodynamic equations are then treated in the long-wavelength 
limit (or, equivalently, in the weakly dispersive regime) where the so-called quantum pressure term,  
$\sim (\nabla^2 \sqrt{\rhon})/\sqrt{\rhon}$, is negligible. 
Qualitatively speaking, one may ignore this term if the so-called ``quantum Reynolds number'' \cite{zak1,zak2} 
(see also the discussion in Refs.~\cite{psh,frmsw}) is $R \equiv an_0 L_{0}^2 \gg 1 $, where $a$ is the scattering length, $n_0$ 
the peak density of the BEC and $L_0$ a minimal scale among all the characteristic spatial scales of 
the condensate wavefunction. In fact, a more rigorous treatment relies on the assumption that the quantum pressure 
term is small, as in the case of the theoretical analysis (and the pertinent experimental results) of Ref.~\cite{ablsw}. 
In any case, since the quantum pressure is reminiscent of viscosity in classical fluid mechanics, this 
weakly dispersive regime suggests the possibility of ``dissipationless shock waves'' in BECs 
(according to the nomenclature of Ref.~\cite{anat}). 

The above mentioned hydrodynamic equations were treated in the limiting case of zero quantum 
pressure in Refs.~\cite{psh,zak1,zak2,damskisw}. In this case, the hydrodynamic equations resulting from the GP equation 
are reduced to a hyperbolic system of conservation laws of classical gas dynamics, 
namely an Euler and a continuity equation (see, e.g., Ref.~\cite{liberman}). In this gas dynamics picture, 
the above hyperbolic system is characterized by two real eigenspeeds (i.e., characteristic speeds of propagation 
of weak discontinuities) $\upsilon^{\pm}=v\pm c$, where $v$ is the velocity of the gas and $c$ is the speed of sound. 
Since the latter depends on the condensate density $\rhon$ [see Eq.~(\ref{chap01:sound})], it is clear that higher values 
of $\rhon$ propagate faster than lower ones and, as a result, any compressive part of the wave ultimately breaks to give 
a triple-valued solution for $\rhon$; this is a signature of the formation of a shock wave. 
In this system, a Gaussian input produces two symmetric shocks at finite time, as was shown 
e.g., in Refs.~\cite{damskisw,damskiswbt} (see also Ref.~\cite{kodama} for a rigorous discussion). 
Importantly, the trailing edge of the shock wave as observed in the simulations and the experiments 
(see, e.g., Refs.~\cite{chap01:dutton,nppsw,ablsw,damskisw,damskiswbt}) can be considered as a modulated train  
of dark solitons \cite{ablsw,anat}.

In the work \cite{frmsw}, a more careful investigation of the regimes of validity of the condition 
$R \gg1$ (which may fail, e.g., in the case of expanding BECs) led to 
an experimentally relevant protocol to produce shock waves in BECs. This protocol is based on the 
use of Feshbach resonance to control the scattering length $a$, namely to make $a$ an increasing 
function of time (by a proper ramp-up procedure), so as to increase the time domain in which the 
quantum pressure is negligible. This way, this ``Feshbach resonance management'' technique 
\footnote{
This technique was suggested in earlier works in various important applications, as e.g., a 
means to prevent collapse of higher-dimensional attractive BECs \cite{ueda03,montesinos,FRM1}, 
to produce periodic waves \cite{chap01:VVK2B}, robust matter-wave breathers \cite{kkp1,kkpz,ourfrm,borinf}, and so on. 
}
was shown to produce multi-dimensional shock waves. On the other hand, in Ref.~\cite{anat} the Whitham averaging 
method \cite{whitham} was used in the weakly dispersive regime to show that dissipationless shock waves are 
emanating from density humps in repulsive BECs. Moreover, the formation of shock waves in BECs confined 
in optical lattices was discussed in Ref.~\cite{menotti}, while in Ref.~\cite{ablsw} an in depth analysis of the 
shock waves appearing in BECs and in gas dynamics (also in connection to relevant experiments of the JILA group) was presented. 
Finally, it should also be mentioned that the above mentioned works chiefly refer to repulsive BECs; an analysis of the 
shock wave formation in attractive BECs can be found in Ref.~\cite{chap01:rab}.

\subsection{Multidimensional solitons and collapse \label{Sec:multi}}

As was highlighted in Sec.~\ref{Sec:dim_reduc}, 
a strong transverse confinement may effectively render the BEC quasi-1D, in which 
case the 1D soliton solutions are physically relevant and are stable. 
On the contrary,  
in the absence of a tight transversal trapping, higher dimensional
extensions of 1D solitons are generally unstable \cite{chap01:sulem}.
However, 
by restricting the transverse direction(s) of the condensate,
it is possible to obtain higher dimensional
soliton solutions that are stabilized for times longer than
the lifespan of the experiments \cite{muryshev99,carr2000e,feder2000}.

Let us now showcase some of the possible higher dimensional
soliton solutions displayed by the GP model. We do not
cover here ``true'' 3D solutions (i.e., solutions that do not have
a 1D equivalent) such as 2D vortices (see previous section),
3D vortex lines \cite{chap01:pittaevskii,Rosenbush02,GarciaRipoll01,Modugno03,Ruostekoski04},
vortex rings \cite{donnelly1991,saffman1992,chap01:dark} or more complicated
topologically charged structures such a skyrmions
\cite{anglin01,skyrme2d,mueller2004,Ruostekoski03,Savage05,Battye02}.

\subsubsection{Dark solitons.}

The trademark of a dark soliton is its phase jump along its
center separating two repelling phases. In a 2D geometry
a dark soliton corresponds to a nodal line separating the
two phases while in 3D it corresponds to a nodal plane.
Both the 2D and 3D dark solitons, respectively called band
(or stripe) and planar
dark solitons, are prone to 
the snaking instability along
their nodal extent~\cite{law1993,mcdonald1993}.
These instabilities result in the nucleation of vortex pairs
in 2D \cite{pelinovsky95,kivshar96}
and pairs of vortex lines and/or vortex rings in 3D.
When the dark soliton is set into motion inside a confining
trap, it suffers bending resulting from the different speeds
of sounds at the edge of the cloud (low density and thus slower
speeds compared to the speed at the center of the cloud)
accelerating the formation of vortices at the trailing edges
\cite{chap01:denschl}.
It is also possible to create dark soliton structures whose
nodal sets, instead of extending linearly, can be wrapped around.
It is therefore possible to create in 2D ring dark solitons and
in 3D spherical shell dark solitons, as the ones discussed in 
Sec.~\ref{Sec:math}.3.2. 
Such structures can be described
by nonlinear Bessel functions (cf.~Ref.~\cite{carr2006d} and
Chap.~7 in Ref.~\cite{BECBOOK}), 
are also prone to the 
snaking instability. 
%
It is worth mentioning that the abovementioned instabilities can 
be weak (slow) enough so that these solitons can be observed in the
experiments \cite{chap01:dark,chap01:dutton,ginsberg05}.

\subsubsection{Bright solitons and collapse.\label{sec:BS_collapse}}

Bright solitons in higher dimensions are prone to a different type
of instability due to the intrinsic collapse of solutions of 
the NLS equation 
\cite{chap01:sulem}.
The first experiments with attractive condensates 
suffered from this collapse instability
\cite{chap01:colexp1,chap01:colexp2,sackett1998,sackett1999}
while the more recent experiments were able to focus on stable regimes and demonstrate bright soliton formation 
\cite{chap01:bright1,chap01:bright2,chap01:bright3}.
In fact, the key feature of these experiments was the quasi-1D nature of the attractive BEC 
realized in anisotropic traps as discussed in 
Sec.~\ref{Sec:dim_reduc}. Thus, the observed bright solitons were
found to be robust, which would not be the case in a higher-dimensional system, as 
%
%
%
they should either collapse or expand indefinitely depending on the
number of atoms and the density profile. The solution that
constitutes the unstable separatrix 
between expansion and collapse is the well-known 
Townes soliton \cite{chiao1964,moll2003}.

In this connection, it is important to mention that even though the experimental condensates 
are never truly 1D,   
fortunately, the tight trapping in the transverse direction(s)
is able to slow collapse to times much longer than
the duration of the experiments. Nonetheless, interactions
of bright solitons in higher dimensions, in contrast with
their 1D counterparts, may be inelastic
\cite{chap01:cqnlsD,chap01:ourpraC,parker2006},
and, furthermore, when two (or more) solitons overlap
their combined number of atoms can exceed the critical
threshold and initiate collapse. The overlap of
higher dimensional solitons, even when the critical
number of atoms is exceeded, might not result in collapse
since one has to take into account the time of the interaction
(depending on the velocities of the solitons and their 
relatives phases, cf.~Chap.~7 in Ref.~\cite{BECBOOK}
and references therein).

Finally, we note that stabilization of higher-dimensional 
bright solitons by means of lower-dimensional optical 
lattices has been proposed in Refs.~\cite{baizakovepl,matuz1,borinf}.
Moreover, stable bright ring solitons carrying
topological charge have been theoretically predicted to exist
\cite{carr2006c,chap01:rup,dodd1996a,dodd1996b,Pu99}
but no experiment has yet corroborated these results.

\subsection{Manipulation of matter-waves}

As discussed in Sec.~\ref{Sec:dim_reduc}, 
one of the most appealing features in BECs is the level of
control over the different contributions on the GP equation.
This includes the possibility to craft almost
any desired external potential (by the appropriate superposition
of multiple laser beams), and to 
change the strength and
sign of the nonlinearity via the 
Feshbach resonance mechanism.
%
This is to be contrasted with
 other contexts where the NLS equation is also a relevant model, 
as, e.g., in nonlinear optics \cite{chap01:kiag}, where
it is extremely difficult and often impossible to demonstrate such control. 
In the BEC context, particularly appealing is the fact that the external potential
and/or nonlinearity can be made to follow in time any
desired evolution. In this section we focus on the use of
appropriately crafted time-dependent external potentials to manipulate
mater-waves in BECs.
%
%
In this section we only consider matter-wave manipulation
by two main types of external
potentials: localized and periodic potentials.
These potentials are experimentally generated by laser beams 
as explained in Sec.~\ref{Sec:GPE}.4. Here, it would be useful to recall that 
the sign of the external optical potential is positive (negative) for
blue- (red-) detuned laser beams.

\subsubsection{Localized potentials.}

The interaction of solitons with localized impurities
has attracted much attention in the theory of nonlinear
waves \cite{pertkm,am,mi1} and solid state physics \cite{marad,kos1}.
In 1D BECs confined by a harmonic trap, bright
and dark solitons perform harmonic oscillations as discussed in 
Sec.~\ref{Sec:math}
as a consequence of the Kohn's theorem (the ``nonlinear analogue'' of the Ehrenfest theorem)  
\cite{kohn,kohn2}, which states that the
motion of the center of mass of a cloud of particles trapped in a
parabolic potential decouples from the internal excitations.
The existence, stability and dynamics of bright solitons
in the presence of the external potential can be analyzed using
perturbation techniques expounded in Sec.~\ref{Sec:math}.
In fact, the combined effects of the harmonic trap
$V_{\rm HT}(x)=\Omega^2 x^2/2$ and an
infinitely localized delta impurity, located at $x_i$, namely
$V_{\mathrm{Imp}}(x)=V_{\mathrm{Imp}}^{(0)}\,\delta(x-x_i)$,
yield the following effective force on a bright
soliton \cite{Herring:PLA:05},
\begin{equation}
\hskip-2cm
F_{\rm eff} = F_{\rm HT} + F_{\rm Imp} =
-2\Omega^2\eta\,\zeta -
2\eta^3V_{\rm Imp}^{(0)}\tanh (\eta (x_i -\zeta))\, \sech^{2}(\eta (x_i -\zeta )),
\label{chap10:Veff1}
\end{equation}
where $\zeta$ and $\eta$ are, respectively,
the center and height of the bright soliton, while 
the first (second) term in the right-hand side corresponds to the harmonic trapping
(localized impurity) potential.
This effective force induces a Newton-type dynamics for the center
$\zeta$ of the bright soliton, as discussed previously.
Note that the force induced by the harmonic trap always points towards
its center, while the direction of the force induced by the impurity
depends on the sign of $V_{\mathrm{Imp}}^{(0)}$.
In the case of an attractive impurity it is possible to not only
pin the bright soliton away from the center of the harmonic trap but
also to adiabatically drag it and reposition it at, almost, any desired
location by slowly moving the impurity \cite{Herring:PLA:05}.
The success in dragging
the soliton not only depends on the profiles of the soliton and the impurity,
but more crucially, on the degree of adiabaticity when displacing the
impurity.

%
A similar study can also be performed in the case of dark solitons. 
The interactions of dark solitons with localized impurities was analyzed in 
Ref.~\cite{vvk1}, by using the so-called direct perturbation theory for dark solitons \cite{vvkve}, 
and later in the BEC context in Ref.~\cite{fr1}, by using the adiabatic perturbation theory 
for dark solitons \cite{KY}. Following the analysis of Ref.~\cite{fr1}, it is possible to show that 
the center $\zeta$ of a dark matter-wave soliton obeys 
%
%
a Newton-type equation of motion 
with an effective potential given by:
\begin{equation}
V_{\rm eff}(\zeta) = V^{\rm eff}_{\rm HT}(\zeta) + V^{\rm eff}_{\rm Imp}(\zeta) =
\frac{1}{2}V_{\rm HT}(\zeta) + \frac{1}{4} V_{\mathrm{Imp}}^{(0)}{\rm sech}^2(\zeta).
\label{chap10:Veff2}
\end{equation}
%
Thus, similarly to the bright soliton case, one can use the above pinning of the
dark soliton to drag it by adiabatically moving the impurity \cite{fr1}.

Finally, it is also possible to pin and drag vortices with
a localized impurity (see Chap.~17 in Ref.~\cite{BECBOOK}
for more details). 
As discussed above, the presence of the harmonic trap induces
vortex precession 
\cite{precession1,precession2,precession3,precession4,precession5,precession6,precession7,precession8}.
Therefore, in order to pin/drag the vortex at an off-center position,
the pinning force exerted by the impurity has to be stronger
than the vortex precession force induced by the harmonic trap
and the impurity has to be deep enough to avoid emission of
sound waves \cite{nol1}. 
An interesting twist to this manipulation problem is the
possibility to snare a moving vortex by an appropriately 
located/designed impurity. 

\subsubsection{Periodic potentials.}

A similar approach as the one described in the previous section can
be devised by using periodic potentials. This method relies, as in
the case of localized impurities, on the pinning properties of
properly crafted periodic potentials.
Specifically, a periodic potential with a wavelength 
longer than the width of the soliton width induces
a periodic series of effective potential minima that
help to pin the soliton.
For instance, in a 1D BEC, a bright soliton subject to a periodic
potential generated by an optical lattice of the form
\begin{equation}
V_{\rm OL}(x) =V_{\mathrm{\rm OL}}^{(0)}\sin^{2}(kx),
\label{1DOL}
\end{equation}
%
behaves (in appropriate parameter ranges) as a quasi-particle inside the
effective potential \cite{st2}: 
%
\begin{equation}
V_{\rm eff}(\zeta) = \eta \Omega^2 \zeta^2 -
\pi V_{\mathrm{\rm OL}}^{(0)} k\, \csch(k\pi/\eta) \cos(2k\zeta).
\label{chap10:Veff3}
\end{equation}
This effective potential possesses minima that, as indicated above,
 can be rigorously
shown to correspond to stable positions for the solitons 
\cite{kks,grillakis1a,grillakis1b,grillakis3a,grillakis3b,pelinovsky,oh1,oh2}.
This in turn can be used, in the
same manner as above for the localized impurities, to
successfully pin, drag and capture bright solitons \cite{st2}.

The case of manipulation of dark solitons by periodic potentials
is more subtle because dark solitons subject to tight
confinements are prone to weak radiation loss, as shown numerically in  
Refs.~\cite{Proukakis1,Proukakis2,Proukakis3,Proukakis4} and 
analytically in Ref.~\cite{fr5}.
Since a dark soliton is an effectively negative mass (density void) 
structure, radiation loss implies that the soliton travels faster
and eventually escapes the local minimum in the effective potential
landscape.
The motion of a dark soliton subject to an external potential
has been treated in detail before \cite{fr1,fr2,huang1,fr3,fr4,fr5,PGK:PRA:03}. 
In the presence of both the magnetic trap and the optical lattice of
Eq.~(\ref{1DOL}),
the motion of the dark soliton depends on the period of the optical lattice
when compared to the soliton's width \cite{dsolA}.
The case of an optical lattice with long-period can be treated as a perturbation
(see Sec.~\ref{nlwd}),
and the dynamics of the dark soliton can adjust itself to the smooth potential.
The short period optical lattice case can be treated by multiple-scale
expansion \cite{dsolA} and it is equivalent to the motion of a dark soliton
inside a renormalized magnetic trap (with no optical lattice) \cite{dsolA}.
For intermediate periods, the optical lattice has the ability to drag/manipulate
the dark soliton as shown in Refs.~\cite{dsolA,gol}. 
However, the effects of radiation loss described above
eventually drive the dark soliton into 
large amplitude oscillations
inside the local effective potential minima leading, eventually, to
its expulsion.

Finally, we briefly discuss vortices under the presence of
periodic potentials. Their stability and dynamics has been 
studied in Refs.~\cite{PGK:JPB:03,morsch1}, while the existence
of gap vortices in the gaps of the band-gap spectrum due to Bragg 
scattering were considered in Ref.~\cite{ost}.
Also, in the presence of deep periodic lattices, where
the discrete version of the GP equation 
(namely the DNLS equation, see, e.g., Sec.~\ref{Sec:DNLS})  
becomes relevant, purely 3D discrete 
vortices can be constructed \cite{discrvort1,discrvort2,discrvort3}.
An interesting twist of the pinning of vortices (which has been observed experimentally \cite{vorpinexp})
is the case when a vortex lattice 
is induced to transition from a triangular Abrikosov vortex lattice
to a square lattice by an optical lattice \cite{nol2,PuBigelow:05}.
Another type of vortex lattice manipulation is the use of 
large-amplitude oscillations to induce structural phase transitions
(e.g., from triangular to orthorhombic) \cite{chap01:latt3}. 


\subsection{Matter-waves in disordered potentials \label{Sec:disordered}}

Recently, there has been much attention focused on the dynamics of matter-waves in disordered potentials. 
Generally speaking, disorder in quantum systems has been a subject of intense theoretical 
and experimental studies.
In the context of ultracold atomic gases disorder may result from the roughness of a magnetic trap \cite{dismt} 
or a magnetic microtrap \cite{dwwang}. However, it is important to note that {\it controlled} 
disorder (or quasi-disorder) may also be created by means of different techniques. These include the use 
of two-color superlattice potentials \cite{rb1,rb2,rb3}, the employment of so-called quasi-crystal 
(i.e., quasi-periodic) optical lattices in 2D or 3D \cite{qcr1,qcr2,qcr3}, the use of impurity atoms 
trapped at random positions in the nodes of a periodic optical lattice \cite{rscat}, random phase masks \cite{raphma}, 
or optical speckle patterns \cite{lens1,clem,lens2}. The latter is a random intensity pattern which is produced 
by the scattering of a coherent laser beam from a rough surface (see, e.g., Ref.~\cite{njpsp} for a detailed discussion). 

The theoretical and experimental investigations of ultracold atomic gases confined in disordered potentials 
pave the way for the study of fundamental effects in quantum systems. Among them, the most famous phenomenon is 
Anderson localization, i.e., localization and absence of diffusion of non-interacting quantum particles, which was 
originally predicted to occur in the context of electronic transport \cite{pwa}. Other important effects, include 
the realization of Bose \cite{bogl1,bogl2} or Fermi \cite{fegl1,fegl2} glasses, quantum spin glasses \cite{qspingl}, 
the Anderson-Bose glass and crossover between Anderson glass to Bose-glass localization, and so on (see 
also the recent review \cite{mlas}). Importantly, as we will discuss below, there exist very recent 
relevant experimental results (and theoretical predictions) towards these interesting directions.

The first experimental results concerning BECs confined in disordered potentials were reported almost 
simultaneously on 2005 \cite{raphma,lens1,clem,lens2}. In the first if these experiments \cite{lens1}, 
static and dynamic properties of a rubidium BEC were studied and it was found that both dipole and quadrupole oscillations 
are damped (the damping was found to be stronger as the speckle height was increased). The suppression of transport 
of the condensate in the presence of a random potential was also reported in Refs.~\cite{clem,lens2}. In these works, 
a strong reduction of the mobility of atoms was demonstrated, as the 1D expansion of the elongated condensate 
along a magnetic \cite{clem} or an optical \cite{lens2} guide was found to be strongly inhibited in the 
presence of the speckle potential. On the other hand, in Ref.~\cite{raphma} a speckle pattern 
was superimposed to a regular 1D optical lattice and, thus, a genuine random potential was created. 
In this setting the possibility of observation of Anderson localization was analyzed in detail, and the crossover from 
Anderson localization in the absence of interactions to the ``screening regime'' (where nonlinear interactions 
suppress Anderson localization in the random potential) was investigated. 

Although the above mentioned results underscore a disorder-induced trapping of the condensate, 
this effect does not correspond to a genuine Anderson localization: for the latter, the correlation length of 
the disorder has to be smaller than the size of the system (see discussion in Ref.~\cite{mlas}). Nevertheless, 
a detailed study \cite{kuhn} has shown that expansion of a quasi-2D cloud may lead to weak and even 
strong localization using currently available speckle patterns. Other works have revealed that 
Anderson localization may occur during transport processes in repulsive BECs \cite{spprl,paul,lens3}. In this case, however, 
and for condensates at equilibrium, the interaction-induced delocalization dominates disorder-induced localization, 
except for the case of weak interactions \cite{wint} (see also earlier work in Ref.~\cite{dkklee}). Another 
possibility is Anderson localization of elementary excitations in interacting BECs, as analyzed in the 
recent works \cite{bipav1,lugan}. Finally, it is worth also
mentioning in passing parallel developments in this area, within the
mathematically similar setting of photonic
lattices, where Anderson localization and transition from ballistic
to diffusive transport were recently observed in the presence
of random fluctuations \cite{fishman}. 

In any case, the above discussion shows that there exists an intense theoretical and experimental effort concerning 
this hot topic of BECs in disordered potentials. Although it seems that relevant experiments have just started, the 
perspectives are very promising: they include not only the possibility of the detection of Anderson localization, but also 
other relevant effects, such as the observation of a Bose-glass phase \cite{lens4}, the possibility of the appearance of a novel 
Lifshits glass phase \cite{lugan2}, and so on.

\subsection{Beyond mean-field description\label{Sec:beyond}}

The GP equation has been extremely successful in describing a wide range of
mean-field phenomena in Bose-Einstein condensation. By construction,
as explained in detail in Sec.~\ref{Sec:GPE}, the
GP equation is the mean-field description of the multi-body quantum
Hamiltonian describing the interaction of a dilute gas of bosonic atoms and it
relies on two main assumptions: 
(a) collisions between atoms are approximated by hard sphere
   collisions with a Dirac delta interatomic potential, and
(b) the gas is at absolute zero temperature where thermal effects 
   are not present. 
Nonetheless, in many BEC settings,
finite temperature effects and quantum fluctuations may play an important
role. The main effect of finite temperature is due to the fact
that a part of the atomic cloud is not condensed (the so-called thermal
cloud) and couples to the condensed cloud.
A microscopic derivation of the mean-field operator for the gas
of bosonic particles reveals that the (standard) condensate mean-field
is coupled to higher order mean-fields (cf.~the insightful review in
Chap.~18 of Ref.~\cite{BECBOOK} and references therein for more details).
Neglecting the exchange of particles between condensed and
non-condensed atoms and taking into account the three lowest 
mean-field orders, the so-called Hartree-Fock-Bogoliubov (HFB)
theory, leads to the generalized GP equation for the
condensate wavefunction $\psi({\bf r},t)$
\cite{HFBEarly1,HFBEarly2,HFBEarly3,HFBEarly4}
\begin{equation}
i \hbar \frac{\partial}{\partial t} \psi({\bf r},t) = \left[ 
H_{\rm GP}
+ g  \left[  2 n'({\bf r},t) \right] \right] \psi({\bf r},t)
+ \tilde{m}_{0}({\bf r},t) \psi^{\ast}({\bf r},t),
\end{equation}
where 
$H_{\rm GP}=-(1/2) \nabla^2 + V_{\rm ext}({\bf r},t) + g|\psi|^2$ 
accounts for the ``classical'' GP terms in non-dimensional units, 
$n'({\bf r},t)$ denotes the non-condensate density and 
$\tilde{m}_{0}({\bf r},t)$ the anomalous mean-field average 
\cite{GriffinHFB,FetterWalecka1,FetterWalecka2,OURREVIEW}.
Furthermore, taking into account that atomic collisions 
happen within the gas (and not in vacuum), one has to modify
the inter-atomic interactions by a contact potential
with a position-dependent amplitude:
$g\rightarrow g [ 1 + \tilde{m}_{0}({\bf r})/\psi^{2}({\bf r}) ]$
\cite{ProukakisGHFB,BurnettLectureNotes,HutchinsonGHFB}.

A similar approach to the above
is to consider the coupling with the thermal
cloud by ignoring the anomalous average and including
a local energy and momentum conservation 
\cite{ZNG,KirkpatrickDorfman1} that yields
\begin{equation}
i \hbar \frac{\partial}{\partial t} \psi({\bf r},t) = 
\left[{H}_{\rm GP} + g \left[ 2 n'({\bf r},t) \right]
- i R({\bf r},t) \right] \psi({\bf r},t), \label{FTGPE}
\end{equation}
where the non-condensate density $n'({\bf r},t)$ is described
terms of a Wigner phase-space representation and a
generalized Boltzmann quantum-hydrodynamical equation
(see Chap.~18 of Ref.~\cite{BECBOOK}). This term includes
collisions between non-condensed atoms and transfer to and
from the condensed cloud. In Eq.~(\ref{FTGPE}), $R({\bf r},t)$
accounts for triplet correlations.

Other approaches to incorporate the thermal cloud include
Stoof's non-equilibrium theory based on quantum kinetic theory 
\cite{StoofJLTP,StoofSummaryPaper,StoofLesHouches,StoofDuine}
and Gardiner-Zoller's quantum kinetic master equation using
techniques borrowed from quantum optics 
\cite{QKTheory1,QKTheory2,QKTheory3,QKTheory4,QKTheory5,QKTheory6,QKTheory7}.
Also, efforts to include stochastic effects have been considered
in Ref.~\cite{StoofStochastic} by considering the thermal cloud
to be close enough to equilibrium that it can be described by
a Bose cloud with chemical potential $\mu_{T}$ at temperature $T$.
In fact, this approach leads, after neglecting noise terms, to the phenomenological
damping included in the GP equation through the damped GP equation
\begin{equation}
i \hbar \frac{\partial}{\partial t} \psi({\bf r},t) = 
( 1 - i \gamma) \left( {H}_{GP} - \mu_{T} \right) \psi({\bf r},t), \label{DampedGPE}
\end{equation}
with damping rate $\gamma>0$. This approach was originally
proposed by Pitaevskii \cite{PitaevskiiPhenomenology} and applied with
a position-dependent loss rate in Ref.~\cite{ChoiPhenomenology}.
A similar phenomenological damping term (but on the left hand side) has been used in 
Refs.~\cite{Ueda02,Ueda03}
to remove the excess energy to dynamically simulate the crystallization
of vortex lattices in rapidly rotating BECs.

It is important to note that recently there has been a lot of activity on the 
study of formation, stability and dynamics of matter-wave solitons beyond the mean-field theory. 
In particular, in the case of attractive BECs, the velocity and temperature-dependent 
frictional force and diffusion coefficient of a matter-wave bright soliton immersed 
in a thermal cloud was calculated in Ref.~\cite{sinha}. Moreover, 
the full set of the time-dependent HFB equations was used in Ref.~\cite{segevbec} 
to show that the matter flux from the condensate to the thermal 
cloud may cause the bright matter-wave solitons to split into two solitonic fragments 
(each of them is a mixture of the condensed and non-condensed particles); these may be 
viewed as partially {\it incoherent solitons}, similar to the ones known in the context 
of nonlinear optics \cite{incsol1,incsol2}. 
Partially incoherent {\it lattice solitons} at a finite temperature $T$ were also predicted to exist 
in Ref.~\cite{ver}; there, a numerical integration of the GP equation, starting with a random Bose 
distribution at finite $T$, revealed the generation of lattice solitons upon gradual switch-on of an 
optical lattice potential. In a more recent study \cite{borisft}, the time-dependent HFB theory was 
used to find inter-gap and intra-gap partially incoherent solitons (composed, as in Ref.~\cite{segevbec} 
by a condensed and a non-condensed part). On the other hand, in 
Refs.~\cite{nppds1,nppds2}, the so-called Zaremba-Nikuni-Griffin formalism \cite{ZNG} was used to 
analyze the dissipative dynamics of dark solitons in the presence of the thermal cloud, as observed in the 
Hannover experiment \cite{chap01:dark1}. Finally, other quantum effects associated to matter-wave solitons have 
been considered as well; these include quantum depletion of dark solitons \cite{quadep1,quadep2,quadep3,quadep4} 
(see also Ref.~\cite{chap01:cqnlsC}),  
formation of a broken-symmetry bright matter-wave soliton 
by superpositions of quasi-degenerate many-body states \cite{kanamoto}, 
quantum-noise squeezing and quantum correlations of gap solitons \cite{leeost}, and so on.

A considerable volume of work in many of the directions
mentioned in this review has continued to emerge both in
preprint and in published form. As two examples thereof,
we can mention the work of Refs.~\cite{ShapiroA,ShapiroB,ShapiroC}
on various higher
dimensional aspects of BECs in random potentials and transitions
associated with Anderson localization (we thank Dr.~B.~Shapiro
for bringing these works to our attention) and the intense
experimental activity on the oscillations and interactions
of dark and dark-bright solitons in Refs.~\cite{Weller,Becker}.

\vspace{10mm}

\vspace{5mm}

{\bf Acknowledgments}

This review would simply not have existed without 
the invaluable contribution of 
our collaborators in many of the topics discussed here: 
Egor Alfimov, Brian Anderson,
Alan Bishop, Lincoln Carr, Zhigang Chen, Fotis Diakonos, Peter Engels,
David Hall, Todd Kapitula,
Yannis Kevrekidis, 
Yuri Kivshar, Volodya Konotop, Boris Malomed,
Kevin Mertes, Alex Nicolin,
Hector Nistazakis, Markus Oberthaler, Dmitry Pelinovsky, Mason Porter,
Nick Proukakis, Mario Salerno, Peter Schmelcher, 
Augusto Smerzi, Giorgos Theocharis 
and Andrea Trombettoni are greatly thanked. 
RCG and PGK both acknowledge support from NSF-DMS-0505663.
PGK is also grateful to the NSF for support through the grants 
NSF-DMS-0619492 and  NSF-CAREER. The work of DJF is partially 
supported by the Special Research Account of the University of Athens.


\end{document}